\documentclass[prd,showpacs,showkeys,floatfix,nofootinbib,eqsecnum,
                  preprint,12pt,tightenlines,fleqn]{revtex4}  
 
\usepackage{amsmath,amssymb,revsymb,graphicx,dcolumn}

\begin{document}
\title{Spin Superfluidity and Magnon BEC}
\author{
Yu.M.~Bunkov$^{(a)}$,   G.E.~Volovik$^{(b,c)}$
}
\affiliation
{$^{(a)}$Institute N\'{e}el, CNRS, Grenoble, France\\
$^{(b)}$Low Temperature Laboratory, Aalto University, Finland\\
$^{(c)}$L.D. Landau Institute for Theoretical Physics, Moscow, Russia}

\date{\today}
\begin{abstract}
 The superfluid current of spins -- spin supercurrent -- is one more representative of
superfluid currents, such as 
the superfluid current of mass and atoms in superfluid
$^4$He; superfluid current of electric charge in superconductors;
superfluid current of hypercharge in Standard Model of particle physics; superfluid
baryonic current and current of chiral charge in quark
matter; etc.  The spin superfluidity is manifested as the spontaneous phase-coherent precession of spins first discovered in 1984 in $^3$He-B, and can be described in terms of the Bose condensation
of spin waves -- magnons. We discuss different phases of magnon superfluidity, including those in magnetic trap;
and  signatures of  magnon superfluidity: (i) spin supercurrent,  which transports the magnetization on a macroscopic distance (as long as 1 cm); (ii) spin current Josephson effect which shows interference between two condensates;  (iii) spin current vortex  --  a topological defect which is an analog of a quantized vortex in superfluids, of an Abrikosov vortex in superconductors, and cosmic strings in relativistic theories; (iv) Goldstone modes related to the broken $U(1)$ symmetry -- phonons in the spin-superfluid magnon gas; etc. We also touch the topic of spin supercurrent in general including
spin Hall and intrinsic quantum spin Hall effects.
 
\end{abstract}


\maketitle
\tableofcontents

 \section{Abbreviations}

NMR -- Nuclear Magnetic Resonance

RF -- Radio-Frequency field (alternating magnetic field with NMR frequency)

FMR, AFMR -- Ferromagnetic Resonance, Antiferromagnetic Resonance.

CW NMR -- Continuous Wave NMR (continuous pumping by RF field).

Pulsed NMR -- excitation of NMR by Pulsed RF field 

ODLRO -- off-diagonal long-range order

BEC -- Bose-Einstein condensation

HPD -- Homogeneously Precessing Domain, the domain with coherently precession of magnetization.

\section{Introduction}

Nature knows different types of ordered states. 

One major class is represented  by equilibrium macroscopic ordered
sta\-tes exhibiting spontaneous  breaking of symmetry.  This class
contains crystals; nematic, cholesteric and other liquid
crystals;  different types of ordered magnets
(antiferromagnets,  ferromagnets, etc.); superfluids,
superconductors and Bose condensates; all types of Higgs fields
in high energy physics; etc. The important sub\-class\-es of this
class contain  systems with ma\-croscopic quantum coherence
exhibiting off-diagonal long-range order (ODLRO),  and/or
nondissipative superfluid currents (mass current,
spin current, electric current, hypercharge current, etc.). The
class of ordered systems is characterized by rigidity, stable
gradients of  order parameter (non-dissipative currents in quantum
coherent systems), and  topologically stable defects (vortices,
solitons, cosmic strings, monopoles, etc.).

A second large class is presented by  dynamical
systems out of equilibrium. Ordered states may emerge under
external flux of energy. Examples are the coherent emission from
lasers; water flow in a draining bathtub; pattern formation in
dissipative systems; etc. 

Some of the latter dynamic systems can be close   to stationary
equilibrium systems of the first class. For example, ultra-cold
gases in optical traps are not fully equilibrium states since
the number of atoms in the trap is not conserved, and thus the
steady state requires pumping. However, if the decay is small
then the system  is close to an equilibrium Bose
condensate, and experiences all the corresponding superfluid
properties. 

 \subsection{BEC of quasiparticles} 

Bose-Einstein condensation (BEC) of quasiparticles  whose
number is not conserved is presently one of the debated phenomena of condensed matter physics. In thermal equilibrium the chemical potential of excitations vanishes and, as a result, their condensate does not form. The only way to overcome this situation is to create a non-equilibrium but dynamically
steady state, in which the number of excitations is  conserved,  since the loss of quasiparticles owing to their decay is compensated by pumping of energy.
Thus the Bose condensation of 
quasiparticles belongs to the phenomenon of second class, when the
emerging steady state of the system is not in a full thermodynamic
equilibrium. 

Formally BEC  requires  conservation of charge or particle number. However, condensation can still be extended to systems with weakly violated conservation. For sufficiently long-lived quasiparticles 
their distribution may be close
to the thermodynamic equilibrium with a well defined finite chemical potential,
which follows from the quasi-conservation of  number of quasiparticles, and the Bose condensation becomes possible.  Several examples of Bose condensation of quasiparticles have been observed or suggested,  including phonons \cite{Kagan2003},  excitons \cite{Butov2001}, exciton-polaritons \cite{Kasprzak2006}, photons \cite{Klaers2011} and rotons \cite{Melnikovsky2011}. The BEC of quasi-equilibrium magnons  -- spin waves -- in ferromagnets has been discussed in Ref. \cite{KalafatiSafonov}  and investigated in 
\cite{Demokritov,Demidov2008,Chumak2009}.

In this review we consider the BEC of magnons and Spin Superfluidity. Magnons are  magnetic excitations in magnetic materials, such as  magnetically ordered systems, like ferromagnets, antiferromagnets, etc., and paramagnetic systems with external magnetic ordering such as  Fermi liquids. The most suitable systems for study the phenomenon of magnon BEC are superfluid phases of helium-3. The absolute purity, long lifetime of magnons, different types of magnon-magnon interactions, well controlled magnetic anisotropy makes antiferromagnetic superfluid phases of $^3$He a basic laboratory of magnon BEC. The first BEC state of magnons was discovered in 1984 in $^3$He-B as a coherent  spin  precession, and it
was baptized as  Homogeneously Precessing Domain (HPD) \cite{HPDexp,HPDtheory}.
 This is the spontaneously  emerging steady state of precession, which 
preserves the phase coherence across the whole sample even in
an inhomogeneous external magnetic field and even in the absence
of energy pumping. This is equivalent to the appearance of a coherent superfluid Bose-Einstein condensate. 

In the absence of energy pumping this HPD state slowly decays,  
but during the decay the system remains in the coherent state of BEC: 
the volume of the Bose condensate (the volume
of HPD)  gradually decreases with time without violation of the
observed properties of the spin-superfluid phase-coherent
state. A ste\-a\-dy state of phase-coherent precession can be
supported by pumping. In particular, the coherence of electron spins 
induced by periodic pumping has been observed in ensemble of (In,Ga)As/GaAs  quantum dots 
\cite{Greilich2007}.  But in case of magnon BEC, the pumping needs not be coherent -- 
it can be chaotic: the system chooses its own (eigen)
frequency of coherent precession, which emphasizes the
spontaneous emergence of coherence from chaos. 

HPD is very close to the  thermodynamic equilibrium of the magnon 
Bose condensate and exhibits all the superfluid properties
which follow from the off-diagonal long-range order (ODLRO) of the coherent precession. 
After discovery of HPD, several other states of magnon BEC have been observed in superfluid phases
of $^3$He, which we discuss in this review, including finite magnon BEC states in magnetic traps.
The very similar BEC states was observed recently in an antiferromagnets with the so-called Suhl-Nacamura interaction, the Long Range Nuclear-Nuclear interaction via the magnetically ordered electronic subsystem \cite{Kazan2011,Kazan2012}.

\subsection{Spin superfluidity vs superfluidity of mass and charge}
\label{Spin superfluidity}

Last decade was marked by the fundamental  studies  of mesoscopic
quantum states of dilute ultra cold atomic gases in the regime where the de Broglie
wavelength of the atoms is comparable with their spacing, giving
rise to the phenomenon of Bose-Einstein condensation (see reviews \cite{revBEC,PitaevskiiStringari2003}). 
The formation of the Bose-Einstein condensate (BEC) 
-- accumulation of the macroscopic number of particles in the lowest energy state -- 
was predicted by Einstein in 1925 \cite{Einstein1925}. 
In ideal gas, all atoms are in the lowest energy state in the zero temperature limit. In dilute atomic gases, weak interactions between atoms produces a small fraction of the non-condensed atoms. 

In the only known bosonic  liquid $^4$He which remains liquid at zero temperature, the BEC is strongly modified by interactions. The depletion of the condensate due to interactions is very strong: in the limit of zero temperature only about 10\%  of particles occupy the state with zero momentum.
Nevertheless, BEC still remains the key mechanism for the phenomenon of superfluidity in liquid 
$^4$He: due to BEC  the whole liquid (100\% of $^4$He atoms) forms a coherent quantum state at $T=0$ and participates in the non-dissipative superfluid flow.

Superfluidity is a very general quantum property of  matter at low
temperatures, with variety of mechanisms and possible nondissipative superfluid
currents. These include supercurrent of electric charge in superconductors and
mass supercurrent in superfluid  $^3$He, where the mechanism
of superfluidity is the Cooper pairing; 
hypercharge supercurrent in the vacuum of Standard Model of elementary 
particle physics, which comes from the Higgs mechanism;
supercurrent of color charge in a dense quark matter in quantum chromo-dynamics; etc.
All these supercurrents have the same origin: the spontaneous breaking of the  $U(1)$
or higher symmetry related to the conservation of the corresponding charge or particle
number, which leads to the so called off-diagonal long-range order.

This spin supercurrent -- the superfluid current of spins -- is one more representative of
superfluid currents. Here the $U(1)$ symmetry is the approximate symmetry of spin rotation, which 
is related to the quasi-conservation of spin. It appears that the finite life-time of magnons, and non-conservation of spin due to the spin-orbital coupling do not prevent the coherence and superfluidity of magnon BEC in $^3$He-B. The non-conservation leads to a decrease of the number of magnons  until the HPD disappears completely, but during this relaxation,  the coherence  of magnon BEC is preserved with all the signatures of  spin superfluidity: (i) spin supercurrent,  which transports the magnetization on a macroscopic distance more than 1 cm long; (ii) spin current Josephson effect which shows interference between two condensates; (iii) phase-slip processes at the critical current; (iv) spin current vortex  --  a topological defect which is an analog of a quantized vortex in superfluids, of an Abrikosov vortex in superconductors, and cosmic strings in relativistic theories; (v) Goldstone modes related to the broken $U(1)$ symmetry -- phonons in the spin-superfluid magnon gas; etc.

 \subsection{Magnon BEC vs equilibrium magnets}\label{BECQP}

The magnetic $U(1)$ symmetry is spontaneously broken also in some static magnetic systems.
Sometimes this symmetry breaking is described in terms of BEC of magnons 
\cite{mag1,mag2,mag3,Giamarchi,KaulMathew2011}.
Let us stress from the beginning that there is  the principal difference between the magnetic ordering in equilibrium and  the  BEC of quasiparticles which we are discussing in this review.

In these magnetic systems, the symmetry breaking phase transition starts when the system becomes softly unstable towards the growth of one of the magnon modes.  The condensation of this mode 
can be used for the description of the soft mechanism of
formation of ferromagnetic and antiferromagnetic states (see
e.g.
\cite{Nikuni}). 
However, the final outcome of the condensation is the true equilibrium ordered state.  
In the same manner, the Bose condensation of phonon modes may serve as a soft mechanism of formation of  the equilibrium solid crystals \cite{Kohn}. 
But this does not mean that the final crystal state is the  Bose condensate of phonons.

On the contrary,  BEC of quasiparticles is in principle a non-equilibrium phenomenon, since quasiparticles (magnons) have a finite life-time. In our case magnons live long enough to form a state very close to thermodynamic equilbrium BEC, but still it is not an equilibrium. In the final equilibrium state at $T=0$ all the magnons will die out. In this respect, the growth of a single mode in the non-linear process after  a hydrodynamic instability \cite{LandauHydro}, which has been
discussed in terms of the  Bose condensation  of the
classical sound or surface waves  
\cite{Zakharov}, is more close to magnon BEC than equilibrium magnets with 
spontaneously broken $U(1)$ symmetry. 

The other difference is that the ordered magnetic states are states with diagonal long-range order. The magnon BEC is a dynamic state characterized by the off-diagonal long-range order (see Sec. \ref{PlanarFerromagnet} below), which is the main signature of spin superfluidity.

\section{Coherent precession as magnon superfluid}
 
\subsection{Spin  precession}

The magnetic subsystem which we discuss is the precessing magnetization. In a full correspondence with atomic systems, the precessing spins can be either in the normal state or in the ordered spin-superfluid state. In the normal state, spins of atoms are precessing with the local frequency determined by the local magnetic field and interactions. In the ordered state the precession of all spins is coherent: they spontaneously develop the common global frequency and the global phase of precession.

\begin{figure}[htt]
 \includegraphics[width=1\textwidth]{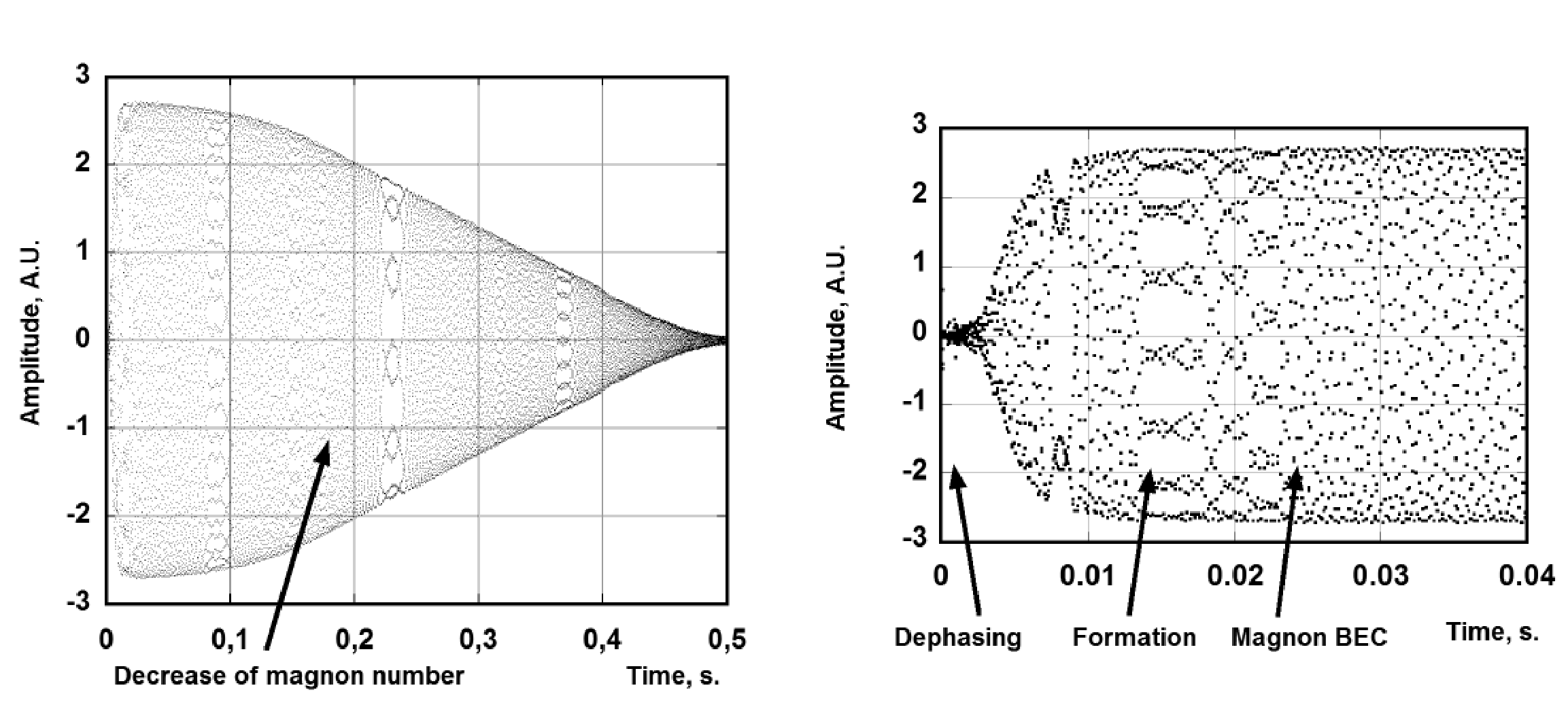}
 \caption{
 The stroboscopic record of  the induction decay signal on a frequency about 1 MHz.
{\it left}: During the first stage of about 0.002 s  the induction signal completely disappears due to dephasing. Then, during about 0.02 s, the spin supercurrent redistributes the magnetization and creates the phase coherent precession, which is equivalent to the magnon BEC state. Due to small magnetic relaxation, the number of magnons slowly decreases but the precession remains coherent.
  {\it right}: The initial part of the magnon BEC signal.
 }
 \label{amplitude}
\end{figure}
 
In pulsed NMR experiments in $^3$He-B the magnetization is created by an applied static magnetic field:   ${\bf M}=\chi {\bf H}$, where $\chi$ is magnetic susceptibility. Then a pulse of the radio-frequency (RF) field ${\bf H}_{\rm RF}\perp  {\bf H}$ deflects the magnetization by an angle $\beta$, and after that the induction signal from the free precession is measured. In the state of the  disordered precession, spins almost immediately loose the information on the original common  phase and frequency induced by the RF field, and due to this decoherence the measured induction signal is very short, of order $1/\Delta \nu$ where $\Delta \nu$ is the line-width coming from the inhomogeneity of magnetic field in the sample.
In the BEC state, all spins precess coherently, which means that the whole macroscopic magnetization of the sample of volume $V$ is precessing.  
\begin{equation}
{\cal M}_x +i {\cal M}_y ={\cal M}_{\perp} e^{i\omega t+i\alpha} ~,~{\cal M}_{\perp}=\chi H V \sin\beta~.
\label{MagnetizationPrecession}
\end{equation}
This coherent precession is manifested as a huge and long-lived induction signal, Fig. \ref{amplitude}. 
It is important that the coherence of precession is spontaneous: the global frequency and the global phase of precession are formed by the system itself and do not depend on the frequency and phase of the initial RF pulse.

\subsection{Off-diagonal long-range order}

The superfluid atomic systems are characterized by the off-diagonal long-range order (ODLRO) \cite{Yang}.
In superfluid $^4$He and in the coherent atomic systems the operator
of annihilation of atoms with momentum ${\bf p}=0$ has a non-zero vacuum expectation value:
\begin{equation}
\left<\hat a_0 \right>={\cal N}_0^{1/2}  e^{i\mu
t+i\alpha}~,
\label{ODLRO}
\end{equation}
where ${\cal N}_0$ is the number of particles in the Bose condensate, which in the limit of weak interactions between the atoms  coincides at $T=0$ with the total number of atoms ${\cal N}$.

Eq. \eqref{MagnetizationPrecession} demonstrates that in the coherent precession the ODLRO is 
manifested by a non-zero vacuum expectation value of the operator
of creation of spin: 
\begin{equation}
\left<\hat S_+ \right>={\cal S}_x +i {\cal S}_y =\frac {{\cal M}_\perp}{\gamma} e^{i\omega
t+i\alpha}~,
\label{spinODLRO}
\end{equation}
where $\gamma$ is the gyromagnetic ratio, which relates magnetic moment and spin.
This analogy suggests that in the coherent spin precession the role of the particle number  ${\cal N}$ is played by the projection of the total spin on the direction of magnetic field ${\cal S}_z$. The corresponding symmetry group $U(1)$ in magnetic systems is the group of the $O(2)$ rotations about the direction of magnetic  field. This quantity ${\cal S}_z$ is conserved in the absence of the spin-orbit interactions. 

The spin-orbit interactions transform the spin angular momentum of the magnetic subsystem to the orbital angular momentum, which causes the losses of spin ${\cal S}_z$ during the precession.  In  superfluid $^3$He, the spin-orbit coupling is relatively rather small, 
and thus  ${\cal S}_z$ is quasi-conserved.  Because of the losses of spin the precession will finally decay, but during its long  life time the precession remains coherent, Fig. \ref{amplitude}. 
This is similar to the non-conservation of the number of atoms in the laser traps: though the number of atoms decreases with time  due to evaporation, this does not destroy the coherence of the remaining atomic BEC. 

\subsection{ODLRO and magnon BEC}

The ODLRO in \eqref{spinODLRO} can be represented in terms of magnon condensation. To view that let us use the Holstein-Primakoff transformation, which relates the spin operators with the operators of creation and annihilation of magnons 
\begin{eqnarray}
\hat a_0~\sqrt{1-\frac{\hbar a^\dagger_0 a_0}{2{\cal S}}}= \frac{\hat {\cal S}_+}{\sqrt{2{\cal S}\hbar}}~,
\label{MagnonAnnih}
\\
\sqrt{1-\frac{\hbar a^\dagger_0 a_0}{2{\cal S}}}~\hat a^\dagger_0=
\frac{\hat {\cal S}_-}{\sqrt{2{\cal S}\hbar}}~,
\label{MagnonCreation}
\\
\hat {\cal N}=\hat a^\dagger_0\hat a_0 = 
\frac{{\cal S}-\hat{\cal S}_z}{\hbar}~.
\label{MagnonNumberOperator}
\end{eqnarray}
Eq. \eqref{MagnonNumberOperator} relates the number of magnons ${\cal N}$ to the deviation of spin ${\cal S}_z$ from its equilibrium value ${\cal S}_z^{({\rm equilibrium})}={\cal S}=\chi H V/\gamma$. In the full thermodynamic equilibrium,  magnons are absent (in $^3$He-B thermal magnons can be ignored, see Sec. \ref{Magnon spectrum and magnon mass}). Each magnon has spin $-\hbar$,  and thus the total spin projection after pumping of ${\cal N}$ magnons into the system by the RF pulse is reduced by the number of magnons, ${\cal S}_z ={\cal S} - \hbar {\cal N}$. The ODLRO in magnon BEC is given by Eq. \eqref{ODLRO}, where ${\cal N}_0={\cal N}$ is the total number of magnons  \eqref{MagnonNumberOperator} in the BEC:
\begin{equation}
\left<\hat a_0 \right>={\cal N}^{1/2}  e^{i\omega
t+i\alpha}=\sqrt{\frac {2{\cal S}}{\hbar}} ~\sin\frac{\beta}{2} ~e^{i\omega
t+i\alpha}\,.
\label{ODLROmagnon}
\end{equation}

Comparing \eqref{ODLROmagnon} and \eqref{ODLRO}, one can see that the role of the global chemical potential in atomic systems $\mu$ is played by the global frequency of the coherent precession $\omega$, i.e. $\mu\equiv \omega$. This demonstrates that this analogy with the phenomenon of BEC in atomic gases takes place only for the dynamic states of a magnetic subsystem-- the states of precession. The ordered magnetic systems discussed in Refs. \cite{mag1,mag2,mag3,Giamarchi} are static, and for them the chemical potential of magnons is always zero. 

There are two approaches to study the thermodynamics of atomic systems:  at fixed particle number $N$ or at fixed chemical potential $\mu$. For the magnon BEC, these two approaches correspond to two different experimental arrangement: the pulsed NMR and continuous wave NMR, respectively. In the case of free precession after the pulse, the number of magnons pumped into the system is conserved (if one neglects  the losses of spin). This corresponds to the situation with the fixed ${\cal N}$, in which the system itself will choose the global frequency of the coherent precession (the magnon chemical potential). The opposite case is the continuous wave NMR, when a small RF field is continuously applied to compensate the losses. In this case the frequency of  precession is fixed by the frequency of the RF field, $\mu\equiv \omega = \omega_{\rm RF}$, and now the number of magnons will be adjusted to this frequency to match the resonance condition.

Finally let us mention that in the approach in which  ${\cal N}$ is strictly conserved and has quantized integer values,  the quantity $<\hat a_0> =0$ in Eq. (\ref{ODLRO}). In the same way the quantity $\left<\hat S_+ \right>=0$ in Eq. (\ref{spinODLRO}),
if spin ${\cal S}_z$ is strictly conserved and takes quantized values. This means that formally there is no precession if the system is in the quantum state with fixed spin quantum number ${\cal S}_z$. However, this does not lead to any paradox in the thermodynamic limit: in the limit of infinite ${\cal N}$ and ${\cal S}_z$,  the description in terms  of  the fixed ${\cal N}$ (or ${\cal S}_z$) is equivalent to the description in terms of with the fixed chemical potential $\mu$ (or frequency $\omega$).

\section{Phenomenology of magnon superfluidity}

We consider this phenomenology using $^3$He-B as an example.

\subsection{Magnon spectrum and magnon mass}

Let us neglect for a moment the anisotropy of spin wave velocity $c$ and the spin-orbit interaction. Then  the magnon spectrum in $^3$He-B has the following form:
\begin{equation}
 \omega(k)= 
  \frac{\omega_L}{2}   +\sqrt{ \frac{\omega_L^2}{4} +k^2c^2} \,,
 \label{MagnonSpectrumextended}
\end{equation}
where $\omega_L=\gamma H$.
At large momentum,  $ck \gg \omega$, this spectrum transforms to the linear spectrum $\omega=ck$ of spin waves propagating with velocity $c$ which is  on  the order  of the Fermi velocity $v_F$.
At small $k$, $ck \ll \omega$, 
this is the spectrum of massive particle
\begin{equation}
E_k=\hbar  \omega(k) ~,~\omega(k)= \omega_L   +\frac{\hbar k^2}{2m_M} ~~,~~ 
 ~,
\label{MagnonSpectrum3}
\end{equation}
where the magnon  mass is:
  \begin{equation}
m_M =\frac{\hbar\omega_L}{2c^2}~.
\label{IsotropicMass}
\end{equation}
Since  in $^3$He-B one has $c\sim v_F$,   the relative 
magnitude of the magnon mass compared to the bare mass $m_3$ of the $^3$He
atom is
  \begin{equation}
\frac{m_M}{m_3} \sim \frac{\hbar\omega_L}{E_F}~,
\label{MassRatio}
\end{equation} 
where the Fermi energy $E_F\sim m_3v_F^2\sim p_F^2/m_3$. 
With the magnon gap $\hbar\omega_L \sim 50~\mu K$ at  $\omega_L\sim 1$MHz, and 
$E_F \sim 1K$, one has $m_M \sim 10^{-4} m_3$.  Small mass of these bosons favors the Bose condensation.  The opposite factor is the small density $n$ of the bosons. 
From \eqref{MagnonNumberOperator} it follows that  the magnon density is
 \begin{equation}
n=\frac{S-S_z}{\hbar}=\frac{\chi H}{\hbar\gamma} (1-\cos\beta)~.
\label{DensityMagnons}
\end{equation} 
In the typical precessing state of $^3$He-B the magnon density $n$ is by the same factor $\hbar\omega_L/E_F$ smaller than the density of $^3$He atoms $n_3=p_F^3/3\pi^2\hbar^3$ in the liquid
     \begin{equation}
\frac{n}{n_3} \sim \frac{\hbar\omega_L}{E_F}~.
\label{DensityRatio}
\end{equation} 

  In $^3$He-B, the typical temperature $T\sim 10^{-3}E_F$. As we shall see below it is small compared to the temperature of Bose condensation, $T< T_{\rm BEC}$. However, $T$ is big compared to the magnon gap, $T\gg \hbar\omega_L$. But this does not make problem, this simply means that according to \eqref{MagnonSpectrumextended},
  thermal magnons are mostly the spin waves with linear spectrum $ \omega(k)=ck$ and with characteristic momenta $k_T\sim T/\hbar c$. The density of thermal magnons is $n_T \sim k_T^3$.  At $^3$He-B temperatures, this density is much smaller then the density of condensed magnons and can be neglected,  $n_T/n \sim T^3/E_F^2\omega_L \ll 1$. 
  
  The smallness of $\omega_L\ll T$ modifies the estimate the temperature of the Bose condensation, compared to the atomic gases. Before we start pumping magnons,  we have an equilibrium system of thermal magnons with $\mu=0$. After pumping of extra magnons with density $n$ we obtain the quasi-equilibrium state in which the number of magnons is temporarily conserved and thus the magnon system acquires a non-zero chemical potential, $\mu\neq 0$. The number of extra magnons which can be absorbed by thermal distribution without formation of BEC is thus the difference of the distribution function at $\mu=0$ and $\mu\neq 0$ at the same temperature:
   \begin{equation}
n= \sum_{\bf k} \left( f(E_k) - f(E_k-\mu)\right) \,.
\label{ExtraNumber}
\end{equation}  
This quantity reaches its maximum value when $\mu=\omega_L$. Since $\omega_L\ll T$ one has:
 \begin{equation}
n_{\rm max}=\omega_L\sum_{\bf k} \frac{df}{dE} \sim \frac{T^2 \omega_L}{c^3}\,.
\label{ExtraNumbeMax}
\end{equation} 
This gives the dependence of BEC transition temperature on the number of pumped magnons
 \begin{equation}
T_{\rm BEC} \sim \left(\frac{nc^3}{\omega_L} \right)^{1/2}  \,.
\label{CondensationT3}
\end{equation} 
At $T< T_{\rm BEC}$ the magnon BEC with $k=0$ must be formed.
In $^3$He-B, $T_{\rm BEC}\sim  E_F$ which is by 3 to 4 orders of magnitude higher than the temperature at which superfluid  $^3$He exists. As a result, in the coherently precessing state of 
$^3$He-B  
practically all the magnons are condensed in the ground state with $k=0$, with negligible amount of thermal magnons, i.e.
the Bose condensation of magnons in $^3$He-B  is almost perfect.

The above estimate also demonstrates that in solid state magnetic systems the BEC may occur even at room temperature, see Ref. \cite{Demokritov,Demidov2008}.

\subsection{Order parameter and Gross-Pitaevskii equation}

As in the case of the atomic Bose condensates the main physics of the magnon BEC can be found from the consideration of the Gross-Pitaevskii equation for the complex order parameter. 
The local order parameter is obtained by extension of Eq. \eqref{ODLROmagnon} to the inhomogeneous case and is determined as the vacuum expectation value of the magnon field operator:
\begin{equation}
\Psi({\bf r},t)=\left<\hat \Psi({\bf r},t)\right>~~,~~ n=\vert\Psi\vert^2~~,~~{\cal N}=\int d^3r ~\vert\Psi\vert^2~.\label{OrderParameter1}
\end{equation}
where $n$ is magnon density.  
To avoid the confusion, let us mention that this order parameter \eqref{OrderParameter1} describes the coherent precession in any system, superfluid or non-superfluid. It has nothing to do with the multi-component order parameter which describes the underlying systems -- superfluid phases of $^3$He
 \cite{VollhardtWolfle}. In other words the mass superfluidity of $^3$He is accompanied  by the antiferromagnetic ordering in the subsystem of nuclear spins. All the magnetic properties, which we are discussing here, are the properties of this magnitically ordered subsystem and are not connected directly to the mass superfluidity of  $^3$He.  

If the dissipation and pumping of magnons are ignored (on relaxation and pumping terms in magnon BEC see Ref. \cite{Malomed2009}), the corresponding Gross-Pitaevskii equation has the conventional form 
$(\hbar=1)$:
  \begin{equation}
-i \frac{\partial \Psi}{\partial t}= \frac{\delta {\cal F}}{\delta \Psi^*}~,
\label{GP}
\end{equation}
where ${\cal F}\{ \Psi\}$ is the free energy functional,  which plays the role of the effective Hamiltonian of the 
spin subsystem.
In the coherent precession, the global frequency is constant in space and time (if dissipation is neglected)
\begin{equation}
\Psi({\bf r},t)=\Psi({\bf r})  e^{i\omega t}~,
\label{OrderParameter2}
\end{equation}
and the Gross-Pitaevskii equation transforms into the Ginzburg-Landau equation with $\omega=\mu$:
  \begin{equation}
 \frac{\delta {\cal F}}{\delta \Psi^*}- \mu\Psi=0~.
\label{GL}
\end{equation}

The important feature of the magnon systems is that that their number density is limited 
\begin{equation}
n<n_{\rm max}=\frac{2S}{\hbar}
\,.
\label{limitation}
\end{equation}
For small $n\ll n_{\rm max}$ the Ginzburg-Landau  free energy functional has the conventional form
  \begin{eqnarray}
  {\cal F} -\mu {\cal N}=\int d^3r   
   \left(
 \frac{\left|\boldsymbol{\nabla}\Psi \right|^2}{2m}  +(\omega_L({\bf r})-\omega)
  \vert\Psi\vert^2 +F_{\rm so}(\vert\Psi\vert^2)\right).
  \label{GLfunctional}
\end{eqnarray}
Here  $\omega_L({\bf r})=\gamma H({\bf r})$  is the local Larmor frequency, which plays the role of external potential $U({\bf r})$ in atomic condensates. The last term $F_{\rm so}(\vert\Psi\vert^2)$ contains nonlinearity which comes from the spin-orbit interaction. It is analogous to the 4-th order term in the atomic BEC, which describes the interaction between the atoms. 

\subsection{Spin-orbit  interaction as interaction between magnons}

In the magnetic subsystem of superfluid $^3$He, the interaction term in the Ginzburg-Landau  free energy is provided by the  spin-orbit interaction  -- interaction between the spin and orbital degrees of freedom.  
Though the structure of  superfluid phases of $^3$He is rather complicated and is described by the multi-component superfluid order parameter  \cite{VollhardtWolfle}), the only output needed for investigation of the coherent precession is the structure of the spin-orbit interaction term  $F_{\rm so}(\vert\Psi\vert^2)$, which appears to be rather simple. The  spin-orbit interaction provides the effective interaction between magnons, which can be attractive or repulsive, depending on the orientation of spin and orbital orbital degrees with respect to each other and with respect to magnetic field. The orbital degrees of freedom  in  the superfluid phases of $^3$He are  characterized by the direction of the orbital momentum of the Cooper  pair  $\hat{\bf l}$, which also marks the axis of the spatial anisotropy of these superfluid liquids. 
 By changing the orientation of $\hat{\bf l}$ with respect to magnetic field one is able to regulate the interaction term in experiments. 

In superfluid $^3$He-B,  the spin-orbit interaction has a very peculiar properties.  The microscopic derivation (see \eqref{gammaGeneral}) leads to the following form \cite{BunkovVolovik1993a}:
\begin{eqnarray}
  F_{\rm so}(s,l,\gamma)= \frac {2}{15} \frac {\chi}{\gamma^2} \Omega_L^2 [ (s l- \frac{1}{2} +
  \frac{1}{2}\cos{\gamma}(1+s)(1+l))^2+
 {\nonumber}
 \\
   \frac{1}{8}(1-s)^2(1-l)^2    + ( 1-s^2)(1-l^2)(1+\cos{\gamma})]
~.
  \label{FD}
  \end{eqnarray}
It is obtained by averaging of the spin-orbit energy over the fast precession of spins. Here  $s=\cos\beta$, while  $l=\hat{\bf l}\cdot \hat{\bf H}$ describes the orientation of the unit vector $\hat{\bf l}$ with respect to the direction $\hat{\bf H}$ of magnetic field.  The parameter $ \Omega_L$ is the so-called  Leggett frequency, which characterizes the magnitude of the spin-orbit interaction
and thus the shift of the resonance frequency  from the Larmor value caused by spin-orbit interaction. In typical experimental situations,  $\Omega_L^2\ll \omega^2$, which means that the frequency shift is relatively small. Finally $\gamma$ is another angle, which characterizes   the mutual orientation of spin and orbital degrees of freedom. At not extremely low low temperatures ($T\geq  0.2 T_c$), it is a passive quantity: it takes the value corresponding to the minimum of $F_{\rm so}$ for given $s$ and $l$, i.e. $\gamma=\gamma(s,l)$.

To obtain $F_{\rm so}(\vert\Psi\vert^2)$ in \eqref{GLfunctional} at fixed $\hat{\bf l}$,  one must express $s$ via $|\Psi|^2$: 
 \begin{equation}
1-s=1-\cos\beta = \frac{\hbar\vert\Psi\vert^2}{S}~,
\label{1-s}
\end{equation}
where $S=\chi H/\gamma$ is spin density. Since Eq. \eqref{FD} is quadratic in $s$, the spin-orbit interaction contains quadratic and quartic terms in $|\Psi|$. While the quadratic term modifies the potential $U$ in the Ginzburg-Landau free energy, the quartic term simulates the interaction between magnons.

 \begin{figure}[htt]
 \includegraphics[width=0.7\textwidth]{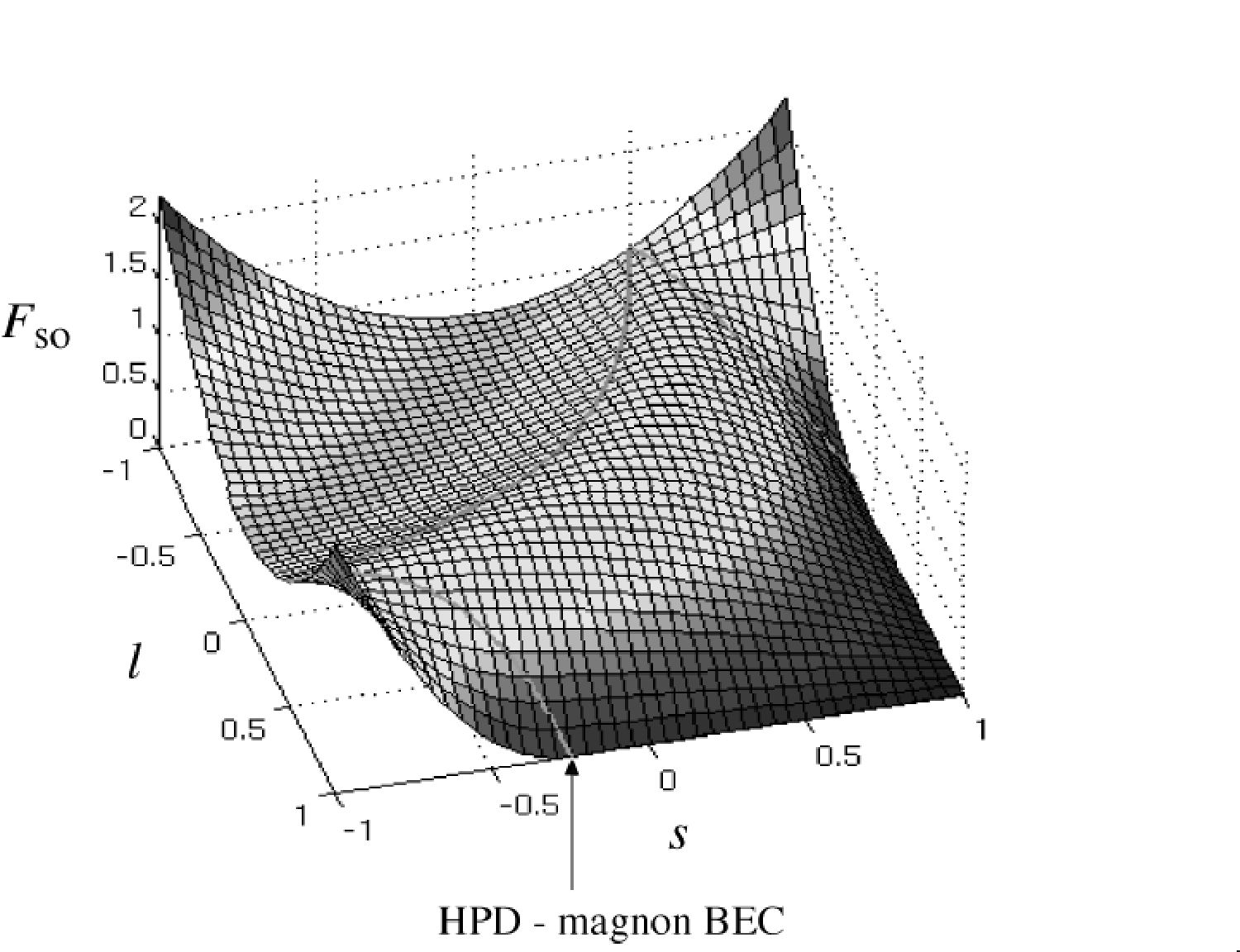}
 \caption{The profile of the spin-orbit energy as a function of $s=\cos\beta$ and orbital variable $l$, where 
 $\beta$ is the tipping angle of precession and $l$ is the projection of the orbital angular momentum of a Cooper pair on the direction of magnetic field.
Spontaneous phase-coherent precession emerging at $l=1$ and $s\approx -1/4$ is called Homegeneously Precessing Domain (HPD). HPD is one of the representatives of magnon BEC in 
$^3$He-B.}
 \label{profile}
\end{figure}
 
The profile of the spin-orbit interaction   $F_{\rm so}(s,l,\gamma(s,l))$ shown in Fig. \ref{profile} determines different states of coherent precession and thus different types  of magnon BEC in $^3$He-B, which depend on the orientation of the orbital vector $\hat{\bf l}$.
The most important of them, which has got the name HPD,
has been discovered about 30 years ago \cite{HPDexp,HPDtheory}.

\section{Magnon superfluids in $^3$He-B}

In this chapter we shall describe the main properties of magnon superfluidity in $^3$He-B. Some more detailed descriptions can be found in an original papers as well in a few review articles
\cite{R1,R2,R3,R4,R5,R6}.

\subsection{HPD as unconventional magnon superfluid}

In the right corner of Fig. \ref{profile}, the minimum of the free energy occurs for  $l=1$, i.e. for the orbital vector $\hat{\bf l}$ oriented along the magnetic field.   This means that if there is no other orientational effect on the orbital vector  $\hat{\bf l}$, the spin-orbit interaction orients it along the magnetic field, and one automatically obtains $l=\cos\beta_L=1$. The most surprising property emerging at such orientation is the existence of the completely flat region in Fig. \ref{profile}. The spin-orbit interaction is identically zero in the large range of the tipping angle $\beta$ of precession, for $1> s=\cos\beta >-\frac{1}{4}$
 \cite{VollhardtWolfle}:
   \begin{eqnarray}
F_{\rm so}(\beta)_{l=1}=0~,~ \cos\beta> -\frac{1}{4},
   \label{SO1}
      \\
   F_{\rm so}(\beta)_{l=1}=\frac{8}{15}\frac {\chi}{\gamma^2}\Omega_L^2\left(\cos\beta
+\frac{1}{4}\right)^2 ,~
 \cos\beta< -\frac{1}{4}.
 \label{SO2}
\end{eqnarray}
 Using Eq.~\eqref{1-s} one obtains the Ginzburg-Landau potential in \eqref{GLfunctional} with:
  \begin{eqnarray}
    F_{\rm so} \left(\vert\Psi\vert^2\right)=0~,~ \vert\Psi\vert^2<n_c=\frac{5}{4}S,
      \label{less104}
    \\
  F_{\rm so} \left(\vert\Psi\vert^2\right)=\frac{8}{15}\frac{\chi}{\gamma^2}\Omega_L^2 \left(
\frac{\vert\Psi\vert^2}{S}-\frac{5}{4}\right)^2 ~,~ \vert\Psi\vert^2>n_c=\frac{5}{4}S.
    \label{larger104}
\end{eqnarray}
 Eqs. \eqref{less104} and \eqref{larger104} demonstrate that when the orbital momentum is oriented along the magnetic field, magnons are non-interacting for all densities $n$ below the threshold value 
  $n_c=(5/4)S/\hbar$. This is a really unconventional gas.
 
  \begin{figure}[htt]
 \includegraphics[width=0.7\textwidth]{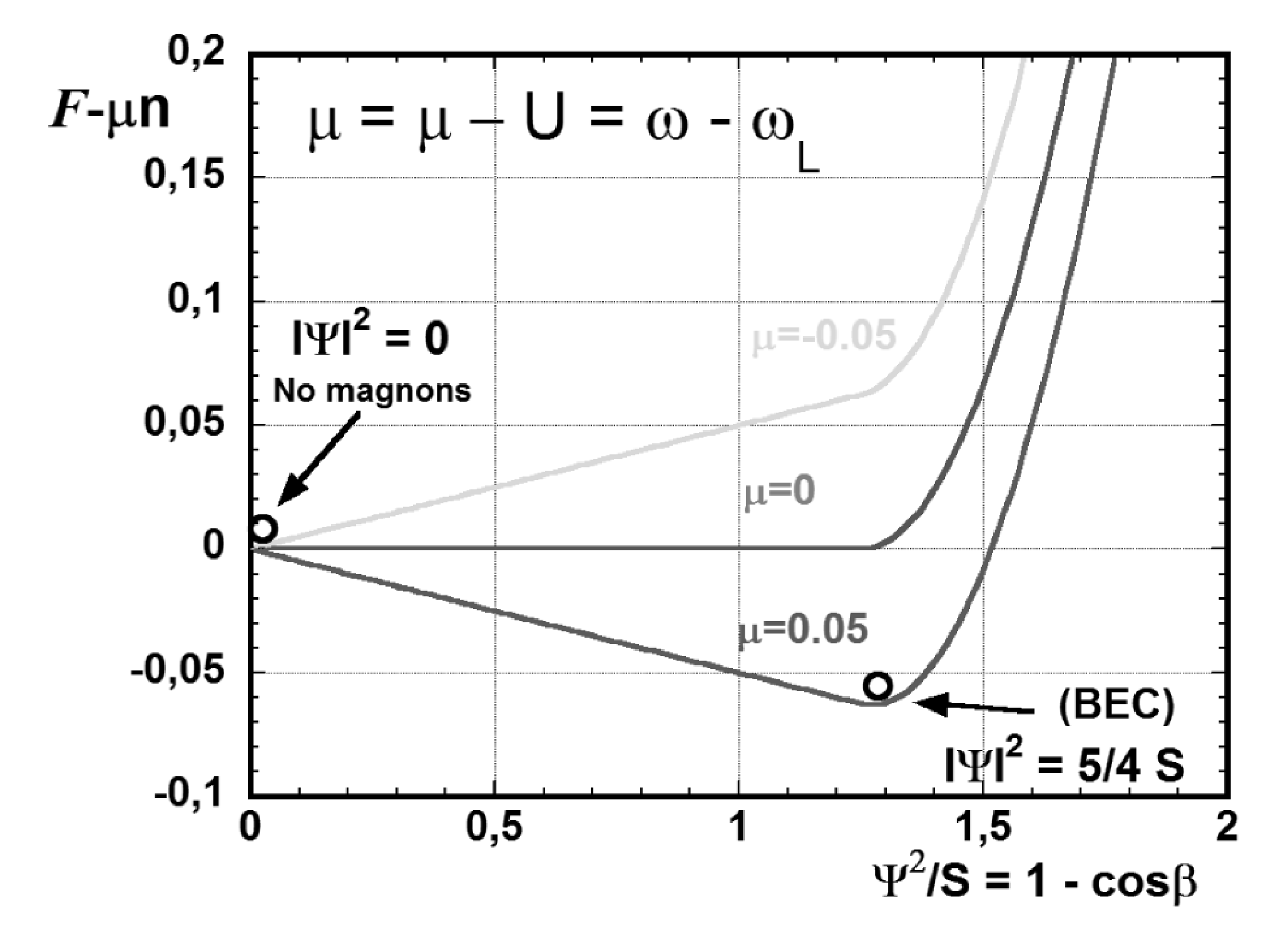}
 \caption{$F-\mu n$ for different values of the chemical potential
$\mu\equiv \omega$ in magnon BEC in $^3$He-B. For $\mu<U$, i.e. for $\omega<\omega_L$, the minimum of $F-\mu n$ corresponds to zero number of magnons, $n=0$. It is the static state  without precession. For $\mu=U$, i.e. for $\omega=\omega_L$, the energy is the same for all densities in the range $0\leq n \leq n_c$.
 For $\mu>U$, i.e. for $\omega>\omega_L$, the minimum of $F-\mu n$ corresponds 
the magnon BEC with density $n\geq n_c$. This corresponds to the coherent precession of magnetization with tipping angle $\beta > 104^\circ$.}
 \label{F-muN}
\end{figure}
 
The energy profile of $F-\mu n$ is shown in Fig. \ref{F-muN} for different values of the chemical potential
$\mu\equiv \omega$. For $\mu$ below the external potential $U$, i.e. for $\omega<\omega_L$, the minimum of $F-\mu n$ corresponds to zero number of magnons, $n=0$. It is the static state of $^3$He-B without precession. For $\mu>U$, i.e. for $\omega>\omega_L$, the minimum of $F-\mu n$ corresponds 
to the finite value of the magnon density:
 \begin{equation}
n=n_c \left(1 +\frac{3}{4} \frac{(\omega-\omega_L)\omega_L}{\Omega_L^2}\right)~.
\label{MagnonDensityBphase}
\end{equation}
This shows that the formation of HPD starts with the discontinuous jump from zero density of magnons to the finite density $n_c=5S/4\hbar$, which corresponds to coherent precession
with the large tipping angle  --  the so-called magic Leggett angle, $\beta_c\approx 104^\circ$ ($\cos\beta_c=-1/4$).
This is distinct from the standard Ginzburg-Landau energy functional (see Eq. \eqref{FDl=1} for magnon BEC in $^3$He-A-phase below), where the Bose condensate density smoothly starts growing from zero and is proportional to $\mu -U$  for $\mu>U$. 

  \subsection{Two-domain precession}
  
    \begin{figure}[htt]
 \includegraphics[width=0.8\textwidth]{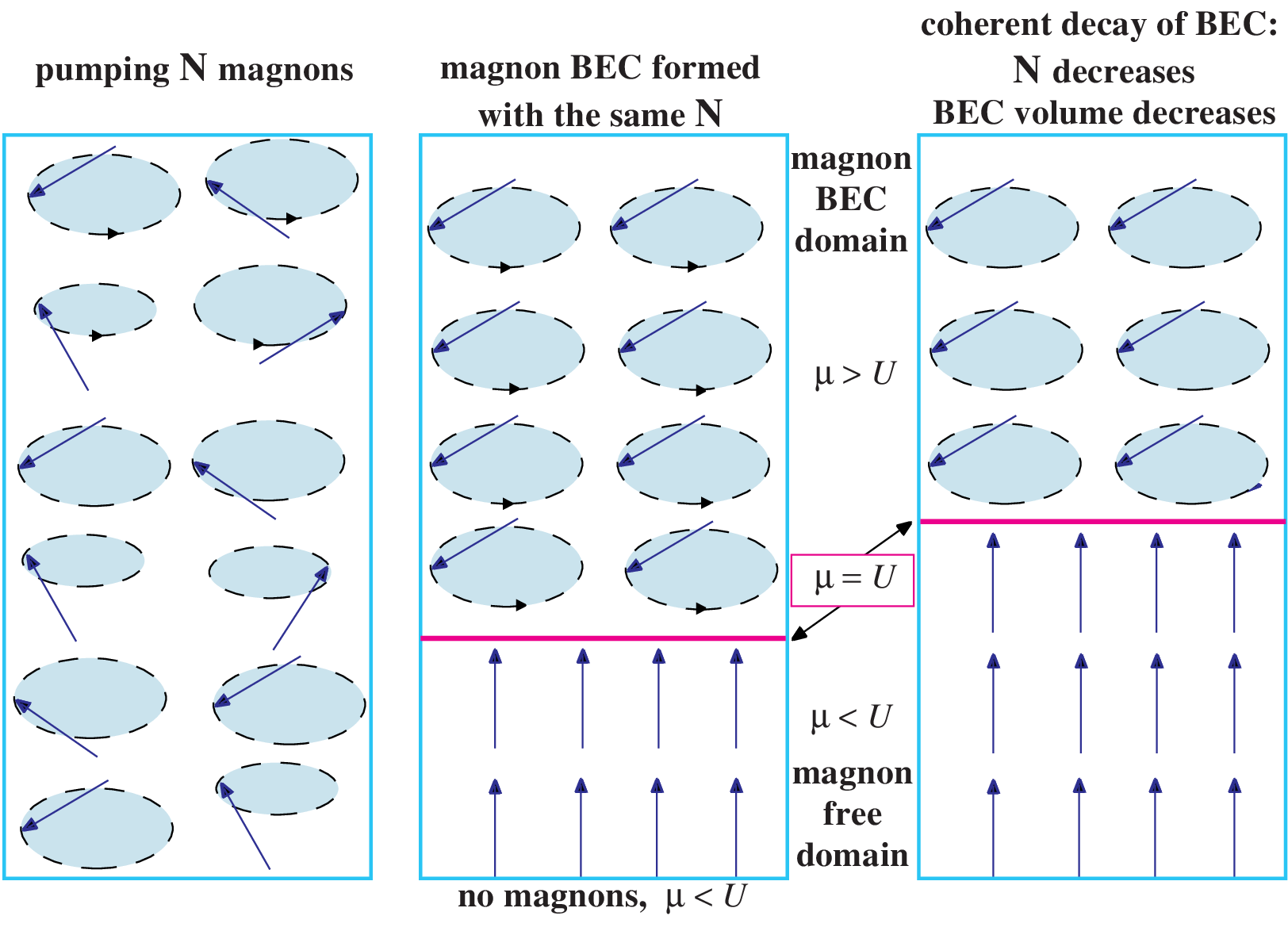}
 \caption{
  Two domains emerging in $^3$He-B in the pulsed NMR experiments. {\it left}: Incoherent spin precession after the pulse of the RF field deflects magnetization from its equilibrium value. The total number of magnons pumped into the system is ${\cal N}=({\cal S}-{\cal S}_z)/\hbar$ magnons. {\it middle}:
Formation of two domains. All ${\cal N}$ magnons  are concentrated in the part of the cell with lower magnetic field, forming the BEC state there. The volume of this state is determined by the magnon density in BEC, $V={\cal N}/n$, where $n\approx n_c$. This volume determines the position $z_0$ of the domain boundary $z_0=V/A$, where $A$ is the area of the cross-section of the cylindrical cell. The position of the interface in turn determines the global frequency of presession, which is equal to the local Larmor frequency at the phase boundary, $\mu\equiv \omega=\omega_L(z_0)$.
{\it right}: Decay of magnon BEC.  The number of magnons decreases due to spin and energy losses.
Since the magnon density in BEC is fixed (it is always close to $n_c$), the relaxation leads to the decrease of the volume of the BEC domain. However within this domain  the precession remains fully coherent. While the phase boundary slowly moves down the frequency of the global precession gradually decreases, Fig. \ref{frequency} {\it left}.
}
 \label{2Domain}
\end{figure}
  
The two states with zero and finite density of magnons, resemble the low-density gas state and the high density liquid state, respectively. Gas and liquid can be separated in the gravitational field: the heavier  liquid state will be concentrated in the lower part of the vessel. For the magnon  BEC, the role of the gravitational field is played by the gradient of magnetic field:
\begin{equation}
\boldsymbol{\nabla} U \equiv \nabla\omega_L= \gamma \boldsymbol{\nabla} H~.
\label{FieldGradient}
\end{equation}
Thus applying the gradient of magnetic field along the axis $z$, one enforces phase separation, Fig. \ref{2Domain}.
The static thermodynamic equilibrium state with no magnons is concentrated in the region of higher field, where 
$\omega_L(z)>\omega$, i.e. $U(z)>\mu$. The magnon superfluid -- the coherently  precessing state --  occupies the low-field region, where  $\omega_L(z)<\omega$, i.e. $U(z)<\mu$. This is the HPD state, in which all spins precess with the same frequency $\omega$ and the same phase $\alpha$.
In typical experiments the gradient is small,   and magnon density is close to the threshold value $n_c$.

The interface between the two domains is situated at the position $z_0$ where $\omega_L(z_0)=\omega$, i.e. $U(z_0)=\mu$. In the continuous NMR, the chemical potential is fixed by the frequency of the RF field:
$\mu=\omega_{\rm RF}$, this determines the  position of the interface in the experimental cel.

 In the pulsed NMR, the two-domain structure spontaneously  emerges after the magnetization is deflected by the RF pulse (Fig. \ref{2Domain}, {\it left} and {\it middle}). The position of the interface between the domains is determined by the number of magnons pumped into the system: ${\cal N}=({\cal S}-{\cal S}_z)/\hbar$. The number of magnons is quasi-conserved, i.e. it is well conserved during the time of the formation of the two-domain state of precession. That is why the volume of the domain occupied by the magnon BEC after its formation is $V={\cal N}/n_c$. This determined the position $z_0$ of the interface, and the chemical potential $\mu$ will be agjusted to this position: $\mu=\omega_L(z_0)$.

 \begin{figure}[htt]
 \includegraphics[width=1\textwidth]{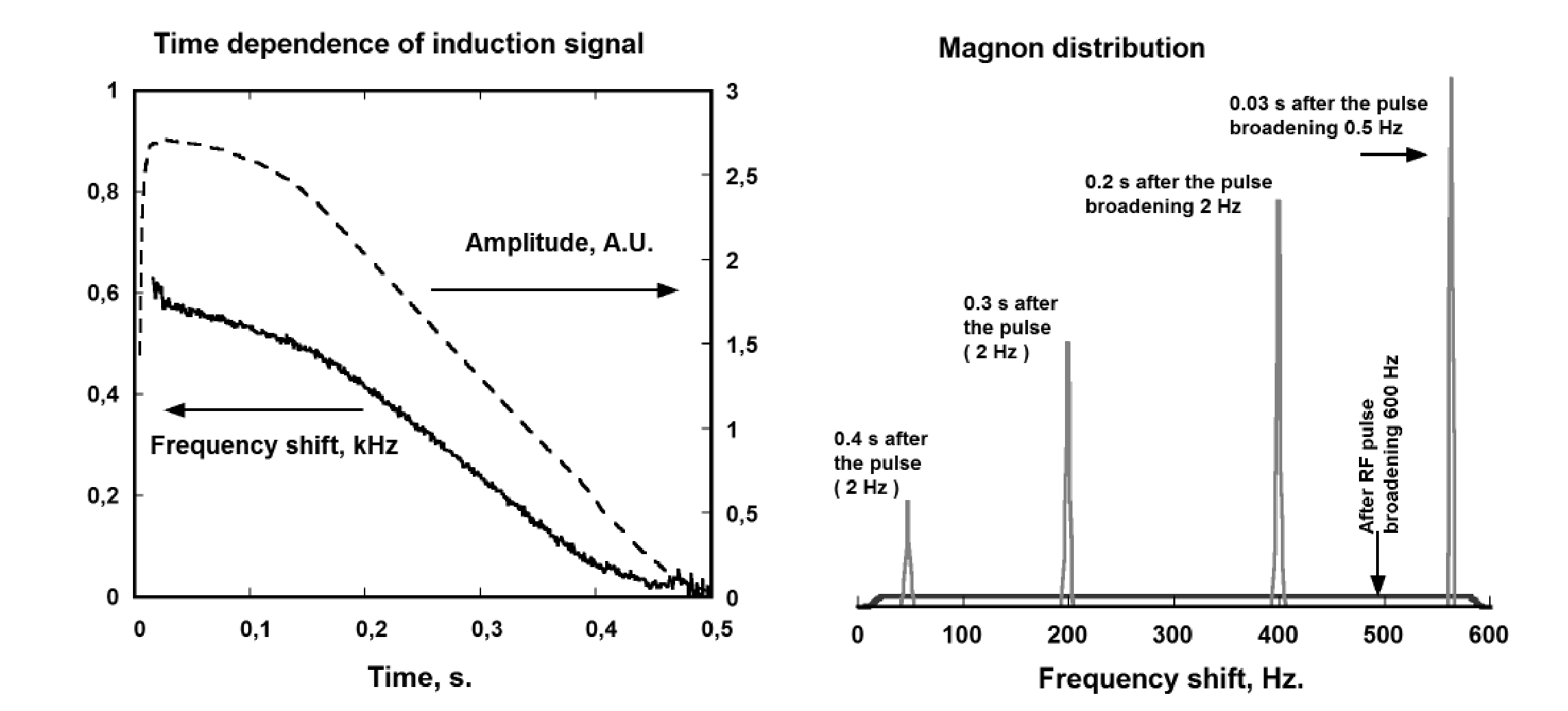}
 \caption{
The amplitude and frequency of the induction decay signal from magnon BEC.
{\it left}: The condensate occupies the domain where the chemical potential $\mu>U$ and radiates the signal
corresponding to the Larmor frequency at the domain
boundary of condensate. With relaxation the number of magnons
decreases, and the chemical potential moves to the region with a lower Larmor frequency.
{\it right}:  The spectroscopic distribution of magnons. Immediately after the RF
pulse each spin precesses with the local Larmore frequency. After  the BEC
formation, all the spins precess with the common frequency $\omega$ and
spontaneously emergent common phase $\alpha$. Due to relaxation the number of magnons decreases,
leading to  the continuously decreasing frequency. The small broadening of BEC state is
due to  relaxation. By comparing the initial broadening of the NMR line of about 600 Hz and
final broadening of about 0.5 Hz we can estimate that about 99.9 \% of the pumped magnons
are in the condensate!
}
 \label{frequency}
\end{figure}

In the absence of the RF field, i.e. without continuous pumping of magnons, the magnon BEC decays due to losses of magnons. But the precessing domain remains in the fully coherent Bose condensate state,  while the volume of the magnon superfluid gradually decreases due to losses and the domain boundary
slowly moves down  (Fig. \ref{2Domain}  {\it right}). The frequency $\omega$ of spontaneous coherence  as well as the phase of precession
remain   homogeneous across the whole domain of magnon BEC,
but the frequency changes with time, since
it is determined by the Larmor frequency at the position of
the interface,  
$\mu(t)\equiv \omega_L(z_0(t))$.  The change of
frequency during the decay is shown in Fig. \ref{frequency} {\it
left}.  This frequency change during the relaxation was the main
observational fact that led Fomin to construct the theory of the
two-domain precession \cite{HPDtheory}.

The details of formation of the magnon BEC are shown in Fig.
\ref{amplitude}, where the stroboscopic record of  induction decay
signal is shown. During the first stage of about 0.002 s  the
induction signal completely disappears due to dephasing. Then,
during about 0.02 s,  the phase coherent precession spontaneously
emerges, which is equivalent to the magnon BEC state. Due to a
weak magnetic relaxation, the number of magnons slowly decreases but
the precession remains coherent during the whole process of
relaxation. The time of formation of magnon BEC is essentially
shorter than the relaxation time, as clearly shown in  Fig.
\ref{amplitude} {\it right}.
 
\subsection{Mass and spin supercurrents in magnon BEC}

As we already discussed, superfluidity is phenomenon arising due to spontaneous breaking of $U(1)$ symmetry, which in our case is represented by the symmetry group $SO(2)$ of spin rotations about direction of magnetic field.
In atomic BEC and in helium superfluids such symmetry breaking leads to a non-zero value of the  superfluid rigidity --  the superfluid density  $\rho_s$ which enters the non-dissipative  supercurrent  of particles and thus to mass supercurrent. The same takes place for magnon BEC. But since magnons has both mass $m_M$ and spin $-\hbar$, they  carry both the mass and spin supercurrents.  The corresponding Goldstone phonon mode of the magnon BEC has been experimentally observed: it is manifested as twist oscillations of  the precessing domain in $^3$He-B   \cite{5} (see the Section below). 

For small density of magnon condensate $n\ll n_c$ the mass current of magnons is given by the traditional equation:
\begin{equation}
{\bf J}= \rho_s {\bf v}_s~~,~~{\bf v}_s =\frac{\hbar}{m_M}\boldsymbol{\nabla}\alpha~~,~~\rho_s(T=0)=nm_M ~.
 \label{MassCurrent}
 \end{equation}
 In translationally invariant systems, where the mass current coincides with density of linear momentum,  Eq.(\ref{MassCurrent}) can be obtained directly from the definition of linear momentum density in spin systems:
\begin{equation}
{\bf P}=(S-S_z)\boldsymbol{\nabla}\alpha=  n \hbar \boldsymbol{\nabla}\alpha\,,
\label{MassCurrentMagnon}
\end{equation}
where we used the fact that $S-S_z$ and $\alpha$ are canonically 
conjugated variables. As in conventional superfluids, the superfluid density of the magnon liquid is determined by the magnon density $n$ and magnon mass $m_M$. 

To avoid the confusion let us mention that in magnon BEC the superfluid density describes the coherent precession in magnetic subsystem and has nothing to do with the superfluid density of the underlying system -- the superfluid $^3$He. In magnon BEC the superfluid mass current \eqref{MassCurrent} carries magnons with  mass $m_M$, while in atomic superfluids  the superfluid mass current carries atoms. The mass current generated by precessing
 magnetization in magnon BEC is similar to electric current generated by  precessing
 magnetization in ferromagnets \cite{Volovik1987}, which is now used in spintronics (see e.g. the recent review \cite{Brataas2012}).

It is important that the proper atomic analog of magnon BEC is actually the A$_1$ phase of $^3$He. Both systems are spin polarized: magnons have spin $-\hbar$ while  in $^3$He-A$_1$ the atoms only with one spin polarization experience superfluidity \cite{VollhardtWolfle}. As a result, 
both the superfluid current of magnons in magnons BEC and superfluid current of atoms  in $^3$He-A$_1$ are necessarily accompanied by the superfluid spin current. 
Since each magnon carries spin $-\hbar$, the magnon mass supercurrent is  accompanied by the magnetization supercurrent -- the supercurrent of the $z$-component of spin:
\begin{equation}
 J_i^z=- \frac{\hbar}{m_M}~ J_i=- n \frac{\hbar^2}{m_M}\nabla_i\alpha \,.
  \label{SpinCurrent}
  \end{equation}
  The same is valid for $^3$He-A$_1$ where the spin current is  $J_i^z=- \frac{\hbar}{2m_3}$.
  
  Let us note that  the  density of linear momentum of the spin subsystem is not well defined globally. While the total momentum of the system is conserved, the canonical momenta of the spin and orbital subsystems are not conserved separately
\cite{Volovik1987,Stone1996,Wong2009}. For the particular choice of the  linear momentum density in \eqref{MassCurrentMagnon}, ${\bf P}$  is not defined at the points where $\beta =\pi$, because at $\beta =\pi$ spins are stationary and thus the  spin precession angle $\alpha$ is ill defined. 
This is another interesting feature of the magnon superfluids, which becomes important for the magnon BEC emerging in normal (non-superfluid) $^3$He. The latter represents a coherently precessing  structure at the interface between the equilibrium domain with $\beta=0$ and the domain with the reversed magnetization, i.e. with  $\beta=\pi$ \cite{Normal3He}.  

  \subsection{London limit: hydrodynamics of magnon BEC}
\label{SupercurrentLinearMomentum}

  In HPD state of magnon BEC, the magnon density is comparable with the limiting value,  $n\approx (5/8)n_{\rm max}$, as a result  the kinetic term in the Ginzburg-Landau energy becomes complicated. However, theory of HPD becomes simple in the London limit, where the magnon superfluidity is described by the hydrodynamic energy functional  written in terms of density $n$ and superfluid velocity ${\bf v}_s$. This hydrodynamic energy functional is similar to that in superfluid liquids and atomic BEC, but with some important differences. One of them is the anisotropy of magnon mass, which leads to anisotropy of superfluid density even at $T=0$. Another one is the presence of the symmetry breaking term which depends explicitly on $\alpha$, and which gives rise to the mass of the Goldstone boson.
  
 The hydrodynamic energy functional is expressed in terms of the canonically conjugated variables --  number density  $n$ (magnon density) and the superfluid velocity ${\bf v}_{\rm s}$, which is expressed via the gradient of the phase of the Bose condensate  $\alpha$ (the phase of the coherent precession of magnetization). It has the following general form:
 \begin{equation}
F= \frac{1}{2}\rho_{{\rm s}ij}(n)v_{{\rm s}i} v_{{\rm s}j}  +\epsilon(n) -\mu n + F_{\rm sb}(\alpha,n)~.
\label{HydrodynamicEnergy}
\end{equation}
Here $\mu$ as before is the chemical potential;  $\rho_{{\rm s}ij}$ is the tensor of anisotropic superfluid density and  $v_{{\rm s}i} $ is the superfluid velocity of magnon superfluid: 
\begin{equation}
\rho_{{\rm s}ij}(n)=nm_{ij} ~~,~~ v_{{\rm s}i}  =\hbar \left(m^{-1}\right)_{ij}  \nabla_j \alpha \,,
\label{SuperfluidDensity}
\end{equation} 
 where  the matrix of magnon masses $m_{ij}(n)$ depends on magnon density $n$ and tilting angle of precession. For the magnons propagating along the field and in the transverse directions, their mass  depends on the tilting angle in the following way:  
\begin{eqnarray}
\frac{1}{m^\parallel(n)}=2
 \frac{c_\parallel^2 \cos\beta+ c_\perp^2(1-\cos\beta)}{\hbar\omega_L} \,,
\label{parallel_mass}
\\
\frac{1}{m^\perp(n)}=\frac{c_\parallel^2(1+\cos\beta)
+ c_\perp^2(1-\cos\beta)}{\hbar\omega_L}   \,,
\label{transverse_mass}
\end{eqnarray}
where the parameters $c_\parallel$ and $c_\perp$ are  on  the
order  of the Fermi velocity $v_F$.   
The mass supercurrent is however isotropic, when it expressed via $\alpha$:
\begin{equation}
J_i= \frac{dF}{dv_{{\rm s}i} }=\hbar n \nabla_i\alpha \,,
\label{MassCurrent2}
\end{equation}   
which is in agreement with equation (\ref{MassCurrentMagnon}) for linear momentum.

Finally  $F_{\rm sb}$ is the symmetry breaking term which depends explicitly on $\alpha$. It arises only in case of continuous wave NMR, where it comes from interaction of the condensate with applied RF field, which is needed  to compensate  the relaxation of magnons. This term explicitly violates the $U(1)$ symmetry and that is why it gives rise to the mass of Goldstone boson which we discuss later on.

The hydrodynamic equations for magnon superfluid are the Hamilton equations for the canonically conjugated variables $n$ and $\alpha$:
\begin{equation}
\dot\alpha= \frac{\delta F}{\delta n}~~,~~ \dot n=- \frac{\delta F}{\delta \alpha} \,.
\label{HydroEq}
\end{equation}

This superfluid spin current is as before determined by the spin to mass
ratio for the magnon. But because the magnon mass is
anisotropic, the spin current transferred by the coherent
spin precession is anisotropic too:
\begin{eqnarray}
J^z_z=-\frac{\hbar^2}{m^\parallel(n)}n \nabla_z\alpha~,
\label{SpinCurrent1}
\\
{\bf
J}^z_\perp=-\frac{\hbar^2}{m^\perp(n)}n 
\boldsymbol{\nabla}_\perp\alpha~.
\label{SpinCurrent2}
\end{eqnarray}
The anisotropy of the current in Eqs.
(\ref{SpinCurrent1}-\ref{SpinCurrent2}) is an important
modification of the conventional Bose condensation, since it
is absent in the atomic Bose condensates.
The spin supercurrent becomes isotropic, when it is expressed in terms of superfluid velocity:
\begin{equation}
J^z_i= \hbar  \left(m^{-1}\right)_{ij} J_j= \hbar n v_{{\rm s}i}  ~.
\label{SpinCurrent3}
\end{equation}

\subsection{Goldstone mode of coherent precession -- sound in magnon BEC}

In atomic superfluids, sound is the Goldstone mode of the spontaneously broken $U(1)$ symmetry. 
The same sound mode exists in magnon BEC.

\begin{figure}[htt]
 \includegraphics[width=1\textwidth]{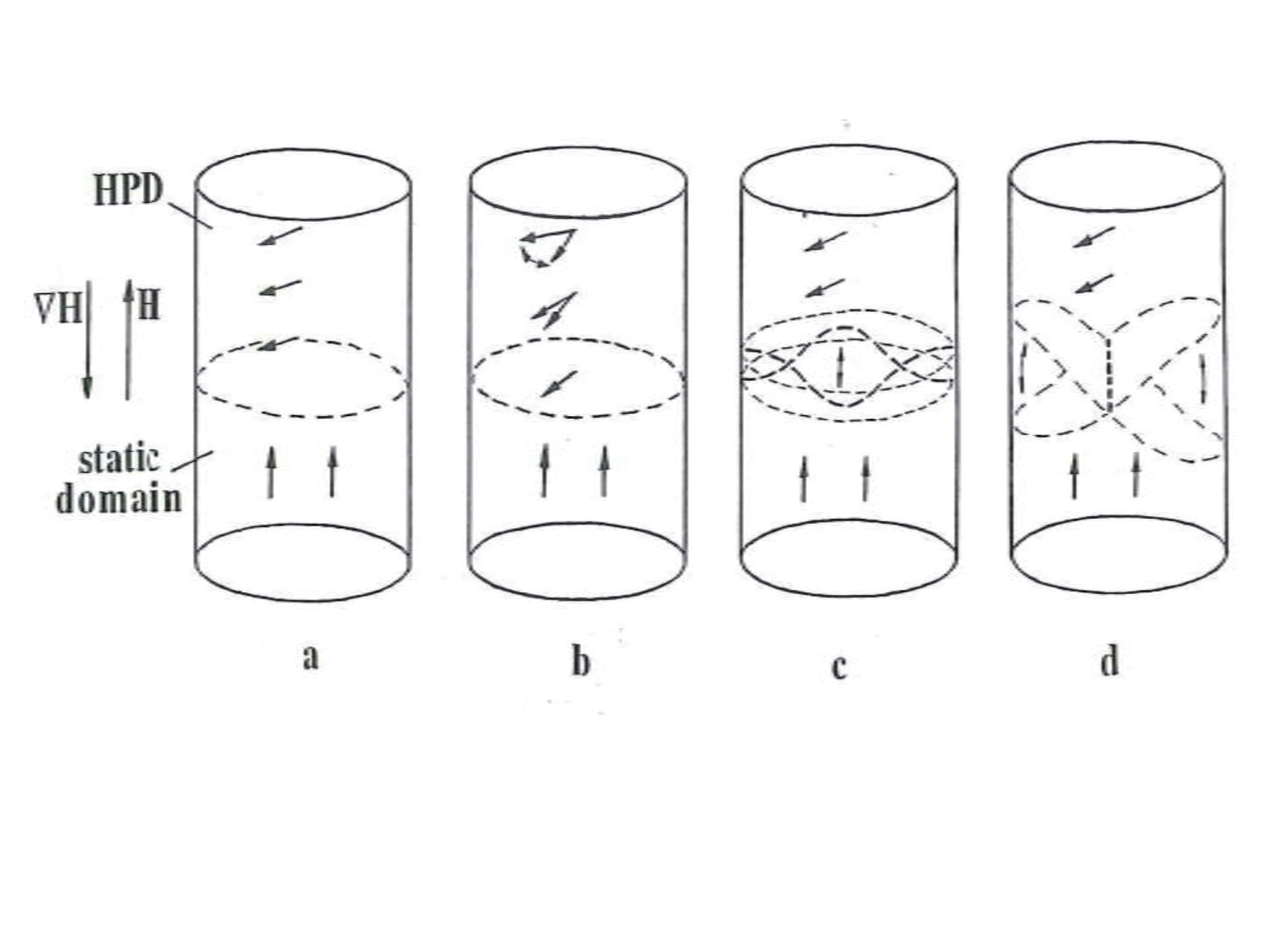}
 \caption{Schematic representation of the precessing magnetization in the rotating frame for an
equilibrium HPD (a); Goldstone mode of coherent precession -- analog of the sound wave in atomic BEC -- is the mode of twist oscillations (b);   surface oscillations -- analogs of gravity waves on the surface of a liquid (c, d).
}
 \label{Oscil}
\end{figure}

The HPD formation corresponds to an energy minimum under the condition of
conservation of the total longitudinal magnetization of the sample. There is a
powerful feedback mechanism that returns the system to the homogeneous precession
state after any disturbance. This is the excitation of spin supercurrent
transport between nonhomogeneous states of the precessing magnetization.
Therefore oscillations of the magnetization distribution near the equilibrium
HPD state can take place. The frequencies of two modes of such oscillation
were calculated first by Fomin \cite{FominSound}. There are torsional oscillations in bulk
and surface oscillations (Fig. \ref{Oscil}). Under experimental conditions both bulk and surface
oscillation modes have been observed \cite{5,Osc,Osc2}. The complete review of this experiments one can found in Ref. \cite{R1}.

The surface oscillations at the domain boundary are analogous to gravity
 waves on the surface of liquids, whose spectrum is $\omega^2=gk$, where $g$ is the gravitational field. The role of the   
gravitational field is played by the gradient of Zeeman energy, while  the kinetic energy of the spin supercurrent 
plays the role of the kinetic energy of the flow of the liquid. For a cylindrical
cell we can visualize these oscillations as the surface waves of water in a
glass. 

Torsional oscillations originate from the degeneracy of the precessing states  with respect to the phase of the precession $\alpha$. This mode,  called the twisting mode, corresponds to spatial oscillations of the
phase of the magnetization precession inside the HPD with spin supercurrent feedback
response, and thus represents the Goldstone mode of the spontaneously broken $U(1)$ symmetry.
It is analogous to a sound wave in atomic superfluids. The sound mode in magnon subsystem is obtained from linearization of the hydrodynamic equations \eqref{HydroEq}. It has been calculated in Ref.\cite{FominSound}  and identified experimentally in Ref. \cite{5}.

 There are several conditions required for the existence and stability of the magnon BEC. Some of them are the same as for the conventional atomic BEC, but there are also important differences, which we discuss later. One of the conditions is that the compressibility $\beta_M$ of the magnon gas must be positive:
\begin{equation}
\beta_M^{-1} = n\frac{dP}{dn}=  n^2\frac{d^2\epsilon}{dn^2} >0~.
\label{compressibility}
\end{equation}
This condition means that the fourth order term in the Ginzburg-Landau free energy should be positive, i.e.
the interaction between magnons should be repulsive.
The magnon interaction energy $\epsilon(n)$  is provided by spin-orbit (dipole-dipole) interaction. It has a very peculiar form for HPD in $^3$He-B:
 \begin{equation}
\epsilon(n)\equiv E_{\rm so}(n) =\frac{8\chi \Omega_L^2}{15\gamma^2} \left(
\frac{\hbar n}{S}-\frac{5}{4}\right)^2 \Theta \left(
\frac{\hbar n}{S}-\frac{5}{4}\right)~,
     \label{FHPD}
  \end{equation}
  where $ \Theta(x)$ is Heaviside step function; $\Omega_L$ is Leggett frequency (we assume that  $\Omega_L\ll \omega_L$).
 This means that in this state of magnon BEC a stable coherent precession  occurs only at large enough magnon density $n>5S/4\hbar$,  where $d^2E_{\rm so}/dn^2>0$. This corresponds to $\cos\beta<-1/4$.
The magnon BEC state in \eqref{pot2} also satisfy the condition \eqref{compressibility}, while
the magnon condensates in \eqref{pot1} and in bulk $^3$He-A are unstable.

The compressibility of the magnon gas determines the speed of sound propagating in the magnon gas.
Since the magnons mass is anisotropic the phonon spectrum is also  anisotropic:
\begin{equation}
\left(c_s^2\right)^{ij}=\left(m^{-1}\right)^{ij}\frac{dP}{dn}=  n \frac{d^2E_{\rm so}}{dn^2}  \left(m^{-1}\right)^{ij}~.
\label{SpeedSound}
\end{equation}
In the typical experiments with HPD, $\cos\beta$  is close to $-$1/4, i.e. $\cos\beta=-1/4-0$. For such  $\beta$ one has:  
 \begin{equation}
c_{s\parallel}^2= \frac{n}{m_{\parallel}}\frac{d^2E_{\rm so}}{dn^2}=\frac{2}{3}\frac{\Omega_L^2}{\omega_L^2}\left(5c_\perp^2-c_\parallel^2\right)~,
\label{SpeedSoundParallel}
\end{equation}
 \begin{equation}
c_{s\perp}^2= \frac{n}{m_{\perp}}\frac{d^2E_{\rm so}}{dn^2}=\frac{1}{3}\frac{\Omega_L^2}{\omega_L^2}\left(5c_\perp^2+3c_\parallel^2\right)~.
\label{SpeedSoundPerp}
\end{equation}

  Owing the anisotropy of phonon spectra, the spin waves velocities appears differently in the modes of oscillations in Fig. \ref{Oscil}. While the frequency of twist oscillations is proportional to $c_{s\parallel}$, the frequency of surface waves is proporsional to $\sqrt{c_{s\parallel}c_{s\perp}}$. By the experimental investigations of these two modes of oscillations the experimental group in Kapitza Institute was able to measure the spin wave velocities $c_\parallel$ and $c_\perp$  \cite{Osc2,R1}.
  
  \subsection{Mass of phonons in magnon superfluid}

As distinct from the conventional superfluids, 
 in magnon BEC one may introduce experimentally the symmetry breaking field which smoothly violates the $U(1)$ symmetry and induces a small gap (mass) in the phonon spectrum. This mass has been  measured.

The symmetry-breaking term appears in continuous wave NMR, when the relaxation of magnon BEC is compensated by RF field. It describes the interaction $F_{\rm sb}(\alpha,n) = -\gamma {\bf H}_{\rm RF}\cdot{\bf S}$ of the precessing magnetization with  the RF field ${\bf H}_{\rm RF}$, which is transverse to the applied constant field ${\bf H}$. In continuous wave NMR experiments the RF field prescribes the frequency of precession, $\omega=\omega_{\rm RF}$, and thus fixes the chemical potential $\mu$; while in the state of free precession the chemical potential $\mu$ is determined by the number of pumped magnons. The symmetry-breaking term depends explicitly on the phase of precession $\alpha$ with respect to the direction of the RF-field in the precessing frame:
\begin{equation}
F_{\rm sb}= -  \gamma H_{\rm RF} S_\perp \cos\alpha =- \gamma H_{\rm RF} S \sin\beta \left(1-\frac{\alpha^2}{2}  \right) ~.
\label{SymmetryBreakingRF}
\end{equation}

 \begin{figure}[htt]
 \includegraphics[width=0.7\textwidth]{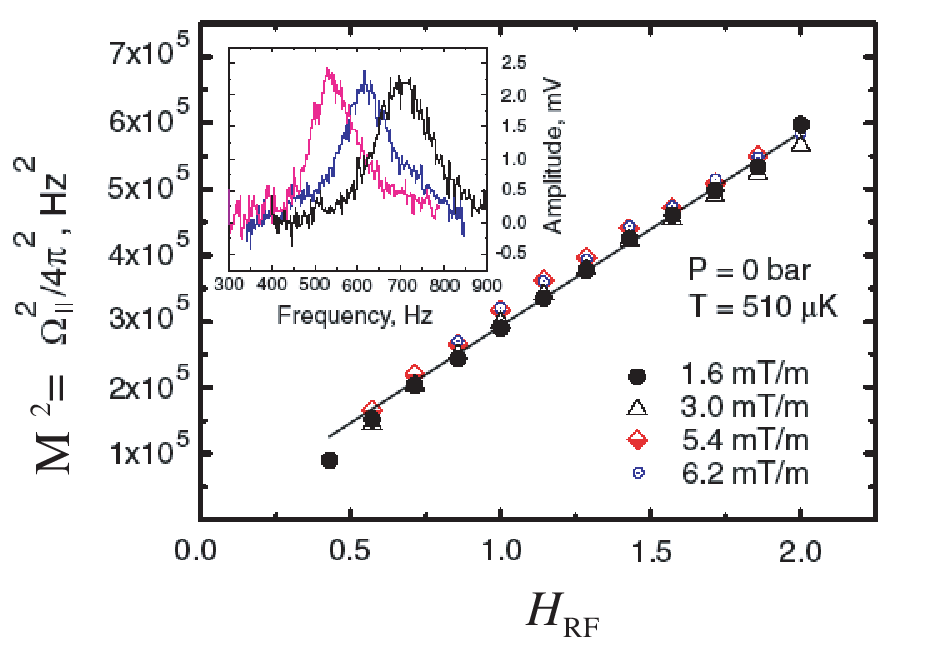}
 \caption{
 Phonon mass in magnon BEC as a function of the symmetry breaking field. From  \cite{Skyba}.
 }
 \label{PhononMass}
\end{figure}

Due to explicit dependence on $\alpha$, this term  generates the mass of the  Goldstone boson (phonon)   \cite{Volovik2008}.  For $\cos\beta=-1/4$ the phonon spectrum becomes:
\begin{equation}
\omega^2_s({\bf k})= \left(c_s^2\right)^{ij} k_i k_j + m_s^2~~,~~ m_s^2=\frac{4}{\sqrt{15}} \gamma H_{\rm RF} \frac{\Omega_L^2}{\omega_L}~.
\label{SoundMode}
\end{equation}
Two experiments with HPD  \cite{PhononMass,Skyba}  reported the gap in the spectrum of the collective mode of the coherent precession. The measured gap is proportional to $H_{\rm RF}^{1/2}$ in agreement with \eqref{SoundMode}.

\subsection{Spin vortex-- topological defect of magnon BEC}

The phase coherent precession of magnetizaton in superfluid $^3$He has all the
properties of the coherent Bose condensate of magnons. The main spin-superfluid properties of HPD have been verified already in the early experiments about 30 years ago, including spin supercurrent which transports the magnetization (analog of the mass current in
conventional superfluids); spin current Josephson effect and
phase-slip processes at the critical current \cite{6,7}, which we shall discuss later. 

\begin{figure}
\centerline{\includegraphics[width=1.0\linewidth]{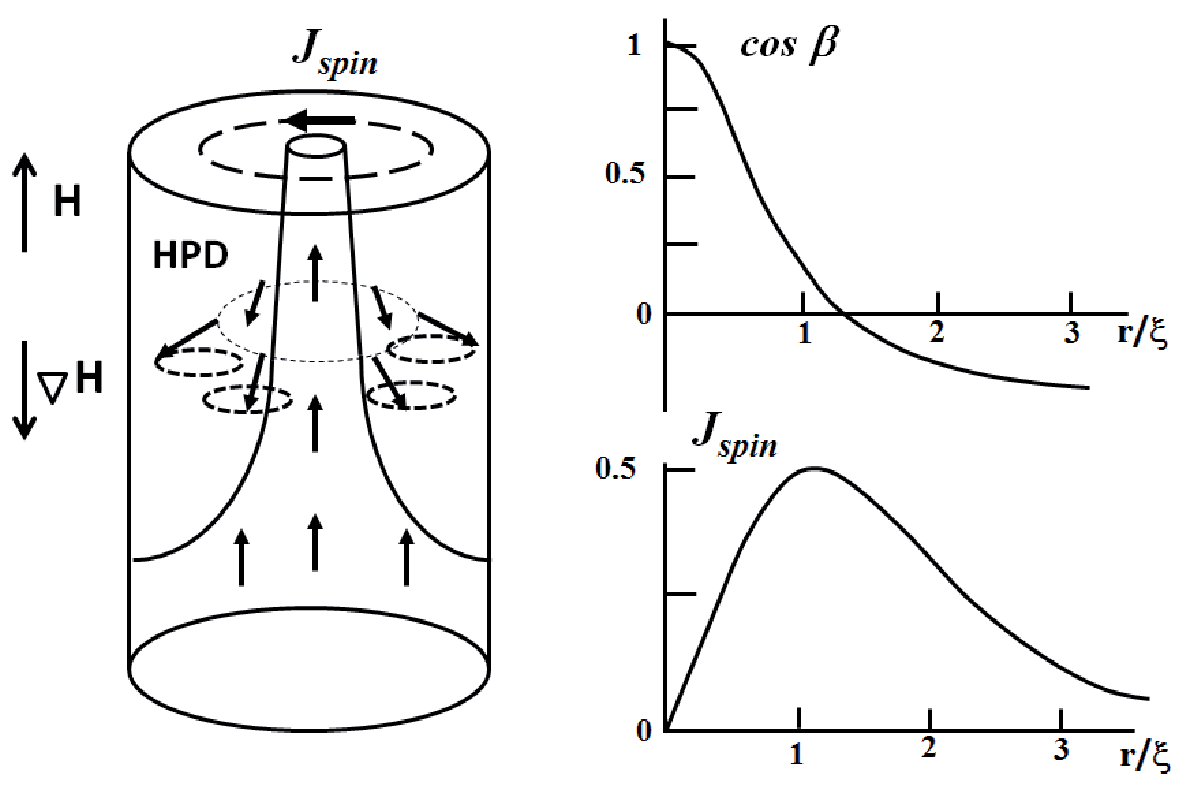}}
  \caption{Spin supercurrent vortex in magnon BEC in $^3$He-B. As in the case of the mass supercurrent  vortex in $^3$He-A, the core of the spin vortex does not have a singularity. The density of spin supercurrent in Eq.(\ref{SpinCurrent2}), which is $J_{\rm spin}\propto (1-\cos\beta)\nabla \alpha$, virtually goes to zero near the core, as was calculated by Fomin.  According to Eq.(\ref{CoreRadiusLarge}) the magnetic coherence lengthe $\xi$ and the size of the vortex core diverge when the HPD domain boundary is approached where the local Larmor frequency $\omega_L=\omega$.}
  \label{VortexFig1}
\end{figure}

Then the spin
current vortex has been observed  \cite{Vortex} --  a topological
defect  which is an analog of a quantized vortex in superfluids
and of an Abrikosov vortex in superconductors. The precession angle $\alpha$ has $2\pi$ winding around the vortex core. In the magnon BEC description, where $\alpha$ is the phase of magnon condensate, this is the mass current vortex, and since magnons are spin polarized this gives rise to spin  current in Eq.(\ref{SpinCurrent2}) circulating  around the vortex core,  see Fig. \ref{VortexFig1}.

   \begin{figure}
\centerline{\includegraphics[width=1.0\linewidth]{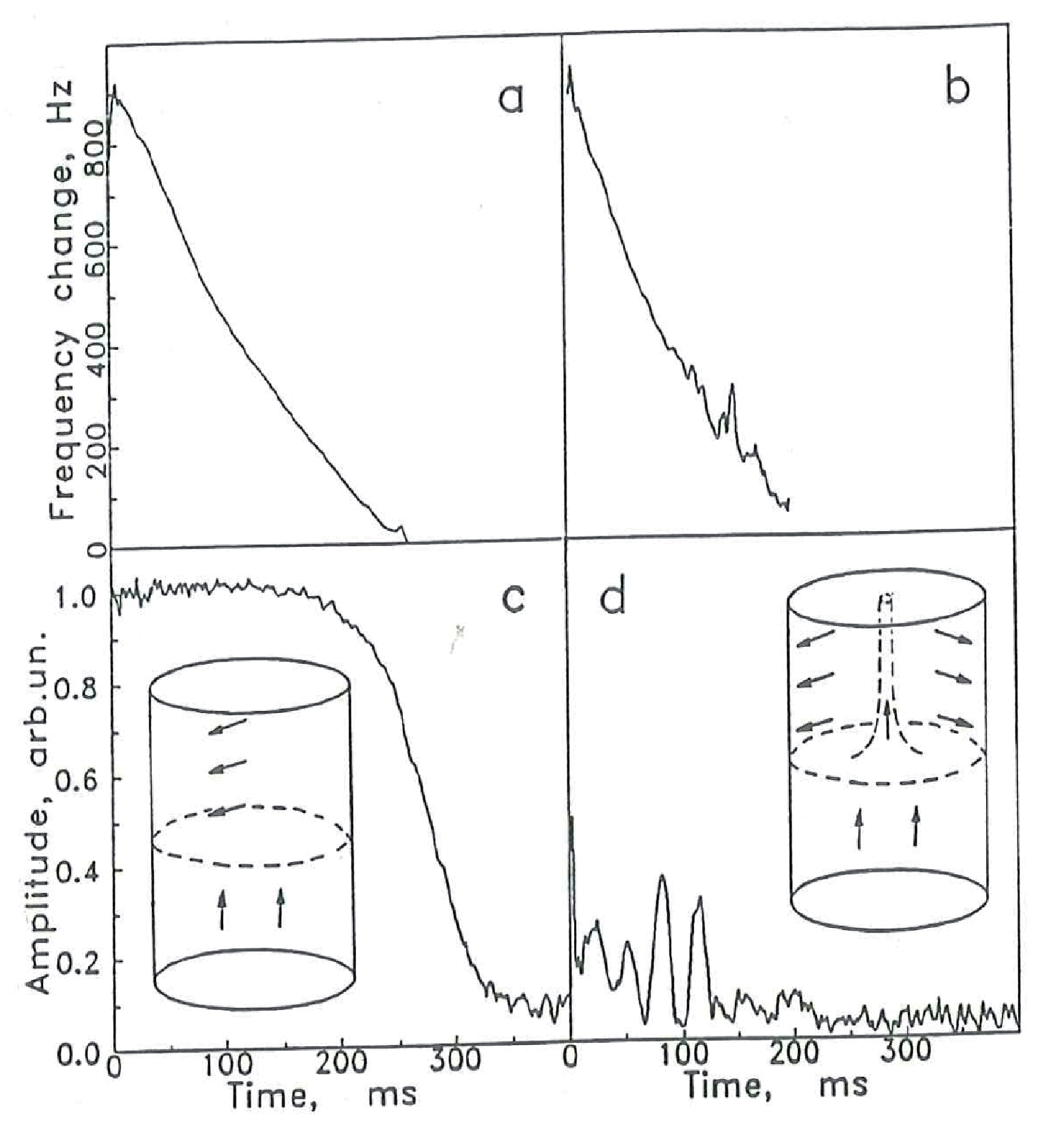}}
  \caption{NMR signature of spin vortex in magnon BEC in $^3$He-B. Frequency (top) and amplitude (bottom) of HPD induction decay measured by a pick-up coil.
 The HPD was maintained with an RF field from parallel connected co1ls (left) and oppositely
(quadrupole) connected coils (right) at $P= 29.3$ bar, $T=0.5 T_c$. After the quadrupole excitation the frequency decays with time in the
same way as in conventional vortex-free precession, while the amplitude of the HPD signal is nearly zero because of compensation of signals from opposite sides of the cells where the phase $\alpha$ of precession differs by $\pi$.}
  \label{VortexFig2}
\end{figure}

 Since in the central part of  cylindrical cell  the
phase $\alpha$  changes by $2\pi$  around the center, 
the transverse magnetization ${\bf M}_\perp$ is opposite on the opposite sides of the cell. In the
central part of the cell, i.e. in the vortex core,  the magnetization remains vertical and does not
precess. The magnon BEC with a spin vortex is  created   by applying the quadrupole RF field.
For this purpose  two parts of the saddle NMR coil are connected in opposite
directions, so that the phase of RF field (and consequently the phase $\alpha$) was opposite at the opposite sides of the cell. By these NMR coils practically the same HPD
signal was observed as in the conventional arrangement with the parallel connection of the coils, though with a slightly reduced  amplitude.Ê This shows that HPD is created   with opposite $\alpha$ on opposite sides of the cell. To verify this  a pair of small pick-up coils  are installed  at the top of the cell connected in usual way. When the RF field is switched off, the pick up coils received a very small RF signal from HPD, while the
frequency of this signal corresponded to the full  HPD signal. This
means that HPD generated the signal with the opposite phase at the two sides of the pickup coils, which nearly compensated each other (see
Fig.~\ref{VortexFig2}).
This corresponds to  HPD with a circular gradient of $\alpha$, as shown in Fig.~\ref{VortexFig1}. The magnetization is oriented vertically in the vortex core. On theÊ 
periphery of the cell it precesses with tipping angle 104$^\circ$,
and with $2\pi$ phase winding around the center. This type of HPD should
radiate at frequency, which corresponds to the Larmor field on the
boundary of HPD, but should not produce any signal in the pick-up coil.
A small signal appears due to asymmetry of the pick-up coil;  oscillations of this signal may correspond toÊ nutations of the vortex
core.

\subsection{Critical velocities, coherence length and the vortex core radius}

In atomic BEC the speed of sound determines the coherence length, the size of the vortex core and
 the Landau critical velocity of flow at which phonons are created:
\begin{equation}
v_{\rm L}=c_s~~,~~ r_{\rm core}\sim  \xi \sim  \frac{\hbar}{m c_s} \,.
\label{LandauVelocity}
\end{equation}
The extension to the magnon BEC would suggest that the coherence length and the size of the vortex core in the HPD state should be on the order of:
\begin{equation}
r_{\rm core}\sim   \frac{\hbar}{m_M c_s} \sim \frac{c}{\Omega_L} ~.
\label{CoreRadius}
\end{equation}
However, this naive extension does not work, and Eq.(\ref{CoreRadius}) gives only the lower bound on the core size. The core is larger due to specific profile of the Ginzburg-Landau (dipole) energy in Eq.(\ref{FHPD}) which is strictly zero for $\cos\beta>-1/4$. This leads to the special topological properties of coherent precession (see  Ref. \cite{MisirpashaevVolovik1992}). As a result the spin vortex created and observed  in Ref. \cite{Vortex}  has a continuous core with broken symmetry, similar to vortices in superfluid $^3$He-A \cite{SalomaaVolovik1987}. The size of the continuous core is determined by the proper coherence length   \cite{Fomin1987} which can be found from the competition between the first two terms in the Ginzburg-Landau free energy in Eq.(\ref{GLfunctional}):
\begin{equation}
r_{\rm core}\sim  \xi \sim  \frac{\hbar}{\sqrt{m_M(\mu -\omega_L)} }  \sim   \frac{c}{\sqrt{\omega_L(\omega -\omega_L)}}   ~.
\label{CoreRadiusLarge}
\end{equation}  
This coherence length determines also the critical velocity for creation of vortices:
\begin{equation}
v_c\sim \frac{\hbar}{m_M r_{\rm core}}\,.
\label{CritVelocity}
\end{equation}  
It is smaller than the Landau critical velocity for creation of phonons, which in the case of isotropic sound is
\begin{equation}
v_{\rm L}=c_s \,.
\label{LandauVelocity}
\end{equation} 
The Landau criterion for onset of the phonon radiation in the case of anisotropic speed of sound has been derived in Ref. \cite{Sonin1987}.

For  large tipping angles of precession  the symmetry of the vortex core is restored: the vortex becomes singular with the core radius
$r_{\rm core}\sim c/\Omega_L$ in \eqref{CoreRadius}  \cite{SoninVortex}.

Other topological defects possible in the coherent precession beyond the Ginzburg-Landau model of magnon BEC are discussed in \cite{MisirpashaevVolovik1992}. In BEC of excitations, the topological defects have been detected  in exciton-polariton condensate \cite{Lagoudakis2008}. Among them the
half quantum vortices -- vortices with half of the circulation quantum.  Half quantum vortices are topologically stable in the superfluid $^3$He-A \cite{VolovikMineev1976}, but they still remain elusive there.

\subsection{Spin supercurrent transport}

The next step in investigations of the magnon BEC  was the experimental studies of spin supercurrent between two independent HPD states, connected by a channel which was either perpendicular  \cite{6,62,R1} or parallel to magnetic field \cite{63}.

\begin{figure}[htt]
 \includegraphics[width=1\textwidth]{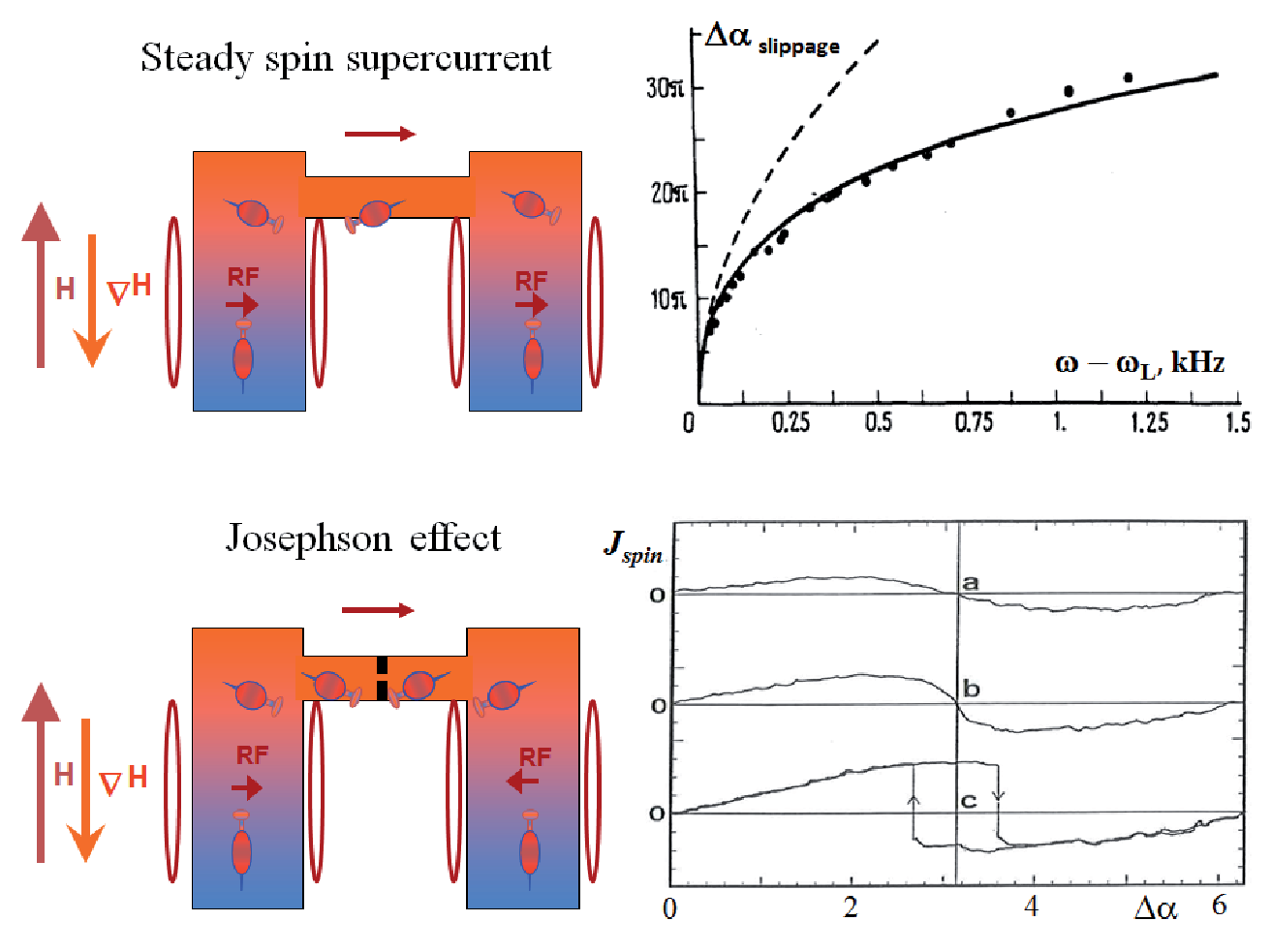}
 \caption{ Illustration of experimental observation of spin supercurrent between two HPD states ({\it top left})
 and the Josephson effect ({\it bottom left}).  The measured critical spin current in the channel as a function of $\omega-\omega_L$ and thus as a function 
 of the coherence length $\xi$ in  \eqref{CoreRadiusLarge} ({\it top right}). For observation of the dc and ac Josephson effects the orifice of diameter about 0.4 mm was installed ({\it bottom left}). The  Josephson effect for magnon BEC  demonstrates the interference between two magnon condensates. Spin current as a function of the phase difference across the junction,
 $\alpha_2-\alpha_1$, where $\alpha_1$ and  $\alpha_2$ are phases of precession in two coherently precessing domains  ({\it bottom right}).
Different experimental records correspond to a different ratio between the diameter of the orifice and the coherence length $\xi$ of magnon BEC.  The pure dc Josephson phenomenon was observed for magnetic coherent length $\xi=1.3$ mm (a)  and the distorted one for $\xi=0.8$ mm (b). The phase slippage processes were observed for $\xi=0.7$ mm (c).
  }
 \label{Jos}
\end{figure}

In the first case the steady state spin supercurrent was created between two magnon condensates formed in two different cells. The two cells were connected by a channel of a diameter 1.4 mm. The HPD states were formed in both cells by a CW NMR.  The frequency of RF field $\omega$ was chosen to be slightly above the local Larmor frequency $\omega_L$ in the channel, this determines the coherence length $\xi$ of magnon BEC in \eqref{CoreRadiusLarge}. The HPD penetrates in a channel as shown in Fig. (\ref{Jos}). Then one  slightly changes the frequency in one of the cells. The gradient of phase of precession appears which leads to the growing spin supercurrent in the channel, which is proportional to the difference of the phases between two HPD. This is the current of magnons, which transports the magnetization and consequently the Zeeman energy from one cell to another. As a result, in one of the cells the density of magnons decreases and the HPD starts to absorb more RF energy to compensate the extra losses. Contrary, in the other cell the more magnons appear and thus the absorbed energy  decreases. At some conditions the transport of magnons becomes so big, that absorption transforms to radiation from HPD. It means that the apparatus starts to operate as a spin supercurrent transformer, which transports the RF signal from one coil to another. By changing  the amplitude of the RF field one can measure directly the value of spin supercurrent in the channel.  If one removes the difference in frequencies of two HPD states (the difference in chemical potentials of two magnon condensates),  the spin supercurrent remains stable and is proportional to the difference of phases between the condensates. By increasing the phase difference one is able to reach a critical current, at which the spin supercurrent starts loosing the 2$\pi$ windings. These phase slippages were measured as a function of $\omega-\omega_L$ in the channel and thus as a function of the coherence length $\xi$, in a good agreement with theory, see Fig. \ref{Jos} {\it top right}.  The 2$\pi$ slippage cases as well a multi 2$\pi$ slippage were observed. In some cases the 2$\pi$ slippage was separated by two independent  slippages, which can be explained by the formation of a spin-current vortex  \cite{R1} or $\pi$-soliton inside the channel as the intermediate state.

\subsection{Spin-current Josephson effect}
\label{SpinJosephson}

The Josephson effect is the response of the current to the phase between two
weakly connected regions of coherent quantum states. It was described by Josephson
\cite{Joseph}  for the case of two quantum states, separated by the potential barrier.
This phenomenon is usually studied for the case of quantum states connected
by a conducting bridge with the dimensions smaller than the coherence
length. In this case the coherent state in the bridge cannot be established so there
is no phase memory, which determines the direction of the phase gradient. As a
result the supercurrent is determined only by the phase difference between the
two states.
As the dimensions of the conducting bridge increase, the more complex   current-phase
relation is observed. For bridge dimensions of the order of the
coherence length, a transition to a hysteretic scenario with phase slippage appears.

In the case of mass and electronic supercurrents the coherence length is a
function of the temperature. In the case of spin supercurrents, however, the Ginzburg-
Landau coherence length $\xi$ is not only a function of temperature,
but also a function of the difference between the HPD precession frequency
and the local Larmor frequency, according to Eq.(\ref{CoreRadiusLarge}). This 
quantity can be varied experimentally
with a magnetic field gradient or position of the domain boundary. As a
result one is able to change the coherence length in the region of the orifice in
the channel and observe the change  from the canonical current-phase relation to
phase slip behavior.
This experiment made in Kapitza Institute    \cite{7,72,R1}  is schematically presented in Fig. \ref{Jos}.  The orifice, of diameter 0.48 mm, was placed in the
central part of the channel. The current-phase characteristics, observed in this
experiment are represented  for different positions of the domain
boundary related to the orifice. One can easily see that the current in Fig. (a)
corresponds to the canonical current-phase relation, which transforms  to the nonlinear 
relation in Fig. (b) and then  to a phase slip phenomenon in Fig. (c).

\subsection{Other states of magnon BEC in  $^3$He-B}

Recent experiments in $^3$He-B allowed to probe the BEC states that emerge in the valley on the other side 
of the energy barrier in Fig. \ref{profile}. This became possible  by  immersing the superfluid $^3$He in a very porous material called aerogel. By squeezing or stretching the  aerogel sample, one creates the global anisotropy which captures  the orbital vector $\hat{\bf l}$. This  allows to orient the orbital vector $\hat{\bf l}$ in the desirable direction with respect to magnetic field  \cite{JapGren,GrenB}. 

For the transverse orientation of  $\hat{\bf l}$, i.e. for $l=0$, two new BEC states have been identified. One of them exists at $\vert\Psi\vert^2< S/\hbar$ and has the following form of spin-orbit interaction obtained from Eq. \eqref{FD} (we omit for simplicity the constant term):
 \begin{equation}
F_{\rm so} \left(\Psi\right)_{l=0}=- \frac{\chi}{4\gamma^2}\Omega_L^2
 \left(\frac{\vert\Psi\vert^2}{S}-\frac{4}{5}\right)^2
~, ~ \vert\Psi\vert^2<\frac{S}{\hbar}.
 \label{pot1}
\end{equation}
This state has an attractive interaction between magnons, and  is unstable since the compressibility $\beta_M$ of the magnon gas in \eqref{compressibility} is negative: $d^2\epsilon/dn^2 <0$.

The other state exists at $\vert\Psi\vert^2> S/\hbar$ and has the following form of spin-orbit interaction 
\begin{equation}
F_{\rm so} \left(\Psi\right)_{l=0}= \frac{\chi}{20\gamma^2}\Omega_L^2  
\left(\frac{\vert\Psi\vert^2}{S}-2\right)^2
~,~ \vert\Psi\vert^2>\frac{S}{\hbar}~. 
\label{pot2}
\end{equation}
This state has repulsive interaction between magnons and is stable. The magnon BEC formation  under these conditions has been observed \cite{New}.

\section{Magnon BEC in $^3$He-A}

As in the case of $^3$He-B, all the information on the $^3$He-A order parameter needed to study the coherent precession  is encoded in the spin-orbit interaction. 
 
 \subsection{Instability of magnon BEC in bulk $^3$He-A}

For $^3$He-A, the spin-orbit interaction averaged over the fast precession has the following form 
\cite{BunkovVolovik1993}:
\begin{eqnarray}
  F_{\rm so} \left(\vert\Psi\vert\right)= \frac {\chi\Omega_L^2}{4\gamma^2}\times 
  \nonumber
  \\ 
  \left[ -2\frac{\vert\Psi\vert^2}{S} +
  \frac{\vert\Psi\vert^4}{S^2}    +
    \left( -2+4 \frac{\vert\Psi\vert^2}{S}  -
  \frac {7}{4}\frac{\vert\Psi\vert^4}{S^2}\right)(1-l^2)\right]
 \label{FDA}
  \end{eqnarray}
 In a static bulk $^3$He-A,
when $\Psi = 0$,  the spin-orbit energy $F_{\rm so}$ in  Eq.(\ref{FDA}) is minimized when the orbital vector $\hat{\bf l}$
is perpendicular to magnetic field, i.e. for $l= 0$. Then one has
\begin{equation}
  F_{\rm so} \left(\vert\Psi\vert, l=0\right)= \frac {\chi\Omega_L^2}{4\gamma^2}
  \left[-2+ 2\frac{\vert\Psi\vert^2}{S} -  \frac {3}{4}
  \frac{\vert\Psi\vert^4}{S^2}    \right],
  \label{FDl=0}
  \end{equation}
with a negative quartic term. The attractive interaction between
magnons destabilizes the BEC, which means that homogeneous
precession of magnetization in $^3$He-A becomes unstable. This instability  
predicted by Fomin \cite{Fomin1979} was experimentally confirmed \cite{InstabAB}.

\begin{figure}[htt]
 \includegraphics[width=0.7\textwidth]{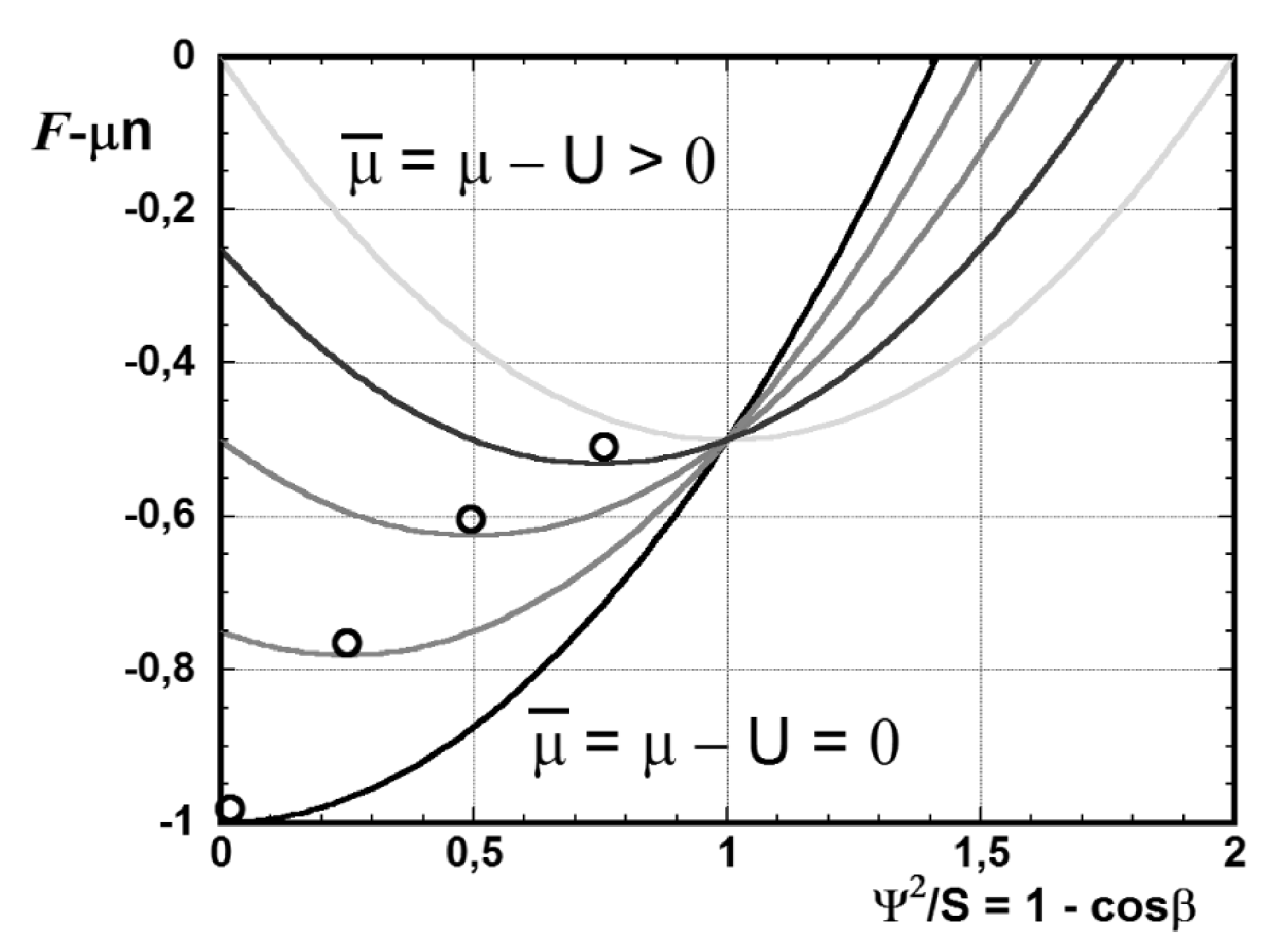}
 \caption{$F-\mu n$ for different values of the chemical potential
$\mu\geq U$ in magnon BEC in $^3$He-A. Magnon BEC in $^3$He-A is similar to BEC in atomic gases.}
 \label{3He-A}
\end{figure}

However, as follows from \eqref{FDA}, at sufficiently large magnon
density $n=\vert\Psi\vert^2$
 \begin{equation}
 \frac{8+\sqrt{8}}{7}~S> n> \frac{8-\sqrt{8}}{7}~S~,
\label{PositiveCondition}
\end{equation}
the factor in front of $l^2$ becomes negative. Therefore
it becomes energetically favorable to  orient the orbital momentum $\hat{\bf l}$
along the magnetic field, $l=1$. 
For this  orientation one obtains the Ginzburg-Landau free energy with
\begin{equation}
  F_{\rm so} \left(\vert\Psi\vert, l=1\right)= \frac {\chi\Omega_L^2}{4\gamma^2}  
  \left[ -2\frac{\vert\Psi\vert^2}{S} + \frac{\vert\Psi\vert^4}{S^2}  \right]~.
  \label{FDl=1}
  \end{equation}
It corresponds to the conventional  Ginzburg-Landau free energy in atomic BEC. The quadratic term modifies the potential $U$; the  quartic term is now positive.

In the language of BEC, this means that, with increasing the density
of Bose condensate, the originally attractive interaction between
magnons should spontaneously become repulsive when the critical
magnon density $n_c= S(8-\sqrt{8})/7$ is reached. If this happens, the magnon BEC becomes
stable and in this way the state with spontaneous coherent precession 
could be formed \cite{BunkovVolovik1993}. This self-stabilization
effect is similar to the effect of $Q$-ball, where bosons create the
potential well in which they condense (we shall discuss the $Q$-ball phenomenon in magnon BEC later on in Sec. \ref{Qball}). 
However, such a self-sustaining BEC with
originally attractive boson interaction has not been achieved
experimentally in bulk $^3$He-A, most probably because of the large dissipation, due
to which the threshold value of the condensate density has not been
reached.

\subsection{Magnon BEC of $^3$He-A in deformed aerogel}

Finally the fixed orientation of the orbital  vector $\hat{\bf l}$ has been
achieved in $^3$He-A confined in aerogel -- the material with high
porosity, which is about 98\%
 of volume. Silicon strands of aerogel play the role of impurities with local
anisotropy along the strands.  According to the Larkin-Imry-Ma effect, the
random anisotropy  suppresses the orientational long-range order of the orbital vector $\hat{\bf l}$; however, when the aerogel sample is deformed the
long-range order of $\hat{\bf l}$ is restored
\cite{VolovikAerogel}.  Experiments with globally squeezed aerogel
\cite{JapGren} demonstrated that a uni-axial deformation by about 1\%
is sufficient for global orientation of the vector $\hat{\bf l}$  along the anisotropy axis.
When magnetic field
is also  oriented along the anisotropy axis one obtains the required geometry  with $l=1$, 
at which the magnon BEC in $^3$He-A becomes stable.
The first indication of coherent precession  in $^3$He-A has been
reported in \cite{Sato2008,BunkovVolovik2009b} and confirmed in \cite{Hunger2010}.  Contrary to the
unconventional magnon BEC in the form of HPD in $^3$He-B, the magnon
BEC emerging in the superfluid $^3$He-A is in one-to-one
correspondence with the atomic BEC, see  Fig. \ref{3He-A}.  For
$\mu>U$, the condensate density determined from equation $dF/dn=\mu$
continuously grows from zero  as $n \propto \mu -U$.

For $l=1$ the Ginzburg-Landau free energy acquires the standard form:
\begin{equation}
 F=\int d^3r\left(\frac{\vert\boldsymbol{\nabla}\Psi\vert^2}{2m} +(\omega_L({\bf r})-\mu)\vert\Psi\vert^2+\frac{1}{2}b\vert\Psi\vert^4\right),
\label{GL2}
\end{equation}
where we modified the chemical potential by the constant frequency shift:
\begin{equation}
\mu= \omega + \frac{\Omega_L^2}{2\omega} ~,
   \label{TildeOmega}
   \end{equation}
   and the parameter $b$ of repulsive magnon interaction is
 \begin{equation}
b= \frac{\Omega_L^2}{2\omega S}
 \label{b}
  \end{equation}
At $\mu>\omega_L$, magnon BEC must be formed with
density
\begin{equation}
\vert\Psi\vert^2=\frac{\mu-\omega_L}{b} ~.
   \label{EquilibriunPsi}
\end{equation}
This is distinct from $^3$He-B, where condensation starts with finite condensate density.
Eq. (\ref{EquilibriunPsi}) corresponds to the following dependence of  the frequency
shift on tipping angle $\beta$ of coherence precession:
 \begin{equation}
\omega - \omega_L= - \frac{\Omega_L^2}{2\omega} \cos\beta~.
 \label{freqshiftNoRF}
\end{equation}

The final proof of the coherence of precession in $^3$He-A in aerogel
was the observation of the free precession after a pulsed NMR and also after a switch
off the CW NMR \cite{Hunger2010}.  
In conclusion, in contrast to the homogeneously precessing domain
(HPD) in $^3$He-B,  the magnon Bose condensation in $^3$He-A obeys
the standard Gross-Pitaevskii equation. In bulk $^3$He-A, the Bose
condensate of magnons is unstable because of the attractive
interaction between magnons. In $^3$He-A confined in aerogel, the
repulsive interaction is achieved by the proper deformation of the
aerogel sample, and the Bose condensate becomes stable.

\section{Magnon BEC in magnetic trap and MIT bag}
\label{MIY}

\subsection{Magnon BEC in the form of Q-ball}
\label{Qball}

There are many new physical phenomena related to the  Bose
condensation of magnons, which have been observed after the discovery of HPD. These include in particular compact objects -- coherently precessing states trapped by orbital texture  \cite{Bunkov2005}.
At small number $N$ of the pumped magnons, the system is similar to the Bose condensate of the ultracold atoms in harmonic traps, while at larger $N$ the analog of the $Q$-ball in particle physics develops \cite{BunkovVolovik2007}.

A $Q$-ball is a non-topological soliton solution  in field theories containing a complex scalar field $\Psi$.  $Q$-balls are stabilized due to the conservation of the global $U(1)$ charge $Q$  \cite{Coleman1985}.
They are formed due to suitable attractive interaction
that binds the quanta of $\Psi$-field into a large compact object. In some modern SUSY scenarios $Q$-balls are considered as a  heavy particle-like objects, with $Q$ 
being the baryon and/or lepton number.  For many conceivable
alternatives, $Q$-balls may contribute significantly to the dark
matter and baryon contents of the Universe, as described in review
\cite{Enq}. Stable cosmological $Q$-balls can be searched for in existing
and planned experiments \cite{Kus}.

The $Q$-ball is a rather general physical object, which in principle
can be formed  in  condensed matter systems.
In particular, $Q$-balls were suggested  in the atomic Bose-Einstein
condensates \cite{QBallBEC}.
In $^3$He-B, the $Q$-balls are formed as special
states of  phase coherent  precession of magnetization. The role
of the $Q$-charge is played by the projection ${\cal S}_z$ of the total spin of the
system on the axis of magnetic  field, which  is a rather well
conserved quantity at low temperature, or which is the same the magnon number
${\cal N}$. At the quantum level, this $Q$-ball is a compact object formed by magnons -- quanta of the corresponding $\Psi$-field.

In $^3$He-B the $Q$-balls are formed at low temperatures, when  homogeneous magnon BEC in the form of Homogeneously Precessing Domain (HPD) becomes unstable due to parametric Suhl instability   \cite{Catastropha,Catastroph2,Catastroph3}, which we shall discuss later in Sec. \ref{ParametricInstability}. At low  temperatures the condensate can be formed only in a trap, similar to that in atomic gases \cite{PitaevskiiStringari2003}, and the $Q$ balls are either formed in these traps or dig their own trap.

Experimentally the $Q$-ball in $^3$He-B \cite{BunkovVolovik2007} is manifested as a long-lived ringing  (of up to an hour!) of the free induction decay after a NMR tipping pulse \cite{PS,pers2}. In a steady state it can be maintained by CW RF pumping  \cite{CWGRE1,CWGRE2} (Fig.~\ref{specexamp}), and even by off-resonance excitation \cite{Bunkov2005,PS2}. The detailed experimental  investigations of $Q$-balls formed in the specially
prepared and  the well controlled traps 
were made in  \cite{Autti2012}.

 \begin{figure}[t]
\includegraphics[width=0.6\textwidth]{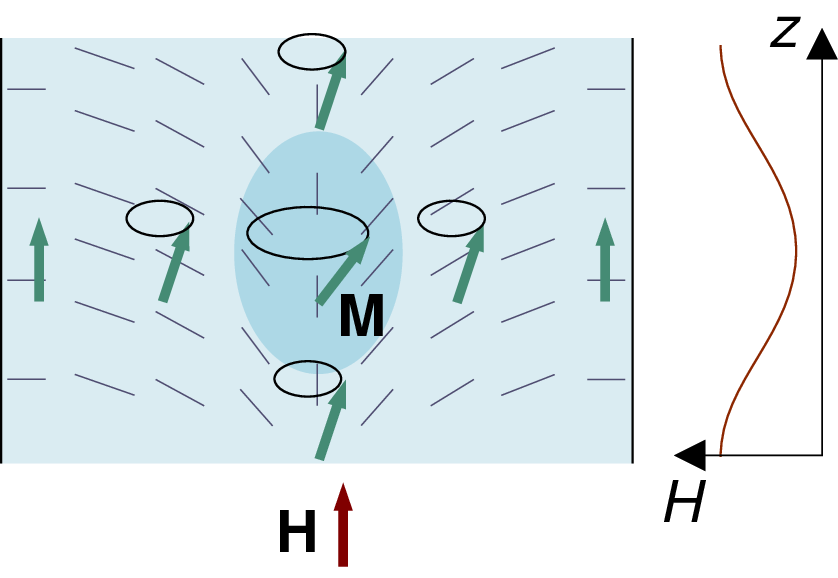}
 \caption{The trapping potential Eq.~\eqref{Potential1} used in \cite{Autti2012} is formed in the cylindrically symmetric ``flare-out'' texture of the  orbital angular momentum ${\bf l}$ (dashed lines) in a shallow minimum of the vertical magnetic field ({\it right}). The arrows represent the precessing magnetization ${\cal M}$, which precesses coherently within the condensate droplet (dark blue). The radial texture
 is manipulated by rotation.
 }.
 \label{InitialTrap}
 \vspace{-5mm}
\end{figure}
 
\subsection{Magneto-textural  trap for magnon BEC in $^3$He-B}

A cylindrically symmetric magnetic trap for magnon BEC in $^3$He-B is schematically shown in Fig.~\ref{InitialTrap} \cite{Autti2012}. The confinement potential  $U_\parallel(z)=\gamma H(z)$ in the axial direction is  produced by local perturbation of magnetic field with a small pinch coil.  In the radial direction the well $U_\perp(r)$ is formed by the cylindrically symmetric flare-out texture of the orbital vector ${\bf l}$. It comes from  the  spin-orbit interaction energy (\ref{FD}), which enters the Ginzburg-Landau functional. The relevant term  in Eq.~(\ref{FD}), which is responsible for the radial potential, is
  \begin{equation}
U_\perp(r)|\Psi|^2= \frac{2\Omega_\mathrm{L}^2}{5\omega_\mathrm{L}}\left(1-l(r)\right)  |\Psi|^2 \,.
\label{OrientationalEnergy}
\end{equation}
Here as before $l=\hat{\bf l}\cdot \hat{\bf H}\equiv \cos \beta_L$ describes the orientation of the unit vector $\hat{\bf l}$ with respect to the direction $\hat{\bf H}$ of magnetic field.
On the side wall of the cylindrical container the orbital momentum $\hat{\bf l}$ is normal to the wall,  while in the center it is parallel to the axially oriented applied magnetic field. This produces a minimum of the potential $U_\perp(r)$ on  the cylinder axis.
So the total confinement potential is
  \begin{equation}
  U({\bf r})= U_\parallel(z) + U_\perp(r)= \omega_\mathrm{L}(z) +
\frac{2\Omega_\mathrm{L}^2}{5\omega_\mathrm{L}}\left(1-l(r)\right) \,.
\label{Potential1}
\end{equation}
 Close to the axis the polar angle $\beta_L$ of
the $\hat{\bf l}$-vector varies linearly with distance $r$ from the axis \cite{SalomaaVolovik1987}. As a result, the
potential  $U({\bf r})$ reduces to that of a usual harmonic trap used for the
confinement of dilute Bose gases \cite{PitaevskiiStringari2003}:
 \begin{equation}
U({\bf r})= U(0) + \frac{m_\mathrm{M}}{2} \left( \omega_z^2 z^2 + \omega_r^2 r^2\right)\,,
\label{Potential2}
\end{equation}
where we take into account that the axial trap is also close to harmonic. This is checked
by measurement of the spectrum of magnons --  standing spin waves in the trap, which has  equidistant levels:
\begin{equation}
 \omega_{\mathrm{nm}}=\omega_\mathrm{L}(0)+   \omega_r(\mathrm{n}+1) + \omega_z(\mathrm{m} +1/2)  \,.
\label{SpinWaveSpectrum}
\end{equation}
The  oscillator frequency $\omega_z$ of the well in the axial direction can be regulated by the field in the pinch coil (the equidistant levels of spin precession localised around the minimum of the external magnetic field
have been derived in Ref. \cite{KupkaSkyba2003} using the full set of equations for spin dynamics
in $^3$He-B). The frequency $\omega_r$ in the radial direction can be adjusted by applying rotation, since the vortex-free superfluid flow or the array of rectilinear vortex lines created by rotation modifies the flare-out texture
$\hat{\bf l}(r)$.

\subsection{Ground-state condensate and self-localization} 

In atomic BEC the condensate is formed in the ground state (0,0) of the trap. When the number of atoms ${\cal N}$ in the ground state increases,  the interactions between atoms become important and the condensate wave function starts deviating from the Gaussian form of an ideal gas. However, as distinct from a system of cold atoms, the peculiarity of the magnon Ginzburg-Landau functional in Eq.~(\ref{FD}) is that  the  prefactor of  the quartic term, which describes the interactions between the magnons, is not a constant, but $\propto(1-l(r))^2|\Psi|^4  \propto r^4|\Psi|^4$. It is  small in the region of the trap and can be neglected. 

Under conditions of  experiment the main effect is caused not by the atom-atom interactions, but by interaction of magnons with $\hat{\bf
l}$-field, which leads to the self-localization discussed in \cite{BunkovVolovik2007}.  At high density of magnons
they start to influence the radial $\hat{\bf
l}$-texture. According to \eqref{OrientationalEnergy} the condensation of $\Psi$ in the trap leads to the preferable   orientation of  $\hat{\bf
l}$ parallel to magnetic field, $l=1$, in the region of the trap.   As a result  the potential well becomes wider and the energy of the level in the trap decreases, and at large ${\cal N}$ the harmonic trap gradually transforms to the box with $\beta_\mathrm{L}\approx 0$ within which magnons are localized. This allows to incorporate more magnons at this same level  by  sweeping the frequency up. 
This is equivalent to effective attractive interaction induced by the exchange of the quanta of the $\hat{\bf
l}$-field. 

In the language of relativistic quantum fields, this is a particular representation of the 
$Q$-ball \cite{Friedberg1976}, in which the self-localization is caused by interaction between the charged field (magnon field  $\Psi$) and neutral field ($\hat{\bf
l}$-field),  where the neutral field  provides the potential for the charged one. In the process of self-localization the charged field modifies locally the neutral field so that the potential well is formed in which the charge is condensed.  We remind that the charge $Q$ corresponds to the spin $S-S_z$ or, equivalently, to the magnon number ${\cal N}$.

\subsection{Localization with formation of a box: analog of electron bubble and MIT bag} 

The phenomenon of self-localization with formation of a box
in \cite{Autti2012} is not unique in nature. Other examples of self-formation of a box-like trapping potential are the electron bubble in liquid helium  and the MIT bag
model of a hadron \cite{Chodos1974}, where the asymptotically free
quarks are confined within a cavity surrounded by the QCD
vacuum.

\begin{figure}[t]
\includegraphics[width=\linewidth]{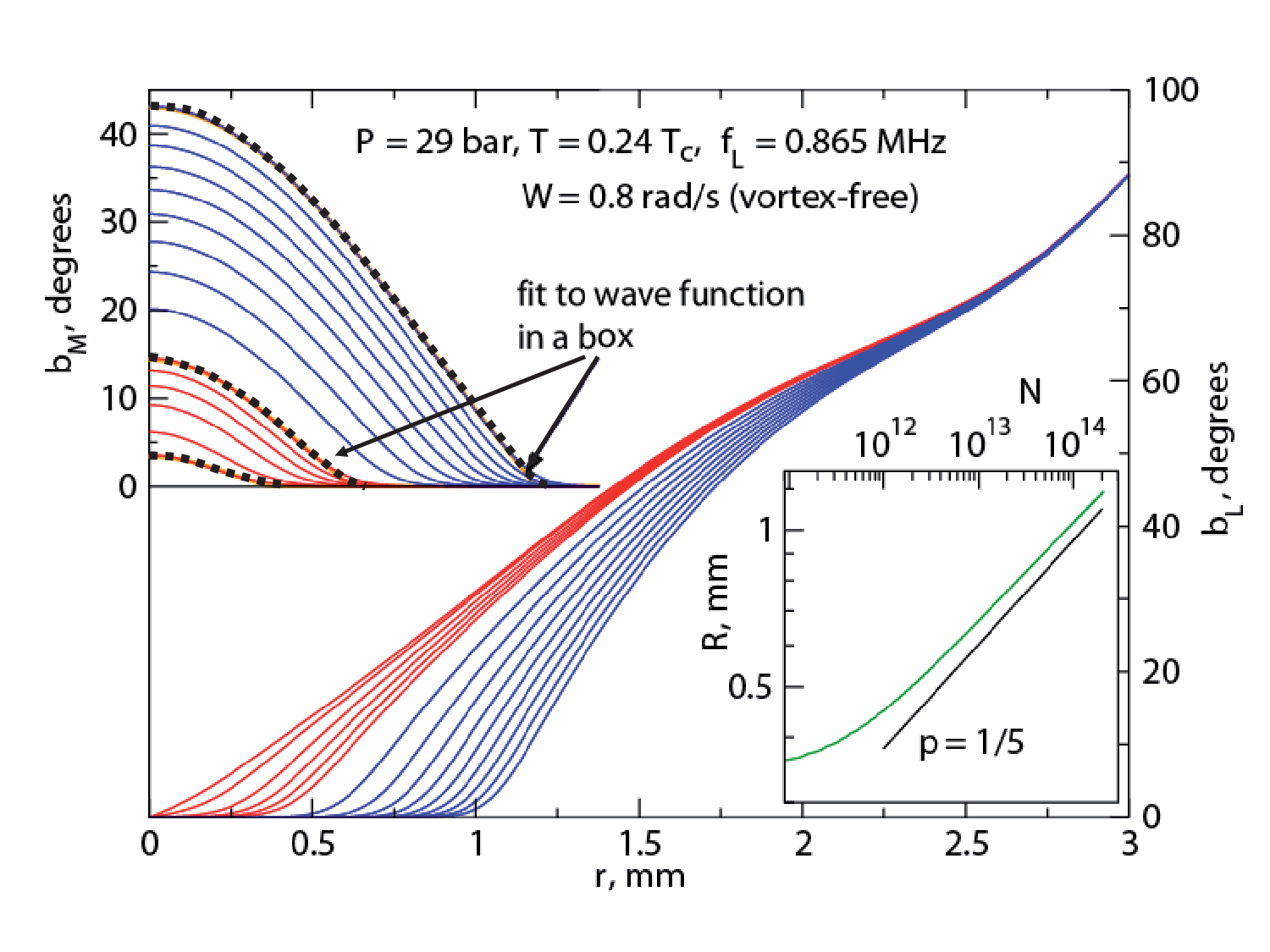}
\caption{Calculations of the multi-magnon bubble in the 2D textural trap using the magnon BEC approximation (see Ref. \cite{Autti2012}).  The  condensate is formed in the ground state, $n_r=0$. The deflection angles of the magnetization $\beta_{\rm M}$ and
  of the textural anisotropy axis $\beta_{\rm L}$ are plotted as a function
  of radius for different condensate
  populations. The population increases from bottom to top in the upper
  plot and from left to right in the lower plot.  When the magnon
  occupation increases, the magnon wave function suppresses the
  orbital texture $\beta_\mathrm{L}$ and the potential well
  transforms towards a box with impenetrable walls. Fit to the wave function of the condensate in a
  box, which is nullified at the box boundary, is shown in the upper plot. The effective radius
  of the box obtained from such fits is shown in the \textit{insert} as a
  function of the magnon occupation number $\cal N$. Slope  $p=1/(k+2)$  from
  Eq.~\eqref{boxrad} with $k=3$ is shown for comparison. }
 \label{magnon_bubble}
\end{figure}

The MIT bag model has been used for construction of different hadrons, including mesons, baryons and even multiquark hadrons, such as tetraquarks \cite{Jaffe1977} and pentaquarks \cite{Strottman1979}. In the MIT bag model, free quarks are forced to move only inside a given spatial region, within which they occupy single-particle orbitals. MIT bag is described by the following energy whose minimization determines the equilibrium radius $R$ of  a given hadron:
 \begin{equation}
E(R)=\sum_a N_a \sqrt{m_a^2c^4 + \frac{\hbar^2c^2x_a^2}{R^2}}  + F(R) ~~,~~ F(R)=B \frac{4\pi R^3}{3} \,.
\label{MITbag}
\end{equation}
Here the first term is the kinetic energy of quarks with masses $m_a$ in the cavity of radius $R$, where parameters $x_a$ are determined by the boundary conditions for fermions on the boundary of the bag
and radial quantum numbers. For the fermion in the ground state  in the box, and  in the ultra-relativistic limit of vanishing fermionic masses, $m_a\rightarrow 0$, the parameter $x = 2.04$. The second term is the potential energy, $B$ is the so-called bag constant that reflects the bag pressure. At zero temperature, the bag constant $B$ is the difference in the energy density between the false vacuum inside the bag (the deconfinement phase) and the true QCD vacuum outside  (the confinement phase).
In the non-relativistic limit, ignoring the term which does not depend on $R$, one gets
 \begin{equation}
E(R)=\sum_a N_a  \frac{\hbar^2 x_a^2}{2m_aR^2}  + F(R) \,.
\label{MITbag_Non-Rel}
\end{equation}
The same equation describes the electron bubble in superfluid $^4$He, where $m$ is the electron mass; $x=\pi^2$ for the ground state level; the potential energy $F(R) = (4\pi/3)R^3P +4\pi \sigma R^2$, with $P$ being the external pressure and $\sigma$  the surface tension.
Extension to the multi-electron bubbles in superfluid $^4$He see in Ref. \cite{SalomaaWilliams1981}. 
 
This consideration is applicable for magnon BEC.
In the harmonic trap for magnons presented in Fig. \ref{InitialTrap} the flexible texture of the orbital momentum $\hat{\bf l}$ of Cooper pairs in our analogy plays either the role of the pion field or the role of the non-perturbative gluonic field depending on the microscopic structure of the confinement phase.
The trap is modified by pumped magnons due to spin-orbit interaction in \eqref{OrientationalEnergy}, which repels the $\hat{\bf l}$-field from the region, where magnons are localized. At large number $N$ of magnons in the trap the systems becomes similar to MIT bag with cavity free from the orbital field, which is occupied by magnons. 
So magnons, like quarks,  dig a hole pushing the orbital field away due to the repulsive interaction,
see Fig. \ref{magnon_bubble}.
The main difference from the MIT bag model is that magnons are bosons and may macroscopically occupy the same energy state in the trap, forming the  Bose-condensate, while in MIT bag the number of fermions on the same energy level is limited by the Pauli principle. The bosonic bag becomes equivalent to the fermionic bag in the limit of large number of quark flavors, when ${\cal N}Ê\gg 1$ quarks 
may occupy the same level.

In experiments, the trap is elongated, so without loosing generality, we may consider the 2D approximation, i.e. the 2D cylindrical trap. In the limit of large ${\cal N}$, the radius $R$ of the cavity filled with ${\cal N}$ magnons occupying the quantum state with  radial number $n_r$ is determined by a balance of two terms in the total energy of the bag \cite{Autti2012}:
 \begin{equation}
E(R,n_r)={\cal N} \epsilon_{n_r}(R)+ F(R)~~,~~ \epsilon_{n_r}(R)=\frac{\hbar^2 \lambda_{n_r+1}^2} {2m_\mathrm{M}R^2} \,.
\label{balance}
\end{equation}
The first term on the rhs is magnon zero-point energy in the radial cavity. It is the magnon number ${\cal N}$
in the Bose condensate times the energy $\epsilon_{n_r}$ of a single magnon on the radial level $n_r$ in the cylindrical box
with impenetrable walls. In NMR experiments, where the homogeneous RF field is used, only the
energy levels with zero azimuthal quantum number are excited, which corresponds to $n=2n_r$ in Eq. (\ref{SpinWaveSpectrum}). They are measured as the shift  of the frequency  of the NMR peak, corresponding to excitation of a magnon,  with respect to the Larmor frequency $\omega_L$:
\begin{equation}
\Delta\omega=\omega-\omega_L=\frac{\epsilon_{n_r}}{\hbar}= \frac{\hbar \lambda_\mathrm{{n_r}+1}^2} {2m_\mathrm{M}R^2}    \,.
\label{R_vs_frequency}
\end{equation}
Here $m_\mathrm{M}$ is  as before the magnon mass, and the parameter $x$ in \eqref{MITbag_Non-Rel} equals the $n_r+1$-th root of the Bessel function, $x=\lambda_\mathrm{n_r+1}$, which corresponds to the proper boundary condition for the magnons populating the radial level $n_r$ in the impenetrable box. The potential energy $F(R)$  in \eqref{balance} corresponds to the pressure exerted to the bag by the field of the orbital texture, which is expelled from the bag. It is the difference in the energy of the orbital field texture with and without the cavity.  

Experimental results for the ground state magnon condensate in \cite{Autti2012} demonstrated that they can be reproduced by the phenomenological equation 
\eqref{balance} if one assumes that there is the scaling law  $F(R)\propto R^k$.  Minimization of the phenomenological equation Eq.~\eqref{balance} with respect to $R$ 
suggests that at large ${\cal N}$ the radius of localization approaches the asymptote:
\begin{equation}
R({\cal N})\sim ~a_r \left({\cal N}/{\cal N}_\mathrm{c}\right)^{1/(k+2)}~~,~~{\cal N} \gg {\cal N}_\mathrm{c}\,,
\label{boxrad}
\end{equation}
where $a_r$ is the harmonic oscillator length in the original radial trap (at ${\cal N} \ll {\cal N}_\mathrm{c}$), ${\cal N}_\mathrm{c}$ is the characteristic number at which the scaling starts.
  In experiments, the  dependence of the transverse magnetization ${\cal M}_{\perp}$ on the frequency shift $\Delta\omega$ is measured. As distinct from  the 
magnon number  ${\cal N}=\int d^2r |\Psi|^2$, where $\Psi$ is the  wave function of magnon condensate, 
 the transverse magnetization density represents the order parameter and is proportional to $\Psi$, see \eqref{spinODLRO}. The total magnetization is thus  ${\cal M}_{\perp}\propto \int d^2r |\Psi| \propto {\cal N}^{1/2} R$. Since $\Delta\omega  \propto 1/R^2$ according to \eqref{R_vs_frequency}, one obtains ${\cal M}_{\perp}\propto (\Delta\omega)^{-1-k/4}$. The measured transverse magnetization suggests that for large ${\cal N}$ the scaling law is approached with $k\approx 3$,  see insert in Fig. \ref{specexamp}. The magnetization estimated in numerical simulations also suggests that under experimental conditions the $k=3$ scaling is the reasonable fit  \cite{Autti2012},   see insert in Fig. \ref{magnon_bubble}, which is just the scaling corresponding to the MIT bag model.
 
In the above approach  the  energy consideration has been used, where the energy potential  is obtained by averaging over fast precession. The numerical simulations of the $Q$-ball, which used the full dynamical equations, can be found in \cite{Bunkov2005}.

\subsection{Comparison with atomic BEC in trap} 

\begin{figure}[t]
\includegraphics[width=\linewidth]{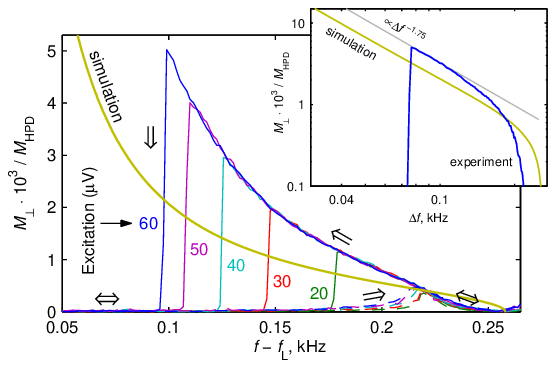}
\caption{Formation of the magnon condensate droplet in the ground state $n_r=0$, which corresponds
to $n=0$ in the original trap in Fig.~\ref{InitialTrap},
in cw NMR measurement. The condensate
  magnetization $M_{\perp}$ precessing in the transverse plane is plotted
  on the vertical axis, normalized to that when the \textit{homogeneously
    precessing domain} (HPD) fills the volume within the detector coil. The arrows indicate the sweep direction of the applied rf
  frequency $f = \omega/(2 \pi)$. $M_{\perp}$ grows when the frequency is
  swept down. Only a
  tiny response is obtained on sweeping in the opposite direction. During
  the downward frequency sweep the condensate is destroyed (vertical
  lines), when energy dissipation exceeds the rf pumping. This point
  depends on the applied rf excitation amplitude, marked at the each line. The lower green
  line represents the result of calculations from
  Fig.~\ref{magnon_bubble}. The calculations have no fitting parameters
  and the difference with the measurements can be
  attributed to the experimental uncertainty in determining $M_{\perp}$ for
  the vertical scaling. \textit{(Insert)} The experimental curve for the
  largest excitation and the numerical curve from the main panel are
  replotted in the double-logarithmic coordinates to demostrate the
  asymptotic limit for large magnon number:
  $M_{\perp} \propto (f - f_\mathrm{L} (0))^{-1.75}$, which corresponds to
  the condensate in a box in Eq. (\ref{boxrad}) with $k=3$.  }
 \label{specexamp}
\end{figure}

Incidentally, for an atomic condensate in harmonic trap the radius $R$ as a function of number of atoms at large $N$ also approaches the scaling  in Eq.~(\ref{boxrad})  with $k=3$ \cite{PitaevskiiStringari2003}. This behavior results from the repulsive inter-particle interactions in the Thomas-Fermi limit.  
However, this similarity in the scaling law $k=3$ both with atomic condensate and with MIT bag model is accidental.
Moreover, the $k=3$ scaling is actually in disagreement with the typical bag models.  If the main contribution to the pressure comes from the bulk, as it happens for the hadron model, then the energy $F(R)$ should be proportional to the volume of the bag, and then for the  two-dimensional radial trap one would expect $F(R) =\pi R^2 P$, i.e. $k=2$.  On the other hand, if the main contribution to the pressure comes from the surface tension, as it happens for electron bubble in liquid helium at $P=0$,  then the energy $F(R)$ should be proportional to the surface area, and for our 2D case this would give $F(R)=2\pi \sigma R$, i.e. $k=1$.  The observed more soft behavior with approximate scaling law  $k\approx 3$ in the 2D case reflects the flexibility of the orbital field, which is inhomogeneous outside the cavity. On general grounds, $F(R)$ depends on several length scales: radius $R$ of the bubble; radius $R_{\rm c}$ of the cylindrical container, where the boundary conditions on the $\hat{\bf l}$-texture are imposed; and the textural healing lengths: magnetic  length $\xi_H$ (the thickness of the layer near the wall of the container
in which the orientation of $\hat{\bf l}$ by magnetic field is restored) and the lengths related to the orientational effects of rotation and vortices on $\hat{\bf l}$.  

Both in experiments and in the numerical simulations in  \cite{Autti2012}, all the length scales were of the same order, and thus no really small parameter was available, which could justify the scaling law.  The true scaling behavior may only appear in some limit cases. For example, in the  vessel rotating with angular velocity $\Omega$ in a vortex-free state in the regime $R_0\gg R \gg \xi_v \gg \xi_H$ (where $\xi_v$ is the healing length related to counterflow $|{\bf v}_s-{\bf v}_n|=\Omega R$) one may expect that the main contribution comes from the orientational effect of  the counterflow, which is removed from the cavity
$F(R)\propto R^2(\Omega R)^2$. The obtained scaling law with $k=4$ gives   $R\sim {\cal N}^{1/6}$ and  ${\cal M}_{\perp}\propto ( \omega -\omega_\mathrm{L})^{-2}$. Such  asymptotic regime, which can be approached in a large vessel, was probed in numerical simulations and is in a reasonable agreement
\cite{Non-ground-state}.

In conditions of the experiment the exponent $k$  in Eq.~(\ref{boxrad}) is close to $k=3$ for atomic condensates in harmonic trap. However,  the physics  of the formation of this exponent  is different (formation of a box inside a flexible texture vs  atom-atom interaction). As a result one has absolutely opposite behavior of 
the analogous quantities --  the frequency shift  $\Delta\omega({\cal N})$ in magnon BEC and the  chemical potential $\mu( {\cal N} )$ in an atomic condensate:
\begin{eqnarray}
 \omega -\omega_\mathrm{L}(0)  \equiv  \mu -U(0)  &\sim&   \omega_r \left( {\cal N}/{{\cal N}_\mathrm{c}}\right)^{-2/5}
~,~~{\rm magnon ~BEC} \, ,
 \label{magnonBEC}
 \\
 \mu -U(0)  &\sim&    \omega_r \left( {\cal N}/{{\cal N}_\mathrm{c}}\right)^{2/5}
 ~,~~{\rm atomic ~BEC}\, .
 \label{atomicBEC}
\end{eqnarray}
As a result in contrast to  atomic condensates, the magnon condensate droplet has negative 
derivative $d\mu/d{\cal N} < 0$. 
This means that with the growing $Q$-ball, its frequency $\omega$ decreases approaching the
Larmor frequency asymptotically, and this behavior determines the way in which the magnon condensate is grown in a cw NMR measurement, as seen in Fig.~\ref{specexamp} for the formation of ground-state BEC in the trap.  The magnons are created when the frequency $\omega$ of the applied RF field is swept down and crosses the ground state level $ \omega_{00}$. When  $\omega$  is reduced further, the number of magnons  follows asymptotically  Eq.~\eqref{magnonBEC}, i.e. ${\cal N} 
\sim (\omega -\omega_\mathrm{L})^{-5/2}$. 

The negative value of  $d\mu/d{\cal N} < 0$ allows also to form the condensates on excited levels under condition when the ground-state condensate 
still does not exist -- the situation which is impossible for the atomic condensates with repulsive interaction, where $d\mu/d{\cal N} > 0$.

\subsection{Formation of magnon BEC on excited state in the trap} 
\label{NonGroundState}

\begin{figure}[t]
\includegraphics[width=\linewidth]{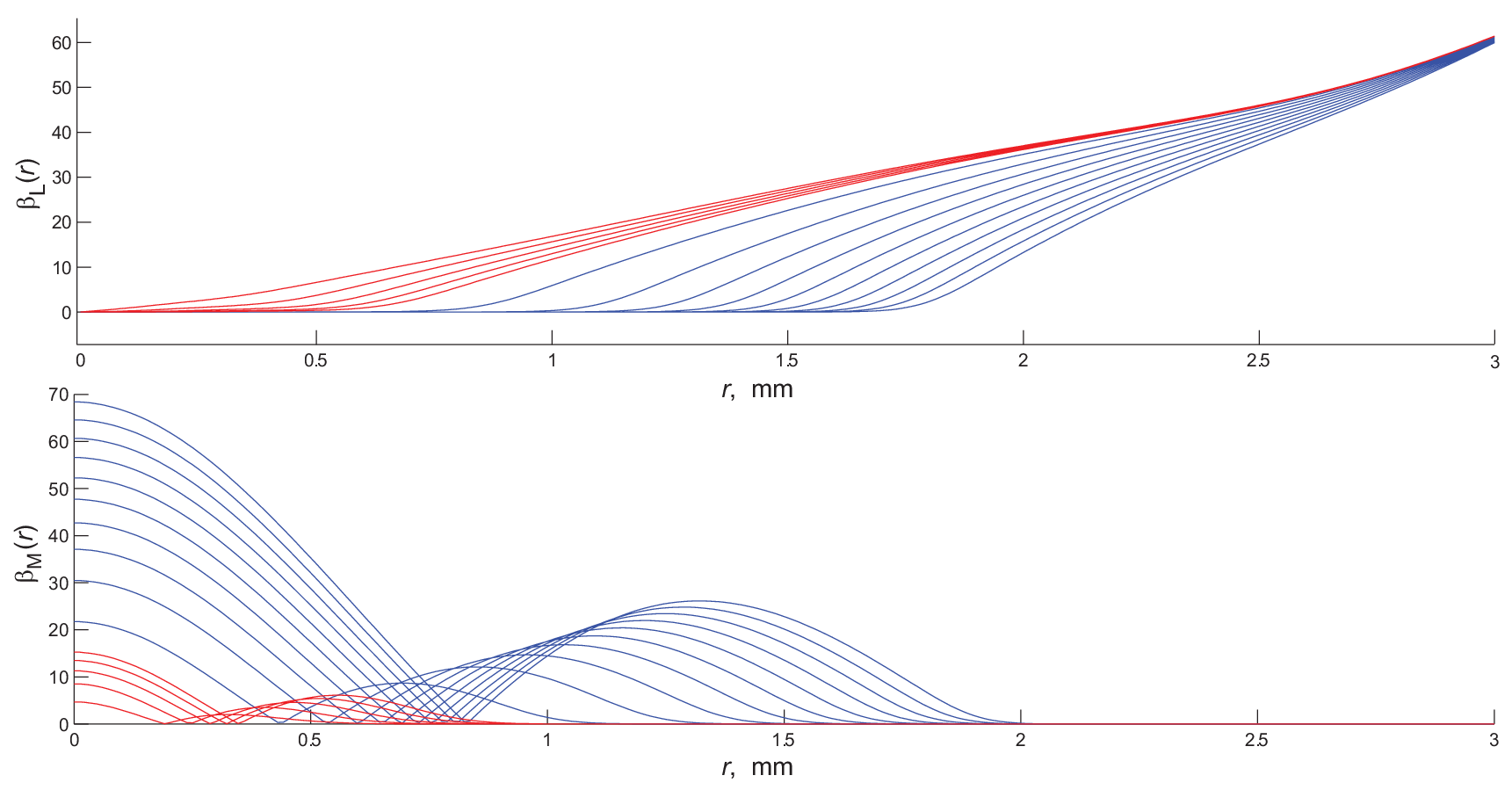}
\caption{Simulation of formation of the multi-magnon bubble with magnon condensate on the first excited level $n_r=1$ (i.e. $n=2$  in  the initially harmonic 2D trap)  \cite{Non-ground-state}.  At large number of magnons, the harmonic trap gradually transforms to the box ({\it top}), while the wave function of the condensate gradually transforms to the Bessel function ({\it bottom}).}
 \label{angles_1stExcited}
\end{figure}

The condensate can be  formed when one starts filling magnons to one of the levels (n,m) in  Eq.~(\ref{SpinWaveSpectrum}). Then one obtains the non-ground-state condensate \cite{Autti2012,Non-ground-state}.
The formation of non-ground-state condensates has been proposed for cold atoms \cite{Bagnato}, but as a dynamic mixture of the ground state and an excited level. It was suggested to use resonant modulation of either the trap potential or the atomic scattering length, \textit{eg.} by applying a temporal modulation of the atomic interactions via the Feshbach resonance technique. In contrast to such schemes the excited states $(n,m)$ of magnon condensate can be populated directly without the original ground-state condensate. This is because the  frequency $\omega_{\mathrm{nm}}({\cal N})$ of excited state condensate also decreases with increasing the magnon number ${\cal N}$. The condensate in the state $(n,m)$ grows when the  frequency of RF field is swept down and crosses the level $\omega_{\mathrm{nm}}(0)$ from above, while at such frequency the ground-state condensate does not exist.
The numerical simulation of formation of the multi-magnon bubble with magnon condensate on the first excited level in  the trap \cite{Non-ground-state} is shown in Fig. \ref{angles_1stExcited}. 

 The condensate in the excited state is metastable: it is supported by continuous pumping at $ \omega_{\mathrm{mn}}({\cal N})$, i.e. it exists in the regime of the controlled chemical potential $\mu$.  After switching off the pumping, i.e. in the regime of the controlled magnon number ${\cal N}$, the excited state condensate decays to the magnon BEC in the ground state.  This quantum process of formation of the ground-state BEC from the pumping of magnons to excited level is similar to formation of magnon BEC by incoherent pumping in yttrium-iron garnet \cite{Chumak2009} (see Sec. \ref{SolidMaterials} below)  and also to the off-resonance excitation of the ground state condensate, observed in
$^3$He-B in Ref.~\cite{PS2}. This demonstrates that formation of magnon BEC is the spontaneous process, in which the precession frequency emerges spontaneously, and is not produced by external rf field, i.e. magnon condensate is not the ``driven condensate''  \cite{Snoke2006}. External rf field if applied is needed for compensation of losses. 

As distinct from the excited-state magnon BEC, the condensate in the ground state can be formed in pulsed NMR measurements after a
large number of magnons ${\cal N}$ are pumped to the cell. This clearly demonstrates the effect
of self-localization: the main part of the pumped charge relaxes but
the rest of ${\cal N}$ starts to concentrate at some place on the axis, digging
a potential well there and attracting the ``charge'' from the other places
of the container.  In earlier experiments \cite{Lan3} 
without the confinement potential  $U_\parallel(z)$ in the axial direction,  $Q$-balls 
were typically formed at the bottom of the cell.  However, $Q$-balls  were often formed on the axis of the flared-out texture, away from the
horizontal walls. This  shows that  $Q$-ball may dig the potential well in different places of the cell.  
In the formation of $Q$-ball  with off-resonance excitation \cite{PS2, Bunkov2005}, the effect
of self-localization also plays a crucial role.

In conclusion, $Q$-balls represent a new phase-coherent state of Larmor precession. They emerge at low $T$, when the homogeneous bulk BEC of magnons  (HPD) becomes unstable. These $Q$-balls are compact
objects which exist due to the conservation of the global $U(1)$ charge $Q=S_z$.
At small $Q$ they are stabilized in the potential well, while
at large $Q$ the effect of self-localization is observed. In terms of relativistic quantum fields the localization is caused by the peculiar interaction between the charged and neutral  fields \cite{Friedberg1976}. The neutral field  $\hat{\bf l}$ provides the potential for the charged field  $\Psi$; the charged field modifies locally the neutral field so that the potential well is deformed and forms a box in which the charge $Q$ is further condensed. In this limit,  the magnon BEC becomes the bosonic analog of the MIT bag with trapped quarks  in QCD or of an electron trapped in the cavity formed in liquid $^4$He.

  \section{Exploiting Bose condensate of magnons}

The Bose condensation of magnons in superfluid $^3$He-B has many 
practical applications. 
In Helsinki, owing to the extreme sensitivity  of the Bose
condensate  to textural inhomogeneity, the phenomenon of Bose
condensation  has been applied to studies of  
supercurrents and topological defects in $^3$He-B in rotating cryostat. The measurement
technique was called  HPD spectroscopy
\cite{HPDSpectroscopy,HPDSpectroscopy2}. 

   \subsection{Observation of Witten string in $^3$He-B}
\label{WittenString}

In particular, HPD
spectroscopy provided direct
experimental evidence for broken axial symmetry in the core of quantized vortices in $^3$He-B \cite{V2vortex}.

  \begin{figure}
\includegraphics[height=7cm]{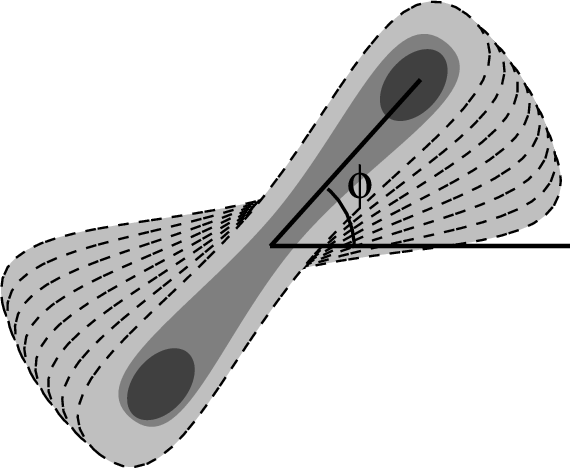}
\caption{Twisted core of non-axisymmetric vortex in $^3$He-B.  The
gradient of the Goldstone field $\nabla \phi$ along the string
corresponds to the superconducting current along the superconducting
cosmic string.  Such twisted core has been obtained and detected using coherent precession of magnetization  \cite{V2vortex}}
\label{TwistedCoreFig}
\end{figure}

The dominating area of the phase diagram of the vortex states in $^3$He-B  is occupied 
by vortices with non-axisymmetric cores, i.e. vortices with the spontaneously broken rotational 
$SO(2)$  of the core calculated in Refs. \cite{CoreTransition1,CoreTransition2}.
The core with broken rotational symmetry can be considered as a pair of
half-quantum vortices, connected by a non-topological soliton
wall  (see Fig. \ref{TwistedCoreFig}). The separation of the half-quantum vortices increases with decreasing
pressure and thus the double-core structure is most pronounced at zero
pressure.

In the physics of cosmic strings, an analogous breaking of continuous
symmetry in the core was first discussed by Witten  \cite{ewitten},
who considered the spontaneous breaking of the electromagnetic gauge
symmetry $U(1)$.
Since the same symmetry group is broken in the condensed matter
superconductors, one can say that in the core of the cosmic string
there appears the superconductivity of the electric charges,
 hence the name `superconducting cosmic strings'.
 
For the $^3$He-B vortices, the spontaneous breaking of
the
$SO(2)$ symmetry in the
core leads to the Goldstone bosons -- the mode in which the degeneracy
parameter, the axis of anisotropy of the vortex core, is
oscillating. The homogeneous magnon condensate, the HPD state, has been used to study  the structure and twisting dynamics of the non-axisymmetric core of the low-temperature vortex in $^3$He-B \cite{V2vortex}. This is because the coherent precession of magnetization excites the vibrational Goldstone mode via spin-orbit interaction. Moreover, due to spin-orbit interaction the precessing magnetization rotates the core around its axis with constant
angular velocity. In addition, since the core was pinned on the top and
the bottom of the container, it was possible even to screw the core (see Fig. \ref{TwistedCoreFig}). Such a twisted core
corresponds to the Witten superconducting string with the electric supercurrent
along the core. The rigidity of twisted core differs from that of the straight core.
This allowed a detailed study of the Goldstone mode of the vortex core resulting from the
spontaneous violation of rotational $U(1)$ symmetry in the core
\cite{V2vortex}. 

\subsection{Observation of spin-mass vortex in $^3$He-B}
 \label{SpinMassVortex} 
  
\begin{figure}
\includegraphics[height=8cm]{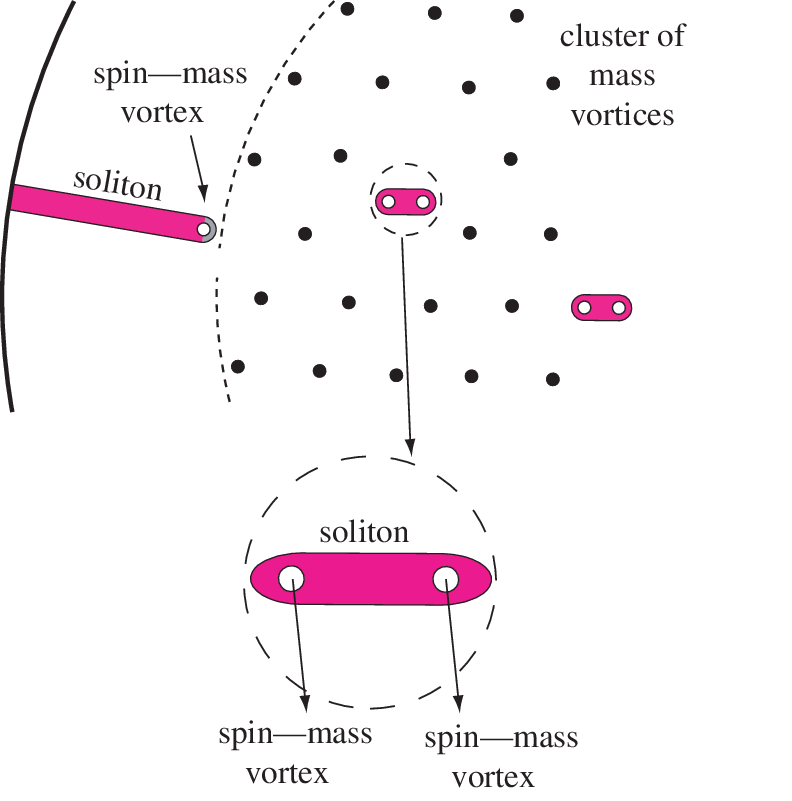}
\caption{ Vortices in rotating $^3$He-B. Mass vortices form a regular
structure like Abrikosov vortices  in
an applied magnetic field. If the number of vortices is less than
equilibrium number for given rotation velocity, vortices are collected  in
the vortex cluster. Within the cluster the average superfluid velocity
$\left<{\bf v}_{\rm s}\right>={\bf v}_{\rm n}$. On the periphery
there is a region void of vortices -- the counterflow region, where $\left<{\bf
v}_{\rm s}\right>\neq  {\bf v}_{\rm n}$.  Spin--mass vortices  
can be created and stabilized in the rotating vessel. The
confining potential produced by the soliton wall is compensated by
logarithmic repulsion of vortices forming the vortex pair -- the doubly
quantized vortex  inside the cluster. A single spin--mass vortex
is stabilized at
the periphery of the cluster by the combined effect of soliton tension
and Magnus force acting on the mass part of the vortex.}
\label{spin_mass_vortices}
\end{figure}

There are different types of the topological defects in the (non-precessing) $^3$He-B. Among them 
there is a $Z_2$ spin vortex -- topological defect of the order parameter matrix $R_{\alpha i}$
in \eqref{LarmorPrecession2}. Due to spin-orbit coupling this defect serves as the termination line of the topological soliton wall, and due to the soliton tension it cannot be stabilized in the rotating vessel.
However, spin and mass vortices attract each other and form the combined defect with common core, the so-called spin-mass vortex, which can be stabilized under rotation (see Fig. \ref{spin_mass_vortices}). The spin-mass vortices also form molecules where the soliton serves as chemical bond. 
These defects -- spin-mass vortex connected with the wall by soliton and bound pairs of spin-mass vortices --
have been observed and studied using  HPD spectroscopy
\cite{27}. 

\subsection{Magnon condensates in aerogel}

HPD  spectroscopy proved to be extremely
useful for the investigation of the superfluid order parameter in
a novel system --  superfluid $^3$He  confined in aerogel 
\cite{Dmitriev,Grenoble1,GrenB,Tokyo1,Sato2008,JapGren}. 

\subsection{Towards observation of Majorana fermions}

The condensate in the magnetic trap can be used to continue NMR measurements in $^3$He-B down to 0.1 -- 0.2\,$T_\mathrm{c}$, where HPD does not exist. Of great current interest are the fermionic states bound to the  core of mass vortices and to the surface of $^3$He-B, especially the still elusive Majorana fermions with zero energy (see the review on Majorana fermions \cite{Beenakker2012}). Recently the topologically nontrivial gapless and gapped phases of matter -- topological insulators,  semimetals, superconductors, and superfluids --  have attracted attention
\cite{HasanKane2010,Xiao-LiangQi2011}. $^3$He-B is the best representative of a  3-dimensional coherent quantum system with time reversal symmetry. Its nontrivial topology gives rise to gapless Andreev-Majorana fermions as surface states  \cite{Schnyder2009a,ChungZhang2009,Volovik2009b}. There is now experimental evidence for Andreev surface states in $^3$He-B at a solid wall \cite{Davis2008,Murakawa2009}, but the Majorana signature of these fermions -- the linear `relativistic' energy spectrum at low energy -- can be observed only at extreme low temperatures, when the thermal quasiparticles in bulk $^3$He-B are exponentially depleted. Majorana fermions, both in the vortex core and at surfaces, are then expected to give the main contribution to thermodynamics and dissipation, with a power-law dependence of the physical quantities on $T$. In rotation or by moving the magnon condensate droplet next to the wall, one will be able to probe Majorana fermion states in $^3$He-B. 

\section{Magnon BEC in other systems}
   
    \subsection{Magnon BEC in normal $^3$He}

A very long lived induction signal  was observed in normal Fermi liquids: in spin-polarized $^3$He-$^4$He solutions 
\cite{Nunes1992} and in normal liquid $^3$He \cite{Dmitriev1995}. It was explained as a coherently precessing  structure at the interface between the equilibrium domain and the domain with the reversed magnetization  \cite{Normal3He}. It would be interesting to treat this type of dynamic magnetic ordering as a new mode of magnon BEC.

   \subsection{Magnons condensation  in solid materials}
   \label{SolidMaterials}

  Recently indication of magnon BEC in terms of coherent spin precession has been reported in 
a solid state material CsMnF$_3$ \cite{Kazan2011}. This condensate is similar to magnon BEC in $^3$He-A in aerogel. The magnon BEC obtained by the parametric pumping of magnons 
has been investigated in yttrium-iron garnet (YIG) films \cite{Demokritov,Demidov2008,Chumak2009}.  
Let us discuss the latter.

Magnons in yttrium-iron garnet have the quasi 2D spectrum:
\begin{equation}
\omega_n(k_x,k_y)= \Delta_n + \frac{k_y^2}{2m_y} + \frac{(k_x \pm k_0)^2}{2m_x}  ~,
\label{2Dspectrum}
\end{equation} 
where magnetic field is along $x$; the gap in the lowest branch $\Delta_0=2.1$ GHz $\equiv 101$ mK at $H=700$ Oe \cite{Demokritov} and $\Delta_0=2.9$GHz at $H=1000$ Oe \cite{Demidov2008}; there are two minima with  $k_0=5\cdot 10^4$ 1/cm \cite{Demokritov}; the anisotropic magnon mass can be probably estimated as 
$m_x\sim k_0^2/\Delta$ with $m_y$ being somewhat bigger, both are of order of electron mass.

In two-dimensional systems the number of extra magnons  -- the difference of the distribution function at $\mu=0$ and $\mu\neq 0$ --    is determined by low energy  Rayleigh-Jeans part of the spectrum:
 \begin{equation}
n=\sum_{\bf k}\left( \frac{T}{E_{\bf k} -\mu} - \frac{T}{E_{\bf k}} \right)~,
\label{ExtraNumber}
\end{equation}   
If one neglects the contribution of the higher levels and consider the 2D gas, the Eq.(\ref{ExtraNumber}) becomes
 \begin{equation}
n_{\rm extra}=  \frac{T}{2\pi \hbar} \sqrt{m_x m_y}~ \ln \frac{\Delta_0}{\Delta_0 -\mu}~,
\label{ExtraNumber2D}
\end{equation} 
In 2D, all extra magnons can be absorbed by thermal distribution at any temperature without formation of Bose condensate. The larger is the number  $n$ of the pumped magnons the closer is $\mu$ to $\Delta_0$, but  $\mu$ never crosses $\Delta_0$. At large $n$ the chemical potential 
exponentially approaches  $\Delta_0$ from below and the width of the distribution becomes exponentially narrow:
\begin{equation}
  \frac{(\delta k_y)^2}{m_y \Delta_0} \sim  \frac{(\delta k_x)^2}{m_x\Delta_0} \sim \frac{\Delta_0-\mu}{\Delta_0}\sim\exp\left( -\frac{2\pi \hbar n_M }{T \sqrt{m_x m_y}}\right) ~.
\label{Width}
\end{equation} 
If one uses the 2D number density  $n=\delta N ~d$ with the film thickness $d=5~\mu$m and 3D number density $\delta N\sim 5\cdot 10^{18}$~cm$^{-3}$, one obtains that at room temperature the exponent is
\begin{equation}
\frac{2\pi \hbar n_M }{T \sqrt{m_x m_y}}\sim  \frac{2\pi  \Delta_0 }{T}   \frac{\delta N~d }{ 10 k_0^2}  \sim  10^2 ~.
\label{Exponent}
\end{equation} 
If this estimation is correct, the peak should be extremely narrow, so that all extra magnons are concentrated at the lowest level of the discrete spectrum. However, there are other contributions to the width of the peak due to:  finite resolution of spectrometer, magnon interaction,  finite life time of magnons and the influence of the higher discrete levels $n\neq 0$.

In any case, the process of the concentration of extra magnons in the states very close to the lowest energy is the signature of the BEC of magnons.  The main property of the  room temperature  BEC in YIG  is that the transition temperature $T_c$ is only slightly higher than temperature, $T_c-T\ll T$; as a result the 
number of  condensed magnons  is small compared to the number of thermal magnons: $n\ll n_T$.  Situation with  magnon BEC in $^3$He is the opposite,  one has $T\ll T_c$ and thus $n\gg n_T$.
In $^3$He-B, the typical temperature is big compared to the magnon gap, $T\gg \hbar\omega_L$, and thus according to \eqref{MagnonSpectrumextended} the  thermal magnons are spin waves with linear spectrum $ \omega(k)=ck$, with characteristic momenta $k_T\sim T/\hbar c$. The density of such  thermal magnons   $n_T \sim k_T^3$  is much smaller than the density of the Bose-condensed magnons and in $^3$He-B they can be neglected.  In yttrium-iron garnet (YIG) situation is opposite:  the temperature is high and number density of the magnons concentrated at small momentum is small compared to the thermal magnons, $n \ll n_T$.

 \subsection{Magnon BEC vs planar ferromagnet}
 \label{PlanarFerromagnet}
 
 Let us compare the HPD state (\ref{ODLRO1}) in $^3$He-B, the  coherent precession in  solid antiferromangnet and BEC in YIG on one side with the equilibrium magnetic states 
 discussed in \cite{mag1,mag2,mag3,Giamarchi,Nikuni} 
on the other side.  In both groups the $U(1)$ symmetry is spontaneously broken, and thus they both belong to the same symmetry class as atomic BEC. At first glance both groups can described in terms of the ODLRO. The spin density in coherently precessing HPD state of $^3$He-B and  in YIG are correpondingly
\begin{equation}
\left<S_+\right>= S_x+iS_y =S_\perp e^{i\omega t+i\alpha} \,,
\label{ODLRO1}
\end{equation}
and
\begin{equation}
\left<S_+\right>= S_x+iS_y =S_\perp \cos(k_0x)e^{i\omega t+i\alpha} \,.
\label{ODLROYIG}
\end{equation}
For the equilibrium planar ferromagnets  one can also express the broken symmetry state in terms
of vacuum expectation value of  spin creation operator \cite{Hohenberg1971,Nikuni}
\begin{equation}
\left<S_+\right>= S_x+iS_y =S_\perp e^{i\alpha}~.
\label{ODLRO2}
\end{equation}
However, as distinct from Eqs. (\ref{ODLRO1}) and (\ref{ODLROYIG}), the Eq.(\ref{ODLRO2})  is time independent, and as a results it can be described by ordinary diagonal long-range order $\left<{\bf S}\right>$ instead of the off diagonal vacuum expectation value $\left<S_+\right>$.
 The phenomenon of ODLRO, which results in the time dependence of the magnetic state manifested by the coherent precession, is the major point which distinguishes between these two phenomena.

For the precessing states the $U(1)$ symmetry is related to the quasi-conservation of the charge $Q$, which is analogue of the number of atoms in atomic BEC, and of the number of electrons in superconductors.  In the magnetic materials the charge $Q$ is played by the spin projection $S_z$, or by the related number $N$ of magnons.  This approximate conservation law gives rise to the non-equilibrium chemical potential $\mu=dE/dN$, which is non-zero only in dynamic states where it coincides with the precession  frequency $\omega$. On the contrary, in a static state one has $\omega=0$, and thus Eq.(\ref{ODLRO2}) does not contain the analogue of chemical potential. This means that the  conservation law  is not  in the origin of formation of the static equilibrium state. While the magnetic field may play the role of external potential, it cannot play the role of magnon  chemical potential, since in a fully equilibrium state  the chemical potential of magnons is always strictly zero, 
$\mu=0$, which results in $\omega=0$. 

For both groups, the underlying $U(1)$ symmetry is approximate due to the spin-orbit interaction, which violates the conservation of $S_z$ and makes the life time of magnons finite.  For the precessing states (\ref{ODLRO1}) and (\ref{ODLROYIG}) this leads to the finite life time of the coherent precession.  To support the steady state of precession the pumping of spin and energy is required. On the contrary,  the spin-orbit interaction does not destroy the diagonal long-range magnetic order of the static states: these are the fully  equilibrium states which do not decay and thus do not require pumping.  That is why 
the approximate $U(1)$ symmetry and its spontaneous violation are not the necessary conditions for the existence of equilibrium magnetic systems. A planar ferromagnet (\ref{ODLRO2}) is just one more equilibrium state of matter with broken time reversal symmetry, in addition to the easy axis ferromagnetic or antiferromagnetic state, rather than the magnon condensate. Formally, in the limit of small spin-orbit interaction the symmetry breaking scheme in these materials belongs to the same $U(1)$ class as conventional superfluids and superconductors, and thus they share many properties of this class, but
except for the ODLRO and the related phenomena.

The property of (quasi)conservation of the $U(1)$ charge $Q$  distinguishes the coherent precession from the other coherent phenomena, such as optical lasers and standing waves.
For the real BEC one needs the conservation of particle number or charge $Q$
during the time of equilibration. BEC occurs due to the thermodynamics, when the number of particles
(or charge $Q$) cannot be 
accommodated by thermal distribution, and as a result the extra part must be accumulated 
in the lowest energy state. This is the essence of BEC.

Photons and phonons can also form the BEC under pumping,  if the lifetime of these excitations is larger than thermalization time. For photons this condition  has been realized, and the photon BEC has been observed \cite{Klaers2011}. These thermodynamic BEC states are certainly 
different from such coherent states as optical lasers and also from the equilibrium deformations of solids.

 \section{Beyond magnon BEC: Suhl instability of coherent precession}
 \label{BeyondBEC}

 \subsection{Catastrophic relaxation of magnon BEC}

The instability of BEC which we discuss here is applicable only to  the BEC of magnon quasiparticles, and is irrelevant for atomic condensates.  For quasiparticles the $U(1)$ symmetry is not strictly conserved. For magnetic subsystem, this is the $SO(2)$ symmetry with respect to spin rotations in the plane perpendicular to magnetic field, and it is violated by spin-orbit interactions.  The magnon BEC is a time dependent process, and it may experience instabilities which do not occur in equilibrium condensates of stable particles. In 1989 it was found that the original magnon condensate -- the HPD  state -- abruptly looses its stability below about 0.4 T$_c$ \cite{Catastropha}. This was called the catastrophic
relaxation. This phenomenon was left unexplained for a long time.

\begin{figure}[t]
\includegraphics[width=\linewidth]{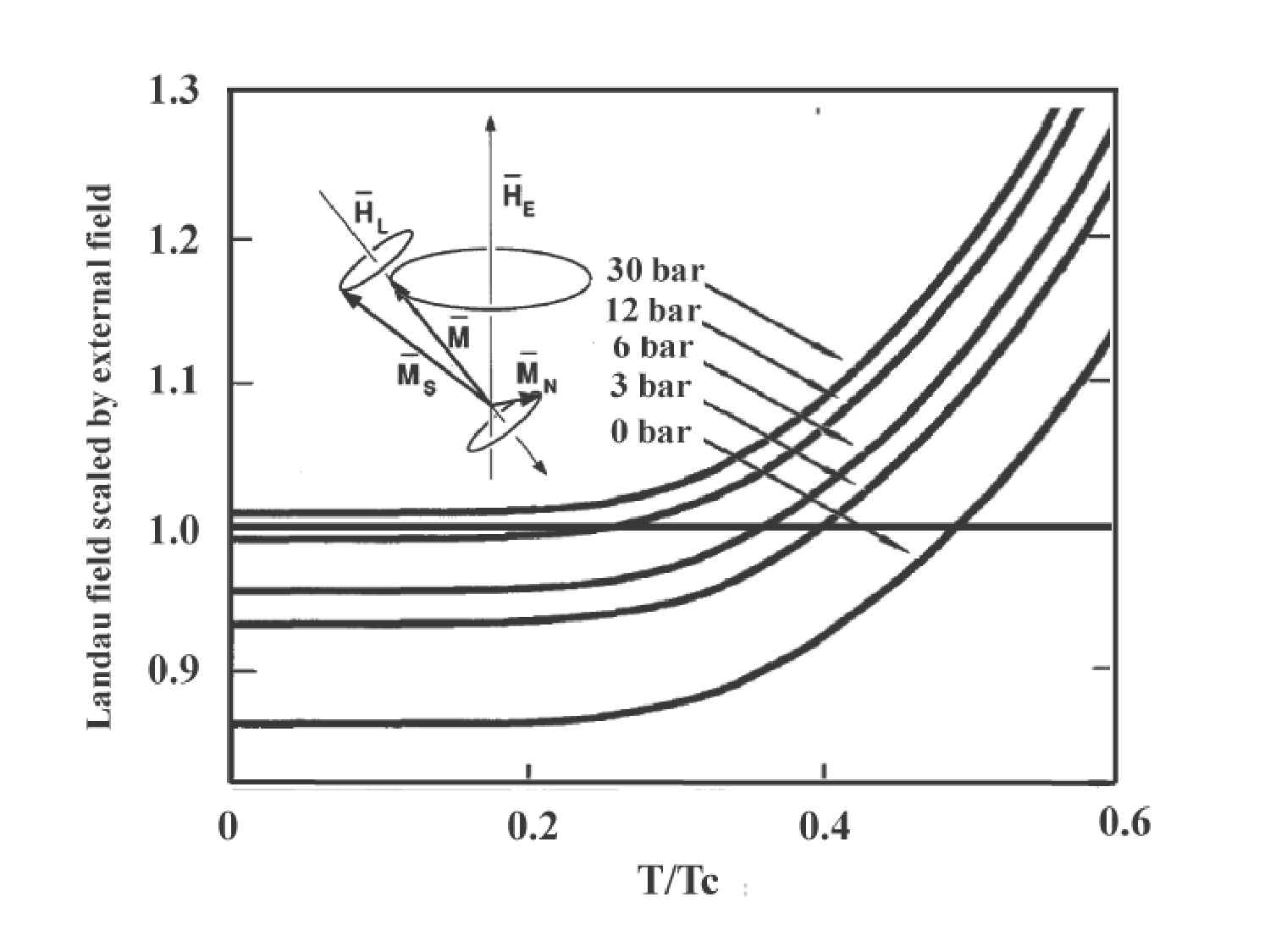}
\caption{Illustration of the concept of the ``molecular''  Landau field: NMR in terms of  the two-component precession. The normal ${\bf M}_{\rm N}$ and superfluid  ${\bf M}_{\rm S}$ components 
of magnetization precess
around the direction of the  Landau field, which in turn is directed along the
full magnetization   ${\bf M}={\bf M}_{\rm N}+ {\bf M}_{\rm S}$, which precesses around external magnetic field. In some range of pressures, there is the transition temperature, at which the magnitude of Landau field crosses the magnitude of external field. The rapid decay of the  NMR signal is observed at this temperature \cite{LandauField}, which justifies the phenomenology of Landau field. However, the concept of Landau field does not explain the catastrophic relaxation. The origin of the latter is the Suhl instability.}
 \label{LandauField}
\end{figure}

One of the first suggestions was very interesting. It was suggested that 
the instability appears due to crossover between the NMR in external 
field and NMR in Landau field \cite{Mark}. The molecular Landau field was 
introduced in phenomenological Landau theory of Fermi liquid. 
According to this concept, the magnetization in superfluid $^3$He can be 
presented as a sum of two components, the superfluid one which comes from the superfluid order
parameter, and the normal one which comes from thermal quasiparticles. 
This is similar to the two-fluid model of superfluidity. In analogy with the second sound,
in NMR both components  precess around the common direction -- the direction
Landau field, which in turn is directed along the total magnetization and thus
does not contribute to the frequency of NMR (see Fig. \ref{LandauField}). This two-component precession  
is described by Leggett-Takagi equations (see 
the book Ref. \cite{VollhardtWolfle}), with each component having 
its own spin-orbit interaction, the phase of precession and the relaxation. 
The value of Landau field is proportional to magnetization and changes 
in superfluid $^3$He with temperature so that in the range of pressures between 0 and 10 
bar there is a crossover temperature when the value of Landau field 
 is equal to the value of external magnetic field. 
At this crossover  temperature the fast relaxation of the conventional NMR signal
was found \cite{LandauField}. It results from the two-mode
precession around the Landau field, which is highly dissipative and pumps out 
the energy from the usual mode of NMR \cite {chaos}. This was the first evidence
that the Landau field is indeed a real molecular 
field and not just a mathematical construction. However, the Landau field
does not explain the catastrophic relaxation phenomena, 
since the crossover temperature has an opposite dependence on 
pressure \cite{LandauField2}.

Finally the reason for catastrophic relaxation was established: in the 
low-temperature regime, where dissipation becomes sufficiently small, 
the Suhl instability destroys the homogeneous precession 
\cite{Catastroph1,Catastroph2,Catastroph3}. This is the parametric 
instability, which leads to decay of HPD due to the parametric 
amplification of an intrinsic mode never discussed previously. The 
latter modes are different from the magnons which we discuss here: they 
represent another branch of the collective modes of superfluid $^3$He-B 
which appears in the regions where the orbital momentum is not parallel to the 
magnetic field \cite{Bunkov2004,Thetamode}. Particularly they are excited near the walls 
of the cell which are oriented along the magnetic field \cite{Nonwet}. 
The instability occurs because the spin-orbit interactions violate the 
$U(1)$ symmetry.

For magnetically ordered systems the instability of homogeneous
precession is a well known phenomenon. Suhl \cite{Suhl} explained
 it in terms of parametric instability of the mode of precession
with respect to excitations of pairs of spin waves satisfying the
condition of resonance:
 \begin{equation}
 n\omega_L=\omega_s({\bf k}) +\omega_s(-{\bf k})~,
\label{resonance}
 \end{equation}
where $\omega_L$ is the precession frequency and $n$ is integer
(see also review \cite{LvovReview}). All the magnetic systems,
where the Suhl instability has been observed, are anisotropic. In
particular, as quantum solids and liquids are concerned, Suhl
instability has been observed in anti-ferromagnetic solid $^3$He
\cite{Mizusaki}, and has been predicted by Fomin for anisotropic
superfluid liquid $^3$He-A \cite{HeA} where it has been observed
later \cite{InstabAB}.  Due to the extreme isotropy of $^3$He-B and
due to the unique symmetry of the spin-orbit interaction, the Suhl
instability was not expected there. 

However, under conditions of
the experiment, the boundary conditions on the wall of container
induce the texture of the order parameter in which the orbital
vector ${\bf l}$ deviates from its symmetric orientation along the
magnetic field ${\bf H}$. The symmetry of the spin-orbit interaction
is violated providing the additional term in the interaction between
the modes, which is dominating in typical experiments with the
catastrophic relaxation \cite{Catastropha,HP1, HP2,Nonwet,Dinamics}.

\subsection{Precessing states and their symmetry}

To describe the interaction of magnon condensate with the other modes
of superfluid $^3$He-B, we must go beyond the magnon BEC description and consider 
all the degrees of freedom of homogeneous free precession in external
magnetic field ${\bf H}$. In liquid $^3$He the spin-orbit
(dipole-dipole) interaction is weak. If it is neglected, we can
apply the powerful Larmor theorem, according to which, in the
spin-space coordinate frame  rotating with the Larmor frequency the
effect of magnetic field on spins of the $^3$He atoms is completely
compensated. This follows from the observation that in dynamics the time derivative of a spin vector
enters equations together with the Larmor frequency vector $\boldsymbol{\omega}_L=\gamma {\bf H}$
\begin{equation}
 D_t{\bf f}=\partial_t  {\bf f}- \boldsymbol{\omega}_L\times  {\bf f}
 \label{D_t}\,.
 \end{equation}
 In other words, the Pauli magnetic field acts on spin vectors as time component of the effective $SO(3)$ gauge field,  \begin{equation}
  {\bf A}_0= \gamma {\bf H} 
 \label{auxiliary_field}\,.
 \end{equation}
  We shall use this equation later 
 in Sec. \ref{SpinSupercurrents} 
 for discussion of spin currents and spin-quantum Hall effect.
 
 Since the magnetic field becomes irrelevant, the
symmetry group of the physical laws in the precessing frame becomes the same as
in the absence of the field. If the spin-orbit interaction is neglected, it is the product of the  $SO_3^{L}$ group of orbital rotations and the
 $SO_3^{S}$ group of spin rotations:
 \begin{equation}
 G= SO_3^{L}\times SO_3^{S} \,.
 \end{equation}
The  difference from the symmetry group $G= SO_3^{L}\times SO_3^{S} $ in the static case
is that the $SO_3^{S}$   rotations are now considered in the precessing frame rather than  in the laboratory frame.
The elements of the latter group ${\bf \tilde g(t)}$ are constructed from the
elements ${\bf g}$ of conventional spin rotations in the laboratory
frame:
 \begin{equation}
 {\bf \tilde g}(t)={\bf O}^{-1}(\hat {\bf z}, \omega_L t)~{\bf g}~{\bf
O}(\hat {\bf z}, \omega_L t)~.
 \end{equation}
Here the matrix $O_{\alpha\beta}(\hat {\bf z}, \omega t)$ describes
the transformation from the laboratory frame into the rotating frame
- this is the rotation about the magnetic field axis $\hat {\bf z}$
by angle $\omega_L t$. 
Now we can find all the degenerate coherent
states of the Larmor precession applying the symmetry group $G$ to
the order parameter in the simplest equilibrium state of the given superfluid phase.
The order parameter in superfluid $^3$He is $3\times 3$, which corresponds to corresponds to the   Cooper pairing with orbital and spin momenta $L=S=1$ \cite{VollhardtWolfle}.
Thus choosing the simplest $3\times 3$
static matrix  ${\bf A}^{(0)}$  in a given superfluid phase, one obtains all the precessing states
in this phase if the spin-orbit coupling is neglected:
\begin{equation}
 {\bf A}(t)={\bf O}^{-1}(t){\bf R}^{(1)} {\bf O}(t){\bf A}^{(0)}({\bf R}^{(2)})^{-1}\,.
 \label{AllStates}
\end{equation}
Here ${\bf R}^{(1)}$ is the arbitrary matrix
describing spin rotations in the precessing frame and ${\bf
R}^{(2)}$ is another arbitrary matrix which describes the orbital
rotations in the laboratory frame. In case of $^3$He-B the simplest state
corresponds to the total
angular momentum of Cooper pair $J=0$ which is described by the isotropic matrix 
\cite{VollhardtWolfle}:
\begin{equation}
A^{(0)}_{\alpha i}=\Delta_B~\delta_{\alpha i}\,,
\end{equation}
 where $\Delta_B$ is the gap in the fermionic spectrum.
 The action of elements of the group $G$ on this stationary state leads
to the following general precession of $^3$He-B with the Larmor
frequency (if the spin-orbit interaction is neglected):
\begin{equation}
A_{\alpha i}(t)=\Delta_B R_{\alpha i}(t)~~, \label{LarmorPrecession1}
 \end{equation}
  \begin{equation}
R_{\alpha i}(t)=O_{\alpha \beta}(\hat {\bf z}, -\omega
t)R^{(1)}_{\beta \gamma}O_{\gamma \mu}(\hat {\bf z}, \omega
t)(R^{(2)})^{-1}_{\mu i}~~ . \label{LarmorPrecession2}
 \end{equation}
The matrix ${\bf R}^{(1)}$ determines the direction of spin density
in the precessing frame:
\begin{equation}
 S_\alpha = \chi R^{(1)}_{\alpha\beta} H_{\beta}~~,
 \end{equation}
where $\chi$ is the spin susceptibility of $^3$He-B. This
corresponds to the precession of spin magnetization with the tipping angle
$\cos\beta_M=R^{(1)}_{zz}$. The matrix ${\bf R}^{(2)}$ determines
the direction of orbital momentum density in the laboratory frame:
\begin{equation}
  L_i =-R_{\alpha i}(t)S_{\alpha }(t)=-\chi R^{(2)}_{i\alpha }
H_{\alpha}~~,
 \end{equation}
with the tipping angle $\cos\beta_L=R^{(2)}_{zz}$.

\subsection{Spin-orbit interaction as perturbation}

The spin-orbit interaction couples the spin and orbital components of the matrix 
$A_{\alpha i}$. For $^3$He-B in \eqref{LarmorPrecession1} one obtains
 \cite{VollhardtWolfle};
 \begin{equation}
F_D=\frac{2}{15}\chi\Omega_L^2 \left(R_{ii}(t)-\frac{1}{2}\right)^2 =\frac{8}{15}\chi \Omega_L^2\left(\cos\theta(t) +\frac{1}{4}\right)^2, \label{SO}
 \end{equation}
where $\Omega_L$ is the so called Leggett frequency -- the frequency
of the longitudinal NMR; $\theta$ is the angle of rotation in the
parametrization of the matrix $R_{\alpha i}$ in terms of the angle
and axis of rotation \cite{VollhardtWolfle}; we shall use here the system of
units in which the gyromagnetic ratio ${\gamma}$ for the $^3$He atom
is 1, hence the magnetic field and the frequency will have same
physical dimension.

 In the general state of the
Larmor precession (\ref{LarmorPrecession1}), the spin-orbit
interaction contains the time independent part and rapidly
oscillating terms with frequencies $\omega_L$, $2\omega_L$,
$3\omega_L$ and $4\omega_L$:
\begin{equation}
  F_D(\gamma)=F_0 + \sum_{n=1}^4F_n\cos(n\omega_Lt)~~.
\label{TotalFD}
 \end{equation}
The time-independent part -- the average over fast oscillations --
gives the spin-orbit potential in \eqref{FD}, which determines the phases of magnon BEC in 
$^3$He-B:
\begin{eqnarray}
 F_0=F_{\rm so}(s,l,\gamma)=
 \nonumber\\
 \frac{2}{15}\chi\Omega_L^2[(sl-
\frac{1}{2}+\frac{1}{2}(1+s)(1+l)\cos\gamma)^2 +
  \nonumber\\
  \frac{1}{8}(1-s)^2(1-l)^2 +(1-s^2)(1-l^2)(1+\cos\gamma)] ~~.
  \label{gammaGeneral}
 \end{eqnarray}
Here $s=\cos\beta_M$ (or simply $\cos\beta$) and $l=\cos\beta_L$ are $z$ projections of unit
vectors $\hat {\bf s}={\bf S}/S$ and $\hat {\bf l}=-{\bf L}/L$; and
$\gamma$ is another free parameter of the general precession.
Altogether the free precession is characterized by 5 independent
parameters coming from two matrices ${\bf R}^{(1)}$ and ${\bf
R}^{(2)}$ \cite{MisirpashaevVolovik1992}: two angles of spin ${\bf S}$, two angles of
the orbital momentum $\hat{\bf l}$, and the relative rotation of
matrices by angle $\gamma$. In the case of the stationary (non-precessing)
magnetization, the $\gamma$-mode corresponds to the longitudinal NMR
mode.

 \subsection{Parametric instability of HPD}
\label{ParametricInstability}
 
 In the simplest description, the dynamics of the $\gamma$-mode is
determined by the following Lagrangian:
\begin{equation}
  {\cal L}= -\frac{1}{2}\chi \left(\dot \gamma^2
-c^2(\boldsymbol{\nabla}\gamma)^2\right) + F_D(\gamma)~~.
  \label{Lagrangian}
 \end{equation}
Here we used the approximation of an isotropic speed of spin waves $c$. In the
time-dependent part of $F_D$ in \eqref{TotalFD} we only consider the first harmonic,
i.e. according to Eq.(\ref{resonance}) we discuss the parametric
excitation of two $\gamma$-modes with $ck\approx \omega_L/2$. The
amplitude of the first harmonic is:
\begin{eqnarray}
 F_1=\frac{4}{15}\chi\Omega_L^2  \sin \beta \sin \beta_L\cos(\gamma/2)
\times \nonumber\\
\left(2sl- 1 +\frac{(1-s)(1-l)}{2} +(1+s)(1+l)\cos\gamma\right) .
\label{FirstHarm}
 \end{eqnarray}
Further we assume that the system is in the minimum of the dipole
energy $F_0$ as a function of $\gamma$. The equilibrium value
$\gamma=\gamma_0$ is
\begin{equation}
\cos\gamma_0=- \frac{(2sl -1) +2(1-s)(1-l) } {(1+s)(1+l)}~~,
  \label{GammaMin}
 \end{equation}
which is valid if the right hand side of Eq. (\ref{GammaMin}) does
not exceed unity, i.e. when $s +l -5sl <2$.

For the discussion of Suhl instability we need the time-dependent
term which is quadratic in $\gamma-\gamma_0$. Then the Lagrangian
(\ref{Lagrangian}) which describes the parametric instability
towards decay of Larmor precession to two $\gamma$-modes with
$kc\approx \omega_L/2$ is (after the shift
$\gamma-\gamma_0\rightarrow \gamma$; neglecting $\Omega_L$ compared
to $\omega_L$; and neglecting the anisotropy of the spin-wave
velocity):
 \begin{equation}
  {\cal L}= \frac{1}{2}\chi \left(-\dot \gamma^2 +c^2(\boldsymbol{\nabla}\gamma )^2+
a  \Omega_L^2 \gamma^2 \cos\omega_Lt \right)  ~,
\label{InstabilityLagrangian0}
 \end{equation}
where, if $s +l -5sl <2$, the parameter $a$ is
\begin{eqnarray}
 a=\frac{4}{15}  \sin \beta \sin \beta_L\left[
\frac{3(s+l-sl)}{2(1+s)(1+l)}  \right]^{1/2} \times \nonumber
\\
\left[(1+s)(1+l) +2(2sl-1) +\frac{35}{8}(1-s)(1-l)\right]~.
\label{a1}
 \end{eqnarray}

Let us rewrite the Lagrangian (\ref{InstabilityLagrangian0}) in
terms of Hamiltonian as function of creation and annihilation
operators $b_{\bf k}$ and $b_{\bf k}^*$:
 \begin{eqnarray}
 \gamma_{\bf k}=\frac {i}{\sqrt{2\chi \omega_s(k)}}  (b_{\bf k}-b_{\bf
k}^*)~,~\omega_s^2(k) =  c^2k^2
\\ p_{\bf k}=\chi\dot\gamma_{\bf k}= \sqrt{ \chi \omega_s(k)/2}
(b_{\bf k}+b_{\bf k}^*)~,
  \label{CreationAnnihilation}
\\
 {\cal H}=\sum_{\bf k} \omega_s(k) b_{\bf k}^*b_{\bf k}
+ \sum_{\bf k}
\frac{a\Omega_L^2}{2\omega_s(k)} \left(e^{-i\omega_L t}b_{\bf
k}b_{-\bf k} +e^{i\omega_L t}b_{\bf k}^*b_{-\bf k} ^*\right)~,
  \label{Hamiltonian}
 \end{eqnarray}
where we neglected $\Omega_L$ compared to $\omega_L$. The spectrum
of the excited mode is
 \begin{eqnarray}
 b_{\bf k}(t)=\tilde b_{\bf k}e^{-i\omega_L/2 t + i\nu_{\bf k} t}~~,
\nonumber
\\
\nu_{\bf k}=\sqrt{(\omega_s(k) -\omega_L/2)^2 - a^2
\Omega_L^4/\omega_s^2(k)}
  \label{SpectrumExcited}
 \end{eqnarray}
At the resonance, i.e. when $\omega_s (k) =\omega_L/2$, the mode
grows exponentially:
 \begin{equation}
 b_{\bf k}(t) \propto e^{\lambda t}~~,~~\lambda=2a \Omega_L^2/\omega_L~.
  \label{ExpGrowth}
 \end{equation}
At finite temperatures this growing is damped by dissipation, but at
low temperature the dissipation becomes small and catastrophic
relaxation occurs. Following Ref. \cite{Catastroph1} one may assume the
spin diffusion mechanism of dissipation. In this case the equation
for temperature $T_{\rm cat}$ below which the instability of the
homogeneous precession towards radiation of spin waves with
$\omega_s(k)=ck=\omega_L/2$ starts to develop is \cite{Catastroph2,Catastroph3}:
\begin{equation}
 D(T_{\rm cat}) = 2\lambda c^2/\omega_L^2~.
  \label{Demp}
 \end{equation}
Here $D(T)$ is the spin diffusion coefficient, which depends on
temperature and decreases with decreasing $T$. This agrees with observations.
 
\section{Beyond magnon BEC: Spin supercurrents and spin Hall effects}
\label{SpinSupercurrents}

Spin current has many faces. Spin can be transferred by convective, diffusive or ballistic motion of particles; it can be also transferred from particle to particle without any particle motion. In real systems these mechanisms compete with each other, so that in the phenomenological description only in a very particular cases it is possible to resolve between them.  In ferromagnetically ordered A$_1$-phase of superfluid $^3$He, the nuclear spin 
is transferred by superfluid mass current. In magnon BEC in $^3$He-B, the nuclear spin is transferred by the superfluid current of magnons (Sec. \ref{SupercurrentLinearMomentum}); in particular, this spin current transfers spins from one experimental cell to another in the spin-current Josephson effect (Sec. \ref{SpinJosephson}).
In magnetically ordered systems the spin current can be also represented in terms of the rigidity of the order parameter to inhomogeneous spin rotations. Example of the non-dissipative spin current arising in inhomogeneous states of these materials is the spin current circulating around the core of spin-mass vortex in $^3$He-B and inside the soliton which terminates on a spin-mass vortex (Sec. \ref{SpinMassVortex}). As its electric counterpart -- the charge current, 
the spin current can be dissipative and non-dissipative, with or without spin accumulation. It may give rise to spin-Josephson effect and also to spin-Hall effect 
with and without external magnetic field, which may be ordinary (Sec. \ref{SpinHall}) and quantized  (Sec. \ref{QSHE}).

\subsection{Microscopic theory of spin supercurrent  in $^3$He-B}
\label{MicroscopicTheory}

Here we give the `microscopic' derivation for the spin supercurrent, which has been discussed on the phenomenological level of magnon BEC. The underlying microscopic physics is the BCS theory of $p$-wave spin-triplet superfluid $^3$He.  
 Superfluid spin currents in $^3$He-B exist even in the absence of magnon BEC. They come from the spontaneous breaking of spin-rotation symmetry. The spin supercurrent  in $^3$He-B is expressed in terms of the   spin superfluid velocities:
\begin{equation}
 \omega_{\alpha i}=\nabla_i \theta_\alpha= \frac{1}{2}e_{\alpha \beta\gamma}R_{\beta j}\nabla_i 
 R_{\gamma j}\,.
\label{SpinSuperVelocity}
\end{equation}
The corresponding gradient energy is
\begin{equation}
F_{\rm grad}= \frac{1}{2}\rho_{\alpha i,\beta j}\omega_{\alpha i} \omega_{\beta j} \,,
\label{GradientEnergy}
\end{equation}
where $\rho_{\alpha i, \beta j}$ is the spin rigidity tensor with spin rigidity parameters
\begin{equation}
 \rho_{\alpha i,\beta j}=\frac{\chi_B} {\gamma^2}~[\tilde
c_{\parallel }^2\delta_{\alpha\beta}\delta_{ij}-(\tilde
c_{\parallel }^2- \tilde
c_{\perp }^2)(R_{\alpha i}R_{\beta j}+R_{\alpha j}R_{\beta i})] \,.
\label{GradientEnergy2}
\end{equation}
From these equations one obtains the spin supercurrent in $^3$He-B:
 \begin{equation}
 J_{\alpha i}=-\frac{\partial F_{\rm grad}}  {\partial \omega_{\alpha i}}=-\rho_{\alpha i,\beta j}\omega_{\beta j} 
\,.
\label{SpinCurrentGeneral}
\end{equation}
This spin current averaged over the fast precession determines the parameters of the phenomenological equations \eqref{SpinCurrent1} and \eqref{SpinCurrent2} for the spin supecurrent emerging in magnon BEC.

\subsection{Spin-Hall effect in $^3$He-B}
\label{SpinHall}

The symmetry properties of the spin superfluid velocity, and thus of the spin supercurrent, allows to couple linearly the spin current with electric field even in the absence of the spin-orbital  interaction. The following term in the action is possible  \cite{MineevVolovik1992}:
\begin{equation}
 F=-\beta  e_{ijk}\omega_{\alpha i}R_{\alpha j}E_k \,.
\label{ElectricFiledVelocityCoupling}
\end{equation}
The parameter $\beta$ is not well defined from the microscopic theory due to the unknown Fermi-liquid corrections involved. The estimate for $\beta$ reported in Ref. \cite{Volovik1984}  is $\beta\sim 10^{-4}$e/cm.
Variation with respect to $\omega_{\alpha i}$ demonstrates that there exists a  linear response of the spin supercurrent on the electric field:
\begin{equation}
J_{\alpha i}=-\frac{\partial F}{\partial \omega_{\alpha i}}=\beta  e_{i j k}R_{\alpha j} E_k  \,.
\label{ElectricFiledVelocityCoupling2}
\end{equation}
This spin current is transverse to the electric field and thus represents the spin current Hall effect.
As distinct from the spin Hall effect predicted by Dyakonov and Perel \cite{DyakonovPerel1971},
this spin-Hall effect occurs in the absence of spin-orbit interaction. The bridge, which connects spin and orbital motion, is provided by the order parameter matrix $R_{\alpha j}$.

\subsection{Electric and magnetic fields as $SU(2)$ gauge fields }

The interaction of the electric and magnetic fields with the order parameter in superfluid phases of $^3$He, can be also found using observation  that ${\bf H}$ and ${\bf E}$ may be considered as temporal and spatial components of the $SU(2)$ gauge field, where   $SU(2)$ is the group of the spin rotations. The  auxiliary
$SU(2)$ gauge field $A_{\mu}^\alpha$ is convenient for the description of the
  effects related to the spin current. In the  spinor representation
  one has the following covariant derivatives coming from the auxiliary $SU(2)$ gauge field
  \cite{MineevVolovik1992,Leurs-Zaanen2008,Berche2012}
\begin{equation}
{\bf D}=\boldsymbol{\nabla} -i{\bf A}^i\frac{\sigma^i}{2}~~,~~D_0=\partial_t +i A_0^i \frac{\sigma^i}{2}\,.
\label{Gauge2} 
\end{equation} 
Some components of the field $A_{\mu}^\alpha$ are physical, being represented
by the real physical quantities which couple to the fermionic charges.
Example is provided by the Pauli magnetic field $H^i$,
which play the role of the component $A_{0}^i$ of the $SU(2)$ gauge field, see \eqref{auxiliary_field}
  \cite{MisirpashaevVolovik1992,MisirpashaevVolovik1992b}, while the spatial components are played by the electric  field which enters the gradient energy via the  covariant spatial derivative 
  \cite{MineevVolovik1992,Goldhaber1989,Leurs-Zaanen2008,Berche2012}:
\begin{equation}
A_0^i =B^i~~,~~A_j^i= e_{jik}E_k\,.
\label{Gauge3} 
\end{equation} 
The electric field enters due to the relativistic spin-orbit interaction of the spin of $^3$He atom with the electric field ${\bf E}$.

The spin current is obtained as variation of the action with respect to the fictituous   $SU(2)$   gauge field: 
\begin{equation}
{\bf J}^i= \frac{\delta S} {\delta {\bf A}^i}\,.
\label{SpinCurrent}
\end{equation}
After the spin current is calculated the values of the
auxiliary fields are made equal to zero or to the  values of the corresponding physical fields which simulate the gauge fields.  

For example, in the presence of electric field, equation \eqref{GradientEnergy} becomes
\begin{equation}
F_{\rm grad}= \frac{1}{2}\rho_{\alpha i,\beta j}(\omega_{\alpha i}-\frac{\gamma} {c}e_{\alpha ik }E_k)(\omega_{\beta j}-\frac{\gamma} {c}e_{\beta jl }E_l)
\,,
\label{extended_gradient} 
\end{equation} 
which demonstrates that electric field enters as the $SU(2)$ gauge field  forming the covariant derivative. This gives another response of the spin cupercurrent on the external electric field:
\begin{equation}
J_{\alpha i}=-\frac{\partial F_{\rm grad}}  {\partial \omega_{\alpha i}}=-\rho_{\alpha i,\beta j}(\omega_{\beta j}-\frac{\gamma} {c}e_{\beta jk }E_k)
\,,
\label{SpinHallEffect} 
\end{equation} 
As distinct from \eqref{ElectricFiledVelocityCoupling2} this spin-Hall effect is governed by the spin-orbit interaction. According to Ref. \cite{72} both spin Hall effects in $^3$He-B should modify the spin current Josephson effect in magnon BEC.  The  supercurrent, induced by electric field,  leads to an additional phase shift  proportional to electric field, which is to be measured.

\subsection{Quantum spin Hall effect}
\label{QSHE}

There are several types of responses of spin and electric currents to transverse forces
which are quantized in 2+1 systems under appropriate conditions. The most familiar is the conventional quantum Hall effect (QHE).  It is quantized response of the particle  current  to the transverse force, say to transverse gradient of chemical potential, ${\bf J}= \sigma_{xy} \hat{\bf z}\times \boldsymbol{\nabla} \mu$. In the electrically charged  systems this is the    
quantized response of the electric current ${\bf J}^e$ to transverse electric field 
${\bf J}^e= e^2\sigma_{xy} \hat{\bf z}\times {\bf E}$.

The other effects involve the spin degrees of freedom. 
An example is the mixed spin quantum Hall effect: quantized response of 
the particle current ${\bf J}$ (or electric current ${\bf J}^e$) to transverse gradient  of magnetic field interacting with Pauli spins (Pauli field in short) \cite{VolovikYakovenko1989,Volovik1992}:
\begin{equation}
{\bf J} = \sigma^{\rm mixed}_{xy} \hat{\bf z}\times \boldsymbol{\nabla} (\gamma H^z) ~~, ~~{\bf J}^e =e{\bf J} \,.
\label{Current_vs_field_gradient}
\end{equation}
The related effect, which is determined by the same quantized parameter  $\sigma^{\rm mixed}_{xy}$, is the quantized response of the spin current, say the current ${\bf J}^z$  of the $z$ component of spin, to  the gradient of chemical potential   \cite{SQHE}.
 In the electrically charged  systems this corresponds  to the   quantized response of the spin current
to transverse electric field:
\begin{equation}
{\bf J}^z = \sigma^{\rm mixed}_{xy} \hat{\bf z}\times \boldsymbol{\nabla} \mu  =e \sigma^{\rm mixed}_{xy} \hat{\bf z}\times {\bf E}  \,.
\label{Spin_current_vs_field}
\end{equation}
  This kind of mixed Hall effect is now used in spintronics \cite{Awschalom2009},
 which exploits the coupling between spin and charge transport in condensed matter.   
  
  Ffinally there is a pure spin Hall effect --  the quantized response of the spin
current to transverse gradient of magnetic field \cite{VolovikYakovenko1989,Volovik1992,HaldaneArovas1995,Senthil1999}:
\begin{equation}
{\bf J}^z = \sigma^{\rm spin/spin}_{xy} \hat{\bf z}\times \boldsymbol{\nabla} (\gamma H^z)  \,.
\label{Spin_current_vs_field_gradient}
\end{equation}  

Let us consider the mixed spin Hall effects in \eqref{Spin_current_vs_field} and 
\eqref{Spin_current_vs_field_gradient}. These two effects are related, since they are described by the same topological Chern-Simons action \cite{VolovikYakovenko1989} and thus by the same parameter $\sigma^{\rm mixed}_{xy}$.
To see this, let us remind that the spin current is obtained as variation of the action over the fictituous   $SU(2)$ or $SO(3)$ gauge field, see \eqref{SpinCurrent}. For example, the current of the $z$-projection of spin is
\begin{equation}
{\bf J}^z= \frac{\delta S} {\delta {\bf A}^z}\,.
\label{SpinCurrent_z}
\end{equation}
The corresponding Chern-Simons term in the action is given by \cite{VolovikYakovenko1989} 
\begin{equation}
F_{\rm CS} =   e\sigma^{\rm mixed}_{xy} e^{\nu\alpha\beta}\int
d^2xdt A_\nu^z \nabla_\alpha A_\beta \,,
\label{ChernSimons}
\end{equation}
where $A_\beta$ is the vector potential of the conventional electromagnetic field, and $ A_\nu^z=(A_i^z, A_0^z)$ represent components of  auxiliary (fictituous) $SU(2)$ gauge
field. Variation of the action with respect to the field $A_i^z$ gives the spin current in 
\eqref{Spin_current_vs_field}. On the other hand, the variation of the action with respect to field $A_i$ gives electric current in terms of the gradients of an  auxiliary gauge field. However, from equation 
\eqref{auxiliary_field} or \eqref{Gauge3} it follows that the role of the auxiliary gauge  field $A_0^z$ is played by magnetic field $H^z$. As a result one obtains equation  
\eqref{Current_vs_field_gradient} for electric current. 

Equation \eqref{ChernSimons} has been originally introduced for a thin film of the so-called planar phase of superfluid $^3$He  \cite{VolovikYakovenko1989}. However, it is better suited for the two-dimensional topological insulators with time reversal invariance (on topological insulators see reviews \cite{HasanKane2010,Xiao-LiangQi2011}). These materials have the same topological structure as the planar phase, which is also time reversal invariant, but the advantage of these materials is that they are insulating and thus the superconductivity does not mask the spin Hall quantization. 

 Discussion of the mixed
Chern-Simons term can be found in Ref.  \cite{MutualCS}. For the related
phenomenon of axial anomaly in particle physics, the mixed action in terms of different
(real and fictituous) gauge fields has been introduced in Ref.
\cite{Zhitnitsky}.

\section{Conclusion}

The superfluid phases of liquid $^3$He at extreme low temperatures are unique states of condensed matter with physical properties which can be compared to the vacuum of relativistic quantum field theories (see Chapter "The Superfluid Universe" in this book). The  A phase ($^3$He-A) belongs to the same symmetry and topology class as the vacuum of the Standard Model of particle physics in its massless (\textit{i.e} gapless)  phase and can also be described as a semi-metal-like system with non-trivial topology. The B phase ($^3$He-B), in contrast,  is similar to the vacuum of the Standard Model in its massive (or gapped) phase and to 3-dimensional topological insulators with time reversal symmetry. In addition to fermionic quasiparticle excitations, superfluid $^3$He has also bosonic quasiparticles, such as magnons  --  quanta of excitations of  the magnetic  subsystem.  These magnon excitations can form long-lived Bose-Einstein condensates both in $^3$He-A and $^3$He-B, and these condensates experience their own superfluidity, which is not related
to superfluidity of the underlying system.

Formally, the phenomenon of superfluidity requires the conservation of charge  or particle number.
However, the consideration can be extended to systems with a weakly violated conservation law, including a system
of sufficiently long-lived quasiparticles - discrete quanta of energy that can be treated as real particles in condensed matter.  The spin superfluidity -- superfluidity in the magnetic 
subsystem of a condensed matter -- is manifested as the spontaneous phase-coherent precession of spins first discovered in 1984 \cite{HPDexp,HPDtheory}.  This superfluid current of spins is one more representative of
superfluid currents known or discussed in other systems, such as 
the superfluid current of mass and atoms in superfluid
$^4$He; superfluid current of electric charge in superconductors;
superfluid current of hypercharge in Standard Model; superfluid
baryonic current and current of chiral charge in quark
matter; etc. The analogy of the dynamical superfluid state of coherent precession with the non-perturbative dynamics of the physical vacuum has been discussed in \cite{Klinkhamer2012}.

Different condensates and thus different states of magnon superfluidity have been created by choosing different experimental arrangements.   At low temperatures the condensate is confined in a magnetic trap which is formed by the order parameter texture of the superfluid state. This produces the analog of atomic BEC in laser traps, but adds some new features, such as formation of the non-ground-state condensate; self-localization and formation of the multi-boson bubble which is analog of the MIT bag model of hadrons;  nonzero mass of the Goldstone bosons, etc. The magnon condensates can be used to probe the quantum vacuum of $^3$He in the limit $\mathbf{\mathit{T}} \rightarrow 0$, where conventional measuring signals become insensitive.

\vspace{2mm}

\textbf{Acknowledgements} 

This work is supported in part by the Academy of Finland, Centers of
excellence program 2006-2011, by the EU's 7th Framework Programme (FP7/2007-2013: grant agreement
\# 228464 MICROKELVIN), and by the collaboration between CNRS and Russian Academy of Sciences (project
\# N16569).

\end{document}